# Design and Test of a
# Carbon-Tolerant Alkaline Fuel Cell

Prof. Mirna Urquidi-Macdonald, Penn State University
Prof. Ayusman Sen, Penn State University
Dr. Patrick Grimes, Grimes Associates
Ashutosh Tewari, Penn State University
Varun Sambhy, Penn State University

This report consists of three major parts, in the order: (I) a foreword from the sponsor; (II) Issues and Approaches; and (III) Findings, Discussion and Recommendations.

## I. Foreword: Sponsor's View of the Importance of This Work

By Dr. Paul J. Werbos, NSF, pwerbos@nsf.gov

The lab-scale tests reported here appear to represent an essential breakthrough in the global effort to protect energy security and reduce $CO_2$ emissions soon enough to matter. Of course, these tests are only a beginning. In a rational world, we would immediately organize and fund new efforts to maximize the probability that this technology actually reaches the mass market for automobiles, at the soonest possible time, on a competitive free-market basis. By replacing the small internal combustion engine in advanced hybrid cars like the Toyota Harrier with this kind of fuel cell, we could hope to raise the efficiency of that part of the system from something like 30 percent or more to something like 60 percent or more, and vastly improve overall fuel use and emissions.

This paper was submitted as the final report to a Small Grant for Exploratory Research (SGER), Number ECS-0342586, funded through myself and Dr. James A. Momoh of the Electrical and Communication Systems (ECS) Division of NSF. The authors have given me permission to make this report public. The views in this foreword are my own personal views, and do not necessarily reflect the views of the authors or the official views of NSF. Likewise, the paper itself does not necessarily reflect the official views of NSF or myself. I have not asked to edit it.

The carbon-tolerant alkaline fuel cell could be used *either* with hydrogen and air, *or* with reformed methanol and air, as the authors stress. However, Momoh and I have become more and more worried about the rapid growth in world dependence on oil from the Middle East, compared with the very very slow progress in transition to a truly pure hydrogen economy. The leaders of the US clearly agree that the problems involving oil and violence in the Middle East today demand the highest possible level of attention and effort, as does the spread of nuclear technology (even as associated with nuclear power plants) in the developing world. Rising gasoline prices are already high on the list of issues which concern the US public – and their fears about fuel supply have already cut deeply into revenues from the sale of cars and trucks. Yet when we look 20 to 40 years into the future, these problems appear to be on a path to becoming ten times as serious as they are today. They may even combine in a way that threatens our very survival. There is no *certain* way to reverse these trends, but Momoh and I believe that new breakthroughs in technology – like the one reported here – do offer serious hope of reversing the trends within 20 to 40 years (hopefully 20!) if we follow up as hard as we can. The breakthrough reported in this paper is not enough by itself to do the job – but it may well be *one* of the essential steps on the path.

For a more complete analysis of how to reverse these trends, see: www.ieeeusa.org/policy/energy_startegy.ppt.

Crudely speaking, most energy policy work (like the development of the Kyoto Treaty or the current Energy Bill) has focused on the question: what can we do *now*, that we are totally sure of, that can give *immediate* results in reducing oil dependency or emissions? However, the best that we can do *now* – as reflected in the Kyoto Treaty itself – adds up to something much smaller than the size of the problem. A 15% reduction does not change much. Conversely, others have asked how we can envision the perfectly $CO_2$-free renewable sustainable society of the future. For some researchers, this has led to a vision of a pure hydrogen economy, although there are other alternatives for a sustainable future. (Again, see our slides.) Momoh and I have focused on a third question: how can we *minimize* the expected delay between now and the time when we reduce oil dependency or $CO_2$ emissions by a large enough factor to save us (a factor of three or four or more)? That is the basis for the energy strategy we have presented through IEEE-USA.

As part of that larger question, we may ask: how can we minimize the expected delay between now and the time when millions of consumers buy these new fuel cell cars? Certainly we have to work on the technology itself, and hedge our bets with regard to the risks involved. But that is not enough. We do not expect that people will buy huge numbers of fuel cell cars *until and unless* the fuel for these cars is available in 10-20% of the gas stations in the region of the US where they live. (This estimate of 10-20% comes from studies reported by Roberta Nichols of Ford and others.) Thus to reach our goal for these cars, we must *both* develop the technology *and* reach this threshold of fuel availability, as soon as possible.

In my view, the fastest way to do this is by focusing on the *methanol* option. Congress (or other regions?) could pass a fuel competition law, requiring that any new car which can use gasoline must also be able to use ethanol or methanol, starting 2-4 years from now. This kind of retooling would be relatively inexpensive, because fuel flexibility is a well-established technology. (Also, there are some new technologies which can reduce the cost of flexibility.) Such a law would unleash the power of market competition between these fuels and between the many sources of such fuels. Methanol, in particular, can be made economically from remote natural gas, from "Clean Coal," and from biological sources. Based on current technology, there is good reason to believe that the free market would supply enough methanol to solve our "chicken and egg" problem at the soonest possible time, without subsidies. Auto companies would benefit, on balance, because reduced insecurity about fuel would have a bigger effect on car sales than the small market-wide effect on the cost of making cars. Oil companies would benefit, because their proved reserves of car fuel would suddenly be enlarged by their methanol production capacity, which they could expand relatively quickly. Again, see the slides and citations in the slides for more analysis.

In the past, some have argued that we should use gasoline itself as a hydrogen carrier. But this idea has not worked out very well, for fundamental thermodynamic reasons, and DOE has recently cancelled research efforts to try to process gasoline to hydrogen on-board a car. By contrast, methanol is an excellent hydrogen carrier, and small-scale efficient steam reformers for methanol have been around for decades. When *intelligent control* is used to integrate such reformers into a larger fuel cell system in an optimal way, the overall system may actually be more efficient than a hydrogen-carrying fuel cell car, if we account for losses in hydrogen storage systems. Nevertheless, this also requires a greater development and use of the kind of intelligent control which really does provide optimization over time in noisy, nonlinear systems; see the *Handbook of Learning and Approximate Dynamic Programming*, Si et al eds, IEEE Press/Wiley, 2004, for more information. It is interesting to consider how that new technology might also be attempted for enhanced oil recovery or hydrate extraction.

      Again, these new efforts and technologies are essential parts of a larger strategy – but they are not enough by themselves to do the whole job. See the slides for other aspects.

# II. Issues and Approaches

# Chapter 1. Introduction

There is considerable interest in the U.S. in becoming a hydrocarbon-independent society. The reasons include: minimizing the production of contaminants to protect the environment, moving to non-OPEC-controlled alternate energy sources, becoming less dependent on foreign sources of energy (oil), and shifting our dependence to "in house" resources such as gas.

As cited by the Intergovernmental Panel on Climate Change (IPCC, Houghton, et al., 1996), "From the point of view of global warming, $CO_2$ is certainly a threat." Although $CO_2$ is the most important greenhouse gas and is the largest emission, the greenhouse effect is a combination of $CO_2$, $CH_4$, and $N_2O$ emissions. The capacity of $CH_4$ and $N_2O$ to contribute to the warming of the atmosphere is 21 and 310 times higher than $CO_2$, respectively.

The exploration of alternate energy sources is the object of increased research. Among the most popular new energy sources are hydrogen and methanol. Hydrogen production may be cheaper than methanol, but both cost about the same to deliver. However, hydrogen is more expensive to store than methanol.

The term "pollution free" is closely related to the fuel used by the fuel cell. The most common "pollution free" fuel used is hydrogen. Among the "pollution free" or "zero emission" fuel cells, the largest research funding has been invested in solving the scientific and engineering problems that impede the wide implementation of PEMFC technologies. These problems can be summarized as: **(a)** designing and producing electrode structures that can produce useful power under near ambient conditions with minimal catalyst loading or with cheap catalysts, and **(b)** immunity inhibition of the anodic electrochemical reactions due to minor contaminants in the fuel stream (e.g., CO, $H_2S$).

## Hydrogen Fuel

It is important to clarify that the "emission free" $H_2$-based fuel cells will not produce $CO_2$, nor $SO_x$ or $N_xO$, emissions where the hydrogen fuel cells are operated. However, hydrogen molecules ($H_2$) do not exist in abundance as a pure molecule on earth. Therefore, hydrogen must be produced other ways: from hydrocarbons, photosynthetic purple non-sulphur bacteria, water pyrolisis, steam reforming, hydrocarbon partial oxidation, or by using different sources of energy, such as plasma or nuclear sources, to electrolyze or decompose water. For example, hydrogen is now made from natural gas (methane), petroleum, coal, various chemical reactions, and biomass (landfill waste, wastewater sludge, and livestock waste). When hydrogen is produced from methane ($CH_4$), $CO_2$ and $CH_4$ are emitted into the air. Although the amount of $CH_4$ emissions is considerably less than the $CO_2$ emissions on a weight basis (10.621 kg of $CO_2$/kg-of-$H_2$-produced versus 0.060 kg of $CH_4$/kg-of-$H_2$-produced), the $CH_4$ has a 21 times higher heat capability than $CO_2$. $CH_4$ emissions from $H_2$ production account for 10.6% of the total greenhouse effect.

Not all methods for producing $H_2$ are damaging to the environment. This is usually true if the hydrogen comes from hydrocarbons. However, until now there was not an existing economically viable method to cleanly and economically produce large volumes of $H_2$. Accordingly, by now, $H_2/O_2$ fuel cells will not have an overall effect of reducing the greenhouse effect or soon becoming economically attractive.

**Methanol Fuel**

The United States provides almost one quarter of the world's supply of methanol. "In 2001, methanol production capacity from U.S. plants totaled over 1.5 billion gallons" (Methanol Institute and the Methanex Corporation). U.S. plants meet about one-half of the U.S. methanol demand, with the remaining supply imported from Trinidad, Chile, Venezuela, and Canada.

The largest market in the U.S. for methanol is the production of methyl tertiary butyl ether, or MTBE. Over 3 million tons of methanol, or 35% of methanol consumed in the U.S., goes into the creation of MTBE. MTBE is blended in clean, reformulated and oxygenated gasoline, serving one-third of the U.S. gasoline market. Accordingly, the need to fabricate MTBE would be considerable reduced if methanol were used as a fuel instead of gasoline.

Methanol can be produced from either hydrocarbons or organic material fermentation. While production through hydrocarbons does not lead to a hydrocarbon independent society or an environmentally friendly society, organic material fermentation does. The $CO_2$ produced through organic material fermentation would be current $CO_2$ contamination and not "fossil" contamination. Trees naturally produce $CO_2$. Cutting trees to transform them into fuel would produce the same amount of $CO_2$ that the tree would produce were it to remain alive. We call this "present contamination" as it does not represent a threat to nature. On the other hand, oil represents "fossil" production of $CO_2$ and release of that "fossil" $CO_2$ in present time greatly impacts the environment.

# Fuel Cells-A Short Review

Several types of Fuel Cells have been developed throughout the years. The original one was an Alkaline Fuel Cells (AFC), followed by the Solid Oxide Fuel Cells (SOFC) and the Proton Exchange Membrane Fuel Cells (PEMFCs). The name of the fuel cell is usually originated by the type of electrolyte used. The cells, in general, have some similar features: **(a)** the electrodes are mostly porous gas diffusion electrodes, **(b)** a metal catalyst (Ni, Pt) is needed at the electrode's site, **(c)** the oxygen and fuel reactions happen at the electrodes, and **(d)** the catalysts have to be in contact with both the electronic and ionic (electrolyte) conductors. The fuel is oxidized at the anode site while the oxygen is reduced at the cathode site.

## H₂-Fuel cells (PEMFC and AFC)

Alkaline fuel cells (AFC) are an attractive, non-conventional source of energy. AFC offer several advantages over the more commonly used and researched PEMFC. The kinetics of the electrode reactions are superior in an alkaline environment (AFC) compared to acidic environment (PEMFC) [1-4]. The inherently faster kinetics of the reactions in an alkaline fuel cell allows the use of non-noble, metal, electro-catalysts, like nickel, silver etc. [5-8]. AFCs also exhibit much higher current densities and electrochemical efficiencies at comparable temperatures over PEMFC. In addition, the liquid KOH electrolyte used in AFCs is much cheaper than the polymer electrolyte (Nafion) used in PEMFCs, which needs constant humidification for proper functioning [9-11]. Water management is also not a major issue with AFCs as with PEMFCs, thus allowing simplicity in design and fabrication. AFCs can be operated at a higher temperature (100°C -120°C), thereby using the Arrhenius effect to the advantage and, hence, obtaining higher efficiencies. PEMFCs cannot be operated above 90°C due to problems with the hydration of the nafion membrane. Moreover, AFC electrodes are stable and not prone to the poisoning caused by carbon monoxide (CO) which poisons the platinum catalyst of the PEMFCs [12]. Therefore, considering the cost and the simplicity of operation, AFCs are more advantageous as compared to PEMFCs and have better prospects in the commercialization of fuel cells.

The reactions that take place at the electrodes when $H_2$ is used as a fuel can be summarized as follows:

**Cathode:**

$$O_2 (g) + 2H_2O (l) + 4e^- \Leftrightarrow 4OH^- (aq) \quad E° = 0.40 \text{ V}$$

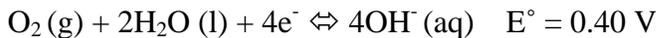

**Anode:**

$$H_2 (g) + 2OH^- (aq) \Leftrightarrow 2H_2O (l) + 2e^- \quad E° = 0.83 \text{ V}$$

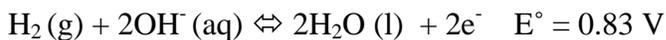

The by-product water and heat have to be removed. This is usually achieved by re-circulating the electrolyte and using it as the coolant liquid.

Table 1.1 shows a comparison in cost and performance between AFCs and PEMFCs, without considering the government and industry research investments. Two main attractions that the AFC presents over the PEMFC are low cost and robustness of the anode to poisoning by impurities.

**Table 1.1:** Performance comparison between AFC v.s. PEMFC [12]

|  | **Alkaline Fuel Cell** | **PEM Fuel Cell** |
|---|---|---|
| **Current Density** | 300-600 mA/cm2 | 600 mA/cm2 |
| **Voltage per cell** | 0.8-0.85 V | 0.65-0.70V |
| **Power Density** | 0.17 KW/Kg | 0.25 KW/Kg |
| **Lifetime** | 4000 hours | ≈5000 hours |
| **Cost per KW** | $ 100 – 150 | $ 500 – 1000 |
| **Catalyst** <br> Anode <br> Cathode |  <br> Nickel <br> Nickel / Silver |  <br> Platinum <br> Platinum |
| **Maximum fuel impuretés** <br> Carbon monoxyde <br> Carbon dioxyde <br> Ammonia |  <br> Not Critical <br> < 100 ppm <br> Not Critical |  <br> < 10 ppm <br> Not Critical <br> Very Critical |
| **Power Density** | ≈ 2 KW/L | ≈ 1.75 KW/L |

## Direct Methanol Fuel Cells (DMFC)

These cells are similar to the PEM cells in the sense that they both use a polymer membrane as the electrolyte. However, in DMFCs the fuel used for oxidation at the anode is methanol instead of hydrogen. Efficiencies of about 40% are expected with this type of fuel cell, which would typically operate at a temperature between 50-100°C. This is a relatively low temperature range, making this fuel cell attractive for tiny to mid-sized applications, such as powering cellular phones and laptops. Higher efficiencies are achieved at higher temperatures. This type of cell produces $CO_2$ as a sub-product.

DMFC's main drawbacks are 1) its high cost due to the expensive catalyst used because they operate in an acidic medium, 2) it is bulky due to the external recirculation pumps and 3) it has a low efficiency when compared to thermal engines. Figure 1.1 shows a schematic representation of this type of fuel cell.

The reactions that take place at the electrodes can be summarized as follows:

**Anode:**

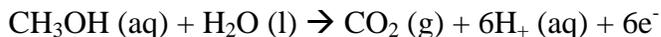

$CH_3OH$ (aq) + $H_2O$ (l) → $CO_2$ (g) + $6H_+$ (aq) + 6e$^-$

**Cathode:**

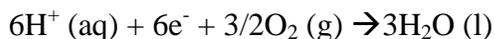

$6H^+$ (aq) + 6e$^-$ + $3/2O_2$ (g) → $3H_2O$ (l)

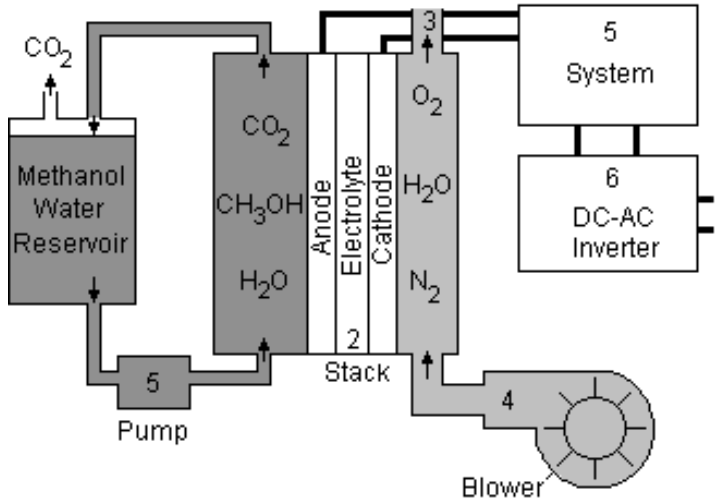

**Figure 1.1:** Schematic representation of the direct methanol fuel cell [13].

## Alkaline Fuel Cells fueled with Methanol

Alkaline fuel cells use fuel, an oxidant and a liquid alkaline electrolyte, usually KOH mixed with water. When methanol is used as the fuel instead of hydrogen, the cell becomes an alkaline fuel cell fueled with methanol. The overall reactions are similar to DMFCs.

This type of cell inherits some of the advantages and disadvantage of the hydrogen fueled alkaline fuel cells and the direct methanol fuel cells. For example, some disadvantages might be cross over problems similar to DMFCs, being heavier and bulkier than PEMFCs, having lower power densities than PEMFCs or AFCs fueled with $H_2$. An advantage might be operating with cheaper catalysts than DMFCs or PEMFCs.

The most important issue to study in this cell is the $CO_2$ poisoning of the alkaline electrolyte. That study may be completely focused on the methanol oxidation in an alkaline media. Over the past few years there has been a resurgence of interest in methanol oxidation. The recent literature concerns the adsorption and oxidation processes occurring at the catalyst surface.

## Issues in Fuel Cells (PEMFCs and AFCs)

**PEMFC:** In general, the $H_2/O_2$ reactions in the fuel cells call for large electrode areas, high temperatures, and expensive catalysts when acidic media is used (polymer membranes). The more acidic the electrolyte and the closer to room temperature the fuel cell operates, the stronger the need to use an expensive catalyst (example platinum, ruthenium) and a highly pure hydrogen fuel and oxygen.

The main problem with PEMFCs is the need to work with very expensive, hard to find catalysts in order to increase their efficiency. Those catalysts are also easily poisoned with minor elements that contaminate the fuel or the oxidant. Moreover the polymer conducting membrane, i.e. nafion, is very expensive and is hard to work with and engineer. The polymeric membrane requires highly controlled conditions in temperature and humidity for optimum performance. Also, PEMFCs cannot operate at temperatures greater than 80°C due to dehydration of the nafion membrane.

**AFC:** AFCs are inherently plagued by the problem of carbon dioxide poisoning, which limits their use as air-breathing energy sources. The poisoning reaction depletes the alkaline KOH electrolyte directly by the following reaction [10-12]:

$$CO_2 + 2KOH\ (aq.) \rightarrow K_2CO_3\ (aq.) + H_2O$$

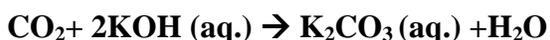

This reaction has the effect of reducing the number of hydroxyl ions available for reaction at the electrodes. This conversion from hydroxide to carbonate also reduces the ionic conductivity of the electrolyte solution. In a very concentrated electrolyte solution, it may also have the effect of blocking the pores of the gas diffusion layer (GDL) by the precipitation of $K_2CO_3$ salt [14]. However, Gülzow et al. reported that, although carbon dioxide poisoning decreases an AFC performance, it does not cause any degradation of the electrodes [15]. Even after thousands of hours of operation in a carbon dioxide rich atmosphere, no additional electrode degradation, like deposition of $K_2CO_3$, was observed. Al Saleh et al. also showed the similar affect of carbon dioxide on a fuel cell performance [16]. It was observed that concentrations of up to 1% carbon dioxide in the oxidant stream of Ag/PTFE electrodes did not affect the cell performance over a period of 200 hours.

Therefore, the most probable reason for the decrease in the cell performance is the change in electrolyte composition. Conversion of the electrolyte, from KOH to $K_2CO_3$, by the absorption of carbon dioxide slows down the rate of oxidation of fuel at the anode [17-18]. The sluggish anode kinetics decreases the performance of a fuel cell as a whole unless the electrolyte is being circulated continuously. In addition, decreased electrolyte conductivity also increases the ohmic polarization leading to lower cell efficiency. Although the harmful effects of carbon dioxide poisoning can be partly reduced by circulating the electrolyte as discussed by Cifrain et al. [19], a permanent solution is sought after to increase the possibility of AFC commercialization.

A lot of effort is being made to rectify the problem of carbon dioxide poisoning in alkaline fuel cells. Molecular sieves and polymeric membranes are being developed (see NIST, NSF, Air Products, etc.) and discussed for $CO_2$ separation from air [20]. However, at the present time, those membranes require large areas of the sieves or membranes and low gas velocity to remove $CO_2$ to ~10ppm.

Another alternative proposed in the literature is $CO_2$ management, which involves the synergistic possibility of using liquid hydrogen to condense the carbon dioxide out of the air. Ahuja and Green [21, 22] discussed this at length and developed a model of the heat exchanger required for this. This solution presents low parasitic energy consumption, but the condensation and re-vaporization of water and $CO_2$ makes the structure complicated. Fyke [23] discussed the possibility of using a solid ionomer alkaline membrane that would enable a cell to run without the possibility of carbon dioxide poisoning, as there would be no free potassium cations to which the carbonate anions could attach. This is an intriguing concept but no progress in solid ionomer alkaline membranes have been reported since Swette et al. [24] discussed this possibility for regenerative fuel cells.

Kordesch and Gunter [25] mentioned that the "Removal of the 0.03% carbon dioxide from the air can be accomplished by chemical absorption in a tower filled with "soda lime". One kilogram of soda lime has the ability to clean 1000 $m^3$ of air from 0.03% to 0.001% $CO_2$. The only method commercially-employed, at the moment, to alleviate the oxidant side carbon dioxide poisoning is $CO_2$ scrubbing using soda lime. Technically, this system works, but is not a strong candidate for commercial systems. This suggests that significant benefits could be obtained from the use of other scrubbing techniques. However, among the present $CO_2$ removal methods, the soda lime $CO_2$ scrubber is the only practical solution for removing $CO_2$ from the cathode air stream, but it is cumbersome due to the periodic maintenance and operational costs over the lifetime of the fuel cells. This may not be suitable for high power applications, as it will require frequent replacement. One can also reduce the carbonate poisoning problem by recirculation of the electrolyte [26].

Another possible remedy of carbon dioxide poisoning is the use of a three-phased fluidized electrode. This electrode has a completely different structure compared with a conventional porous gas-diffusion electrode. It increases the $CO_2$ tolerance of a fuel cell by eliminating the structure of the porous gas diffusion electrode. Also, this electrode can increase limiting current density and decrease concentration polarization due to its large electrode area and high rates of mass transfer to the surface of the electrode particles by the gas bubbling and flow of electrolyte. However, the research results are not impressive, mainly due to the poor contact between the current collector and the electrode particles.

## $CO_2$ Poisoning of AFC

The electrolyte composition changes when $CO_2$ is carried along with the oxygen through the air cathode. The $CO_2$ dissolves in water.

$$CO_2 \text{ (in air)} \Leftrightarrow CO_2 \text{ (dissolved in water)} \quad \text{R1}$$

The solubility of $CO_2$ = 0.76 lt/lt-of-$H_2O$ at 25°C and 1 atmosphere $CO_2$ or 0.76/24.5 = 0.031 mole/liter; where 24.5 liter = volume occupied by gas at 25°C and 1 atmosphere. The dissolved $CO_2$ reacts with water to form carbonic acid.

$$CO_2\text{(in air)} + H_2O \Leftrightarrow H_2CO_3 \text{ (carbonic acid)} \quad \text{R2}$$

$$K = \frac{[H_2CO_3]}{[CO_2][H_2O]} = \frac{0.031}{1} = 10^{-1.5}$$

The H$_2$CO$_3$ undergoes slight dissociation. H$_2$CO$_3$ dissociates stepwise:

$$H_2CO_3 \Leftrightarrow H^+ + HCO_3^- \qquad K_1 = 10^{-6.4} \qquad \text{R4}$$

$$HCO_3^- \Leftrightarrow H^+ + CO_3^{2-} \qquad K_2 = 10^{-10.3} \qquad \text{R5}$$

$$H_2O \Leftrightarrow H^+ + OH^- \qquad K_w = [H^+][OH^-] = 10^{-14} \qquad \text{R6}$$

If an alkaline solution is prepared by adding KOH to the water, H$_2$CO$_3$ and KOH reacts and comes in equilibrium with K$_2$CO$_3$ and water, as follows:

$$H_2CO_3 + KOH \Leftrightarrow K_2CO_3 + 2H_2O \qquad \text{R7}$$

If pH increases, the reaction shifts to the right; while pH decreases, the reaction shifts to the left. K$_2$CO$_3$ can be presented as a salt or a solution in the electrolyte (dissociated). The reaction R7 can be simplified as:

$$CO_3^{2-} + H_2O \Leftrightarrow HCO_3^- + OH^- \qquad \text{R8}$$

because the potassium is abundant and does not change the overall results. Even though reaction R4 uses up most of the H$^+$, enough is grabbed by CO$_3^{2-}$ in reaction R5 to make the solution distinctly basic.

$$K = \frac{[HCO_3^-][OH^-][H^=]}{[CO_3^{2-}][H^+]} = \frac{[HCO_3^-]}{[CO_3^{2-}][H^+]} \times [OH^-][H^+] \qquad \text{R9}$$

$$K_{hyd} = \frac{K_{water}}{K_{diss}} = \frac{10^{-14}}{10^{-10.3}} = 10^{-3.7}$$

<div style="text-align:right">R10</div>

The equilibrium between the ions in the solution is dictated by pH and temperature. Reaction R2 tends to drive the potential to more acidic pHs, while reaction R5 tends to drive the pH to basic pHs. If the pH decreases, reactions R4 and R5 shift to the left, increasing the concentration of H$_2$CO$_3$ (reaction R5). If the pH increases, reaction R8 is driven to the left and the concentration of CO$_3^{2-}$ increases. Figure 1.2 shows a balance between the H$_2$CO$_3$, HCO$_3^-$, and CO$_3^{2-}$.

Accordingly, when KOH is dissolved in water to form an alkaline electrolyte and methanol is add to the electrolyte to be used as a fuel, if CO$_2$ comes in contact with the electrolyte, potassium carbonate (K$_2$CO$_3$) is formed and the precipitation of the potassium in the cells will depend on the relative concentrations of the electrolyte composition (the water, the methanol and the formed potassium carbonate) that dictates the pH and the temperature. From reactions R1, R2, and R7, it is evident that the concentration of $H_2CO_3$ (R2) is very important to determine the concentration of $K_2CO_3$. We can immediately say, by looking at Figure 1.2, that $H_2CO_3$ is the dominant species in all solutions with pH < 6.4 or $H^+$ > 6.4, $HCO_3^-$ is dominant in the pH range 6.4-10.3, and $CO_3^{2-}$ is dominant at a pH above 10.3. From the figure, we can conclude that if the concentration of KOH increases (higher pH), the $K_2CO_3$ will increase (Reaction 7 will shift to the right).

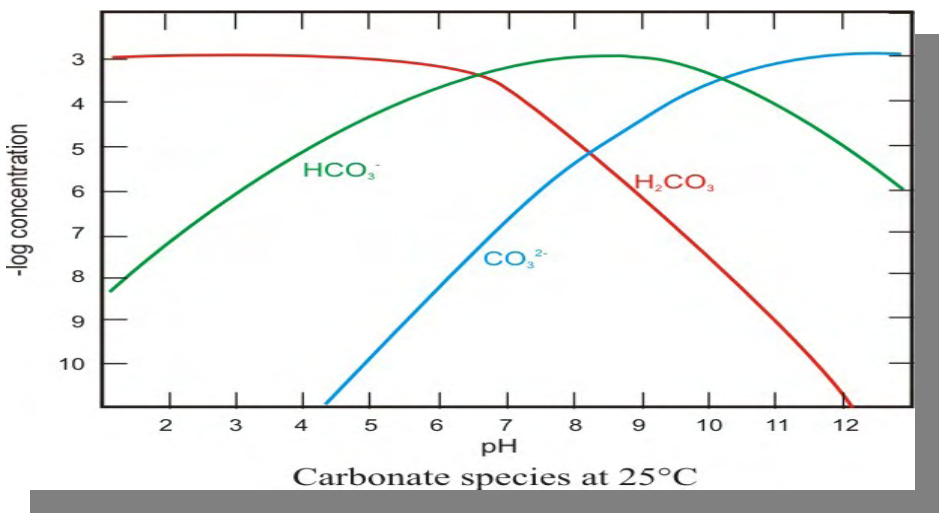

**Figure 1.2:** Concentration in equilibrium of a $CO_2$ alkaline contaminated electrolyte as a function of pH at room temperature

If one or more of the new compounds is insoluble, it will precipitate out of the solution and the reaction will go spontaneously. You need to know the solubility rules for various ions so you can predict which reactions will occur and which ones will not.

## $CO_2$ Poisoning of AFC Methanol fuelled

Keep in mind that the ability of the electrolyte to dissolve $CO_2$ and form $K_2CO_2$ depends on the KOH concentration in the electrolyte. Figure 1.3 shows the triangle phase equilibrium between $K_2CO_3$, methanol, and water. Each vertex corresponds to a pure substance. The substance at the vertex of the triangle concentration reads at the left of the arrow and in the direction of the arrow from 0 to 1 or 0 to 100%. The side opposite to a vertex corresponds to a mixture of the other two substances.

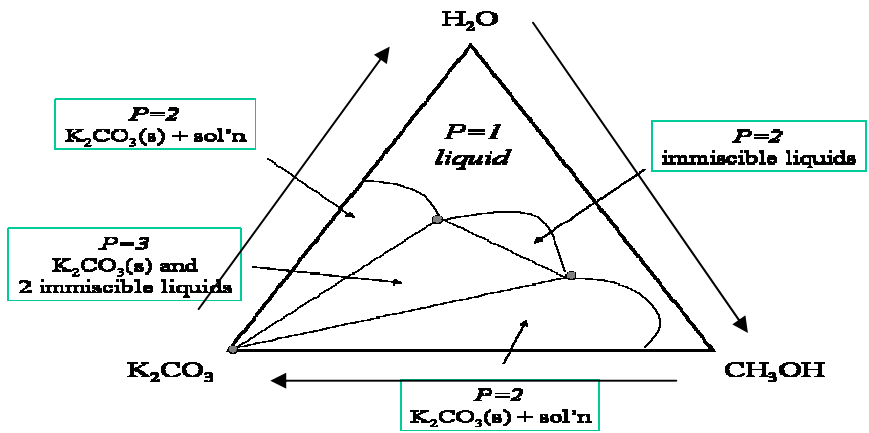

**Figure 1.3:** Tr-phase diagram for a mixture of water, methanol and $K_2CO_3$ (which it is formed when the KOH electrolyte becomes contaminated with $CO_2$.

If A, B, and C are the vertices, the mole fraction of A is proportional to the distance from the opposite side and the 3 mole fractions add to 1. It clearly indicates in what region to operate and where there will not be salt precipitation or formation. It is clear that the area of interest in order of not having salt precipitating in the membranes pores for a fuel cell application is a mixture of the electrolyte from 0.65 up to 1 fraction mole of water, and from 0 to 0.45 of methanol depending on the mole fraction of $K_2CO_3$ as indicated in the phase diagram below. That "area" of interest is placed from the top right side of the triangle.

The phase diagram clearly delineates the areas' mixtures for which no precipitation of salt will occur and the possibility of blocking the electrodes is minimal. However, in those areas for which the mixture remains liquid, the unbalance of the OH- concentration (changes in pH) may have a detrimental effect on the catalyst action.

The phase diagram and the reactions indicated above gives us a general idea of under which conditions the cell's membrane may be blocked by potassium carbonate precipitating on the pores. However, as we investigate, it appears that the precipitation of potassium carbonate is not the most important problem plaguing the loss of efficiency on the alkaline fuel cells when the electrolyte is being $CO_2$ contaminated.

After reviewing the general problems that plague the fuel cells and the historic remedies that have been attempted to remediate those problems, we conclude that none of the cells named above (PEMFC, AFC, DMFC) have demonstrated successful stable operation if air is used instead of pure oxygen. The problem is due to the contaminants that come with air and with the $CO_2$ presence for AFCs. The use of pure oxygen is too costly to consider. Also, none of the cells named above have demonstrated successful and stable operation if they use "dirty" $H_2$ as fuel (i.e. contaminated with $CO_2$ or others contaminants). An interesting proposition will be that of moving to use air instead of pure oxygen, and "dirty" hydrogen instead of hydrogen. We focused our research in extending the life of alkaline fuel cells due to electrolyte $CO_2$ poisoning to come up with a simple methodology that permits engineers to quantify the cell poisoning and take measures (as changing the electrolyte) needed to extend the cell's life and to increase the understanding of the fundamental problems that cause the $CO_2$ poisoning in AFCs.

## Objective:

 **(A).** We propose to study the $CO_2$ poisoning problem on alkaline fuel cells (AFC) by means of a membrane that preferentially lets $O_2$ pass over $CO_2$ when air instead of pure $O_2$ is used at the cathode side. Metal chelates will be used for selective transport of oxygen ($O_2$) from air to the oxygen-catalyst/electrolyte surface.

*To evaluate the performance of the cathode side of the fuel cell and making the experimental procedures easier to handle, we adopted methanol as a fuel instead of $H_2$.*

 **(B)** For contaminate fuel as "dirty" $H_2$, we will use a polymeric membrane for which the pores size is chosen to maximize the mixture gas separation between the $H_2$ and the $CO_2$.

*To study the effect of "dirty" hydrogen on the anode side of the AFC, we will use an AFC fueled with hydrogen and evaluate the effect of $CO_2$ on the cell performance.*

## References
1. H. Lee, J. Shim, M. Shim, S. Kim, J. Lee, Materials Chemistry and Physics, 45 (1996) 238-242.

2. S. Gamburzev, K. Petrov, A.J. Appleby, Journal of Applied Electrochemistry, 32 (2002) 805-809.

3. E. Gülzow, Journal of Power Sources, 61 (1996) 99-104.

4. M. Al-Saleh, S. Gultekin, A. Al-Zakri, A. Khan, International Journal of Hydrogen Energy, 21 (1996) 657-661.

5. M. Schulze, E. Gülzow, G. Steinhilber, Applied Surface Science, 179 (2001) 252-257.

6. J. Shim, H.K. Lee, Materials Chemistry and Physics, 69 (2001) 72-76.

7. A.K. Chatterjee, R. Banerjee, M. Sharon, Journal of Power Sources, 137 (2004) 216-221.

8. M. Schulze, E. Gülzow, Journal of Power Sources, 127 (2004) 252-263.

9. Banerjee, Shoibal; Curtin, Dennis E. Nafion perfluorinated membranes in fuel cells. Journal of Fluorine Chemistry (2004), 125(8), 1211-1216.

10. Antolini, E. Review in Applied Electrochemistry. Number 54: Recent Developments in Polymer Electrolyte Fuel Cell Electrodes. Journal of Applied Electrochemistry (2004), 34(6), 563-576

11. Costamagna, Paola; Srinivasan, Supramaniam. Quantum jumps in the PEMFC science and technology from the 1960s to the year 2000 Part I. Fundamental scientific aspects. Journal of Power Sources (2001), 102(1-2), 242-252.

12. G.F. Maclean, T. Niet, A. Prince-Richard, N. Djilali, International Journal of Hydrogen Energy, 27(2002) 507-526.

13. Fuel Cells ±Fundamentals and Applications By L. Carrette1+, K. A. Friedrich1 and U. Stimming1* FUEL CELLS 2001, 1, No. 1, p. 5-39

14. K.V., Kordesch, Outlook for Alkaline Fuel Cell Batteries, From Electrocatalysis to Fuel Cells, Seattle, WA, 1972, pp.157.

15. E. Gülzow, M. Schulze, Journal of Power Sources, 127(1-2) (2004) 243-251.

16. M. Al-Saleh, S. Gultekin, A.S. Al-Zakri, H. Celiker, Journal of Applied Electrochemistry, 24 (1994) 575.

17. E.H. Yu, K. Scott, R.W. Reeve, Journal of Electroanalytical Chemistry, 547 (2003) 17-24.

18. E.J. Cairns, Handbook of Fuel Cells- Fundamentals Technology and Applications, first ed., 2003 pp 301-309.

19. M. Cifrain, K.V. Kordesch, Journal of Power Sources, 127 (2004) 234-242

20. A.J. Appleby, F.R. Foulkes, Fuel Cell Handbook, Krieger Publishing Company, Malabar, Florida, 1993

21. V. Ahuja, R.K Green, International Journal of Hydrogen Energy, 21 (1996) 415-421

22. V. Ahuja, R.K. Green, International Journal of Hydrogen Energy, 23 (1998) 131-137.

23. Fyke, An investigation of alkaline and PEM fuel cells. Institute for Integrated Energy Systems, University of Victoria, Canada, 1995.

24. Swette L., Kopek J.A., Copley C.C., LaConti A.B., Intersociety Energy Conversion Engineering Conference Proceedings, Atlanta, GA (1993), 1227–1232.

25. Kordesch K., Gunter S., Fuel cells and their applications, Wiley-VCH, Berlin, Germany (1996).

26. M. Cifrain, K.V. Kordesch, Journal of Power Sources, 127 (2004) 234-242

# Chapter 2.  Membranes for Gas Separation Introduction

One way of reducing/eliminating the carbon dioxide poisoning problem is to prevent carbon dioxide from entering the electrolyte.  When "dirty" hydrogen (a mixture of $CO_2$ and $H_2$) is considered as a fuel, the hydrogen molecules are considerable smaller than $CO_2$ molecules and they should be easier to separate if a membrane with the correct pore diameter is chosen.  On the other hand, when air instead of pure oxygen is used as an oxidant, the molecules of $CO_2$ and $O_2$ are similar in size and it is difficult to separate them using a size selective membrane.  However, one can increase the permeance of oxygen across by chemically modifying a membrane, i.e. by using chemical selectivity over size selectivity.

A membrane which is selectively permeable to oxygen and hydrogen and has the ability to limit/stop the passage of carbon dioxide could, in principle, eliminate the carbon dioxide poisoning effect.  Several types of membranes can potentially be used for gas separations.  These include the following

- Polymer based membranes: Polymers containing gas binding units e.g. oxygen binding cobalt complexes.  They have high selectivity, but limited permeability.[1]
- Liquid membranes: Oxygen binding liquid carrier immobilized in a polymer matrix.  They have moderate selectivity, but high permeability.[2]
- Nanoporous carbon membranes (~ 5 Å pore size): The separation depends on flow dynamic and gas particle size.[3]
- Ceramic membranes: La & Gd based ceramics doped with impurities to create oxygen vacancies.  They actively transport oxygen at high temperatures.[4]
- Hollow fiber membranes: Commercialized by Air Products to separate nitrogen from air.[5-7]

In this study we developed and tested several membranes aimed at separating $H_2/CO_2$ and $O_2/CO_2$ to prevent carbon dioxide poisoning of alkaline fuel cells.

## Polymer Based Oxygen Selective Membrane:

Many metal complexes play important roles in oxygen transport by reversibly binding oxygen.  If an oxygen binding metal complex is dispersed in a polymer membrane, oxygen can bind to it in a reversible manner and be transported through the membrane.  Such a polymer-metal complex system can function as a gas separating membrane system exhibiting facilitated transport of oxygen across the polymer membrane.  Several Cobalt (II) porphyrin complexes (CoP), as shown in Figure 2.1, have been widely employed for this purpose  [8-11].  The porphyrin ligand around the cobalt can be readily tuned to optimize oxygen-binding capabilities and facilitate transport.  Moreover, the ligands can be modified such that the oxygen-binding complex shows high solubility in a particular polymer matrix system.

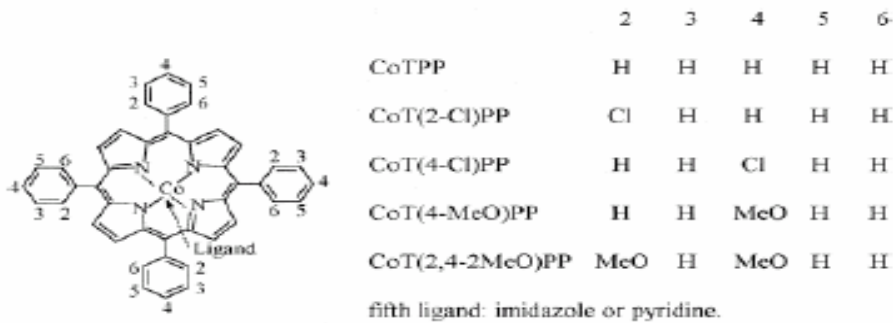

**Figure 2.1:** Structure of oxygen binding cobalt (II) porphyrin complexes

Based on whether the carrier is fixed or mobile, the membranes exhibiting facilitated transport can be broadly classified as Fixed Carrier Polymer Membranes or Mobile Carrier Liquid Membranes. In Fixed Carrier Polymer Membranes, the oxygen-binding complex is either covalently bound to the solid polymer matrix or is physically dispersed in the solid polymer. The stationary porphyrin molecules reversibly bind and unbind oxygen, thereby increasing the solubility of the oxygen in the membrane and, consequently, increasing the relative flux of oxygen through the polymer matrix (see Figure 2.2).

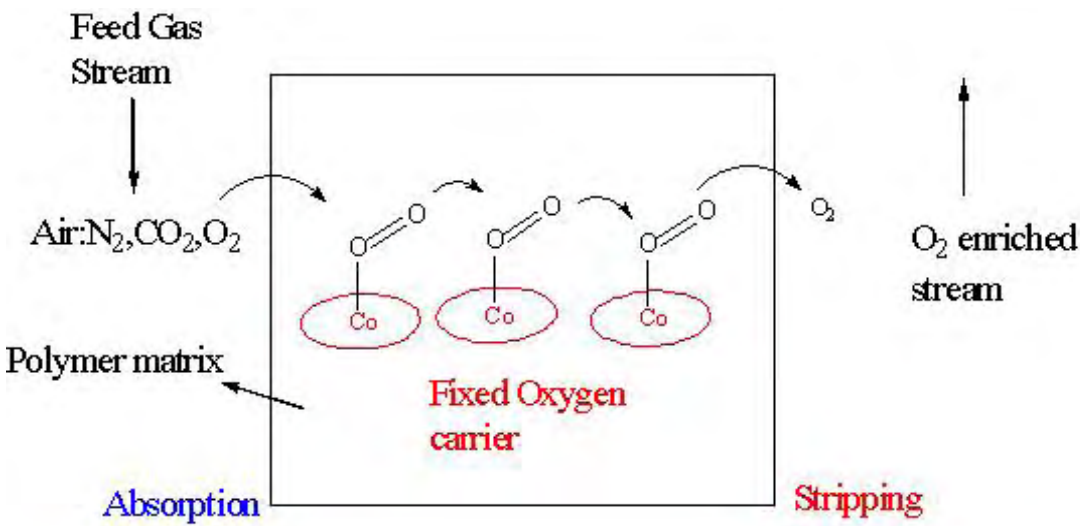

**Figure 2.2:** Schematic of oxygen transport through Fixed Carrier Polymer Membrane

In Mobile Carrier Liquid Membranes, the oxygen-binding complex is dissolved in a liquid medium, which is immobilized in a solid polymer matrix. Since the oxygen-binding complex molecule is mobile, it can transport oxygen molecules from the feed stream side of the membrane to the stripping side (see Figure 2.3) at a higher rate compared to Fixed Carrier Polymer Membranes.

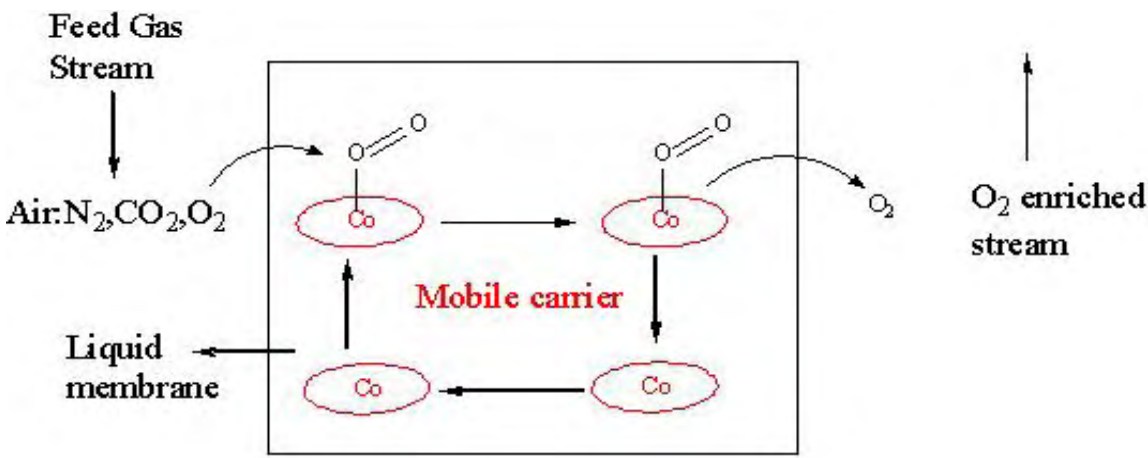

**Figure 2.3:** Schematic of oxygen transport through Mobile Carrier Liquid Membrane

A comparison of the two membrane systems is presented in Table 2.1 below. A large number of membrane systems exhibiting facilitated transport of oxygen over nitrogen by different metal complexes have been reported in multiple publications [12]. A few representative examples are given in Table 2. 2.

## Selection of Polymer Matrix for Immobilizing Oxygen Carrier:

Polymers used for gas separation can be broadly classified as Size Selective or Solubility Selective. The Size Selective membranes are made of stiff, chain, glassy polymers, such as polysulphone, polystyrene, etc. They are more permeable to smaller molecules since they have a lower free volume. The Solubility Selective membranes are made of highly flexible polymers, such as poly(dimethylsiloxane) PDMS, poly(1-trimethylsilyl-1-propyne) PTMSP, etc. They have higher free volumes and exhibit selectivity based on the solubility of different gases in the polymer. The permeability and selectivity of the different gases by the two kinds of polymeric membranes are given in Table 2.3.

It is clear that the nature of the polymer matrix has a large effect on both the permeability and selectivity of the membrane towards different gases. Hence, the oxygen carrier Cobalt complex was incorporated into the range of different polymers (glassy, rigid with low free volume, somewhat rigid with high free volume and highly flexible and rubbery) to optimize the gas diffusion (permeability) and gas separation (selectivity) properties of the membrane.

**Table 2.1:** Comparison of Mobile Carrier versus Fixed Carrier Membranes for gas separation [12]

|  | **Mobile carrier (liquid)** | **Fixed carrier (polymer film)** |
|---|---|---|
| **Requirements** | *Membrane*: low effective thickness<br><br>*Liquid medium*: low viscosity, low volatility, high compatibility with polymeric material<br><br>*Carrier*: high concentration in the liquid medium, high selectivity for $O_2$ | *Membrane*: low thickness<br><br>*Carrier*: high concentration in the polymer matrix, high selectivity for $O_2$, high carrier-oxygen binding constant |
| **Advantages** | *Good selectivity*<br><br>*High diffusivity of the permeant molecule* | *High selectivity* |
| **Disadvantages** | *Loss* of membrane solvent and carrier<br><br>*Low carrier concentration*<br><br>*Carrier inactivation due to oxidation* | *Inactivation* of the carrier after fixation in the solid state<br><br>*Non-uniformity* in chemical reactivity of the fixed carrier<br><br>*Defect formation* in the solid membrane<br><br>*Low diffusivity* of the permeate molecule |

**Table 2.2**: Polymer Carrier systems designed for gas separation [12]

| Facilitate transport by | Solvent / polymer | Carrier | α = Selectivity $PO_2/PN_2$ <br><br> P=Permeability (in barrer*) |
|---|---|---|---|
| Fixed | Polybutylmethacrylate | CoP | α= 3-12 <br><br> P= 6-14 |
| Fixed | Epoxydiacrylate grafted onto cellulose | Co salen | α= 50 |
| Fixed | Styrene-butadiene | Co salen | α= 3.4 <br><br> P= 23 |
| Liquid | 4-methylanisole | Co P Im | α= 20-40 <br><br> P=20-40 |
| Mobile | Butyrolacton/N-methylpyrollidone/DMSO | Co( OMe salen) | α= 30 <br><br> P=1000 |
| Encapsulated carrier in liquid | Polyethersulphone | Co(5-$NO_2$-saltmen) | α= 19.7 |

1 barrer = $10^{-10}$ $cm^3$(STP)cm $cm^{-2}$ $s^{-1}$ $cmHg^{-1}$.
Co = Cobalt(II); P = porphyrin ligand; Im= Imidazole; salen = [N,N'-bis(salicylaldehyde)ethylenediimine]; saltmen= *N,N'*-(1,1,2,2-tetramethyl-ethylene)bis(salicylideneiminato).

**Table 2.3**: Effect of polymer structure on gas permeability of membranes[13, 14]

|  | PSF | PDMS | PTMSP |
|---|---|---|---|
| **Polymer Type** | Glassy, <br><br> Low Free volume | Rubbery | Glassy, <br><br> High free volume |
| **Oxygen Permeability x $10^{10}$** | 1.4 | 770 | 9700 |
| **$O_2/N_2$ Selectivity** | 5.6 | 2.1 | 1.5 |

| n-butane Permeability x $10^{10}$ | .007 | 48.5 x $10^3$ | 78.8 x $10^3$ |
|---|---|---|---|
| n-butane/methane Selectivity | 0.026 | 40 | 5 |

# $H_2/CO_2$ separation and Rationale for using pure polystyrene membranes for gas separation

Any polymeric material will separate gases to some extent. Proper selection of the polymeric material comprising the membrane is extremely important. It determines the ultimate performance of the gas separation module.

Module productivity for a given base material is determined by three factors: partial pressure difference across the membrane, membrane thickness and membrane surface area. The partial pressure difference is typically determined by the specifics of the application.

Figure 2.4 shows the separation factor for several gases-couples as a function of pore diameters in the membrane. This preferential gas separation is performed mainly on basis of molecule or atom size ratios.

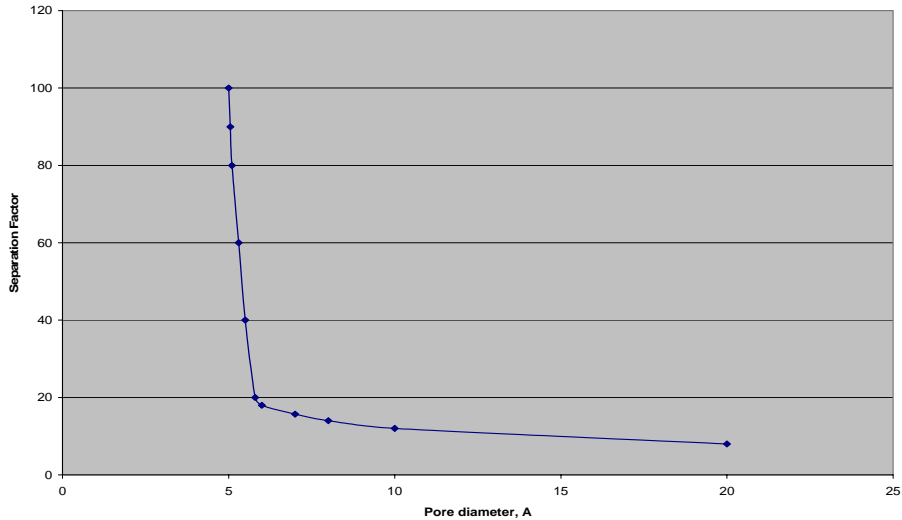

**Figure 2.4.** Separation factor in percentile versus membrane pores for the separation of a mixture of $H_2/CO_2$ [15].

Off the shelf Polystyrene has intrinsic gas permeability selectivity towards $H_2$ and $CO_2$. The permeability of polystyrene towards different gases is given in Table 2.4 below.

**Table 2.4 :** Permeability of pure polystyrene [16]

| Permeant Gas | Permeability ( P X 10 $^{13}$ ) |
|---|---|
| Nitrogen $N_2$ | 0.59 |
| Carbondioxide $CO_2$ | 7.9 |
| Hydrogen $H_2$ | 17.0 |
| Methane $CH_4$ | 0.15 |

P = {quantity of permeant gas X film thickness} / {area X time X pressure drop across film }
Hence units =
Permeability , P = {cm$^3$( at STP = 273.15K & 1.013E5 Pa) X cm }/ { cm$^3$ X s X Pa}

From the table above it is clear that a pure polystyrene membrane would show a $H_2$:$CO_2$ selectivity of 2.15 and would effectively filter $CO_2$ from $H_2$ in a dirty hydrogen fuel stream. Further chemical modifications in polystyrene( see Table 2.5 ) and modifications in membrane engineering could enable one to achieve even higher selectivity

**Table 2.5 :** Permeability of a chemically modified polystyrene i.e. Poly(styrene-co-styerenesulfonic acid) $Mg^{2+}$ polysalt ( 15.2 mol% sulphonic acid) [ 17]

| Permeant Gas | Permeability ( P X 10 $^{13}$ ) |
|---|---|
| Nitrogen $N_2$ | 0.0563 |
| Carbondioxide $CO_2$ | 2.25 |
| Hydrogen $H_2$ | 7.2 |
| Methane $CH_4$ | 0.0525 |

A number of other polymers also exhibit high intrinsic $H_2$:$CO_2$ selectivity and they are listed in Table 2.6 below. These polymeric materials are inexpensive and easy to fabricate and could act as effective gas separation materials for use in alkaline fuel cells to prevent the $CO_2$ poisoning problem.

**Table 2.6** : $H_2$-$CO_2$ permeability of various polymers [18]

| Polymer | $H_2$ permeability ( P X 10$^{13}$ ) | $CO_2$ permeability ( P X 10$^{13}$ ) |
|---|---|---|
| Polyvinyl acetate | 6.84 | 1.52 |
| Poly vinylchloride | 1.3 | 0.12 |
| Poly dimethyl butadiene | 12.8 | 5.63 |
| Nylon 11 | 1.34 | 0.754 |

Hence, a membrane made of pure poly vinyl chloride, which is a very inexpensive and readily available polymer should be able to theoretically improve the performance of an AFC running on dirty hydrogen by a factor of 10.8 times. Once again further chemical and engineering modifications should enable one to achieve higher selectivity and lessen the poisoning of the electrolyte. With this in mind, we decided to use "off the shelf" polystyrene as the membrane matrix to try to separate $H_2/CO_2$ in dirty hydrogen fuel for alkaline fuel cells. The results of the polystyrene interfaced $H_2$/Air fuel cell operating on dirty hydrogen are discussed in the Results section of this report.

# Membrane Fabrication

Four different kinds of polymers with different structures and properties were chosen as a membrane matrix for the described application. They were poly(styrene), poly(methyl methacrylate), poly(4-vinylpyridine) and poly(dimethylsiloxane). Poly(styrene) and poly(methyl methacrylate) were chosen for their toughness and durability/stability. Polyvinylpyridine was chosen for its ability to coordinate to the fifth site on the cobalt complex, thereby homogenously fixing the cobalt complex to the polymer matrix. Polydimethylsiloxane was chosen for its high gas permeability and good membrane forming ability. The oxygen binding cobalt complexes used were 5, 10, 15, 20-Tetraphenyl-21H, 23H-porphine cobalt(II) –N-methylimidazole (CoP) and Cobalt acetylacetonate Co(acac). CoP was synthesized as reported in literature[19]. Co(acac) was purchased from Aldrich and was used without further purification. All solvents used were analytical grade with > 99.9% purity.

The membrane fabrication technique is described as follows:

## Polystyrene

Polystyrene (MW= 280,000), purchased from Aldrich, was dissolved in benzene to yield a 10wt% solution. Co(acac) and CoP solutions in chloroform were added to the polymer solution to yield polymer - chelate solutions having 5wt% and 10wt% of Co(acac) and 10wt% of CoP respectively. The additions were performed under a nitrogen atmosphere. The resulting polymer – chelate solution was stirred vigorously to homogenously distribute the cobalt complex. These solutions were then cast onto microporous steel discs( pore size = 200 micron ) to yield supported membranes. After the initial slow evaporation of the solvent for 24 hours, the polymer coated discs were kept in a vacuum for another 24 hours to remove any solvents present. These supported polymer membranes were used to study gas selectivities, as described later.

## PMMA (Polymethylmethacrylate)

Polymethylmethacrylate (MW=120,000) membranes were fabricated in the same way as polystyrene membranes described above. Microporous steel supported PMMA membranes having 5wt% and 10wt% of Co(acac) and 10wt% of CoP, respectively, were prepared and used in gas selectivity studies, as described later.

## PDMS (poly(dimethylsiloxane))

A commercially available silicone elastomer, Sylgard 184 from Dow Corning, was used as the polymer matrix for the cobalt oxygen carrier complex. Sylgard 184 consists of a silicone base and a curing agent which is mixed in a 10:1 ratio and cured to a solid by heating for an hour at 70°C. We tried to cure the Sylgard 184 elastomer in the presence of the cobalt complex to yield PDMS membranes homogenously embedded with the cobalt complex.

CoP and Co(acac) solutions in benzene were blended with the silicone base to get 1.5wt% and 3wt% of the cobalt complex in the polymer. The cobalt complex-polymer blend was kept under vacuum for 24 hours to remove the solvent from the blend. The curing agent was then added in a 1:10 ratio and the blend was mixed thoroughly. The viscous mixture was then spin coated onto microporous discs. The polymer layer was then heated at 70°C in a nitrogen atmosphere to cure the blend into a solid membrane. However, even after 24 hours of heating, the blend failed to solidify into a solid membrane. Control experiments indicated that the curing agent was reacting with the cobalt porphyrin complex, thereby deactivating and failing to cross link the Sylgard base.

The polymer coated discs were thoroughly wiped clean to remove the adhering viscous polymer blend. It was thought that the viscous polymer blend would fill the 200 micron pores of the microporous disc, thereby effectively yielding a liquid membrane. The microporous discs having pores filled with the liquid PDMS–chelate was used to study gas selectivities, as described later.

## Polyvinylpyridine (PVP)

Polyvinylpyridine (MW=160,000), purchased from Aldrich, was dissolved in nitromethane to yield a 10 wt% solution. The polymer solution was deoxygenated by bubbling nitrogen through the solution for 15 minutes. A 5, 10, 15, 20-Tetraphenyl-21H, 23H-porphine cobalt (II) solution in chloroform was degassed and added to the polymer solution. The contents were then stirred under nitrogen for 3 hours at 50°C to yield a dark red solution. The pyridine side groups of polymer coordinate to the cobalt porphyrin molecules yielding the active oxygen carrier species. This is shown in Figure 2.5. Two solutions having 5wt% and 10wt% CoP w.r.t. polymer were synthesized as described above. These solutions were then cast onto microporous steel discs(pore size = 200 microns). After the initial slow evaporation of the solvent for 24 hours, the discs were kept in a vacuum for another 24 hours to remove any solvents present. The microporous discs, along with the adhering membrane, were used to study selectivity towards different gases.

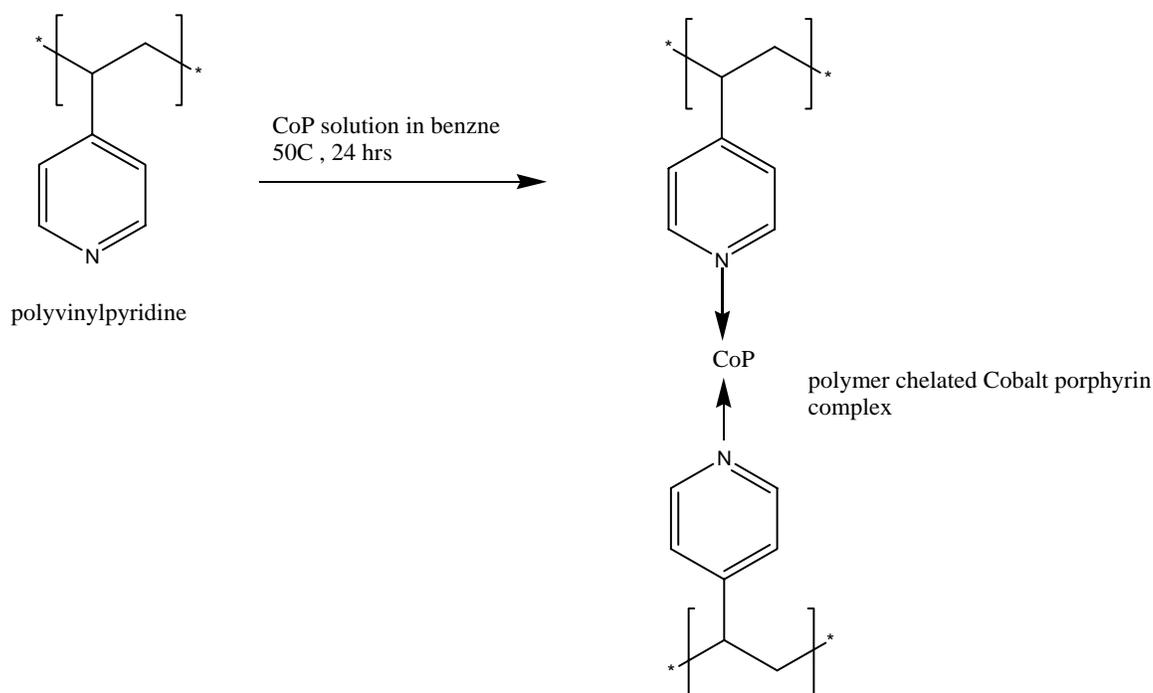

**Figure 2.5.** Synthesis of polyvinylpyridine CoP polymer- chelate

N-Alkylated polyvinylpyridine was also synthesized as the polymer matrix for immobilizing the oxygen-binding cobalt complex. Polyvinylpyridine from Aldrich (MW= 160,000; Tg = 142 $^{o}$C) was partially alkylated using 1-bromopropane to yield the polymer having the desired number of coordinating pyridine groups. Polyvinylpyridine dissolved in nitromethane was heated with 1-bromopropane in a 1:0.8 molar ratio to yield the partially alkylated polyvinylpyridine. This is shown schematically in Figure 2.6. NMR analysis of the polymer showed that 73% of the pyridine groups were alkylated. The polymer solution was degassed and 3wt% cobalt (II) tetraphenyl porphyrin solution in benzene was added to the polymer. The contents were then stirred together at 50°C for 24 hours under nitrogen. The dark red polymer–CoP solution obtained was used to cast membranes as described ahead.

Freestanding membranes were fabricated by spin coating the polymer-CoP solution in nitromethane on a glass disc (500 rpm for 90 seconds) using a Laurel WS-400B series spin coater. The solution on the disc was allowed to evaporate and the spin coating was performed again. After 5 such coatings, the polymer coated disc was dried for 5 hours under vacuum. However, the membranes thus formed proved to be brittle and cracked during peeling from the glass disc.

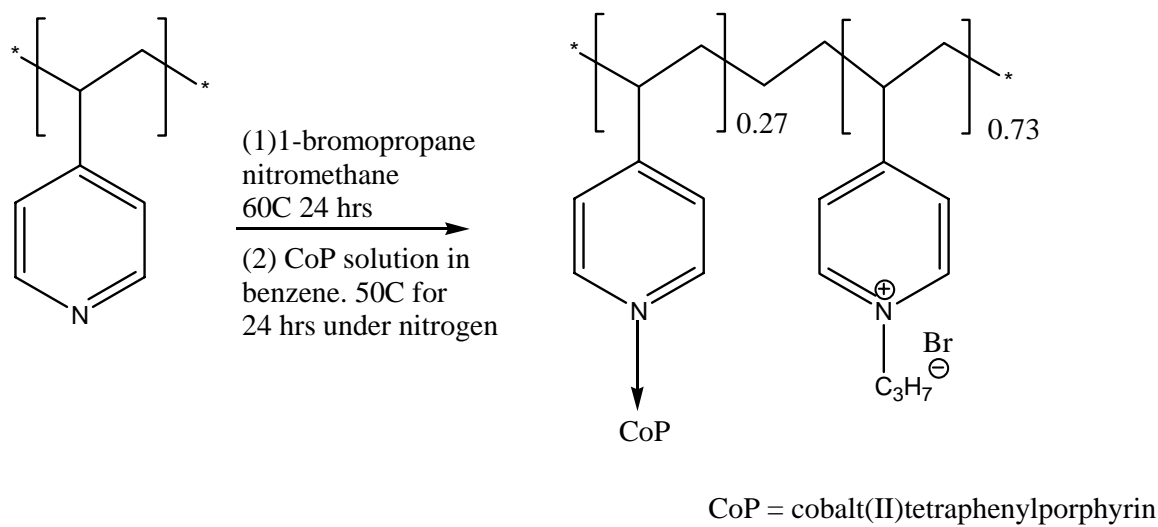

CoP = cobalt(II)tetraphenylporphyrin

**Figure 2.6.** Synthesis of N- alkylated polyvinylpyridine CoP polymer- chelate

To make more robust membranes, the polymer-CoP solution was cast onto a piece of filter paper to yield supported membranes. Qualitative filter paper by VWR (no- 415) was cut into the desired rectangular shape (5cm X 7cm) such that it could be mounted on top of the air cathode of the fuel cell. Small increments of the polymer solution in nitromethane were poured on both sides of the filter paper. The solution was allowed to evaporate slowly at room temperature to yield a dark brown polymer on the filter paper. After an appreciable amount of polymer had been deposited, the supported membranes were dried under vacuum for 5 hours. The thickness of the supported membranes was approximately 0.1 mm. The dark brown supported membranes were interfaced with the fuel cell cathode and electrochemical testing of the cathode-membrane unit was performed to study the effect of carbonate poisoning.

## Membrane Permeation Studies

We used a constant volume variable pressure (CVVP) gas permeation apparatus for our permeation studies. The apparatus consists of an air tight stainless steel chamber divided in two halves by the membrane i.e. upstream and downstream chambers. The target membrane is cast on a micro-porous steel disc of specific dimensions which is designed to fit into the permeation cell. The volume of the downstream chamber is known and remains constant. The upstream side is maintained at a known fixed pressure of the target gas ($O_2$, $CO_2$ and $N_2$) while the downstream chamber (permeate side) is under dynamic vacuum. The dynamic vacuum is then switched off at the start of the measurement. As the target gas permeates from the upstream side to the downstream side through the membrane, the pressure increases in the downstream side chamber. This pressure increase with time is monitored in the downstream chamber using a Baratron absolute pressure gauge. Permeability of different gases can be measured using a simple formula.

$$P = \frac{J \times l}{\Delta p};$$

Where *P* is the permeability, *J* is the flux, *l* is the membrane thickness and Δ*p* is the pressure difference across the membrane. Permeability of these gases can be measured and compared to observe the selective oxygen transport across the membrane.

## References

1. Figoli_, W.F.C. Sager, M.H.V. Mulder ., J.Memb. Sci., 2001,181, 97-110.

2. Maria P. Bernala, Manuel Bardajıb, Joaquın Coronasa, Jesús Santamarıa of journal of Membrane Science 203 (2002) 209–213.

3. M.B.Shiflett, H.C. Foley / Journal of Membrane Science 179 (2000) 275–282 .

4. Dyer, P., Richards, R., Russek, S., Taylor, D., Solid State Ionics, 134, 2000, 21-33.

5. Liu, Shaomin; Gavalas, George R. Oxygen selective ceramic hollow fiber membranes. Journal of Membrane Science (2005), 246(1),

6. Fabiani, Claudio; Bimbi, Luigi; Pizzichini, Massimo; Santarossa, Luigi. Performance of a hollow fiber membrane unit in oxygen-enriched air production. Gas Separation & Purification (1996), 10(1), 75-79.

7. Feng, Xianshe; Ivory, John; Rajan, Varagur S. V. Air separation by integrally asymmetric hollow-fiber membranes. AIChE Journal (1999), 45(10), 2142-2152.

8. Baoqing SHENTU, Hiromi SHINOHARA, and Hiroyuki NISHIDE . High Oxygen Permeation and Persistent Oxygen-Carrying in a Poly(vinylimidazole-co-fluoroalkyl methacrylate)-Cobaltporphyrin Membrane Polymer Journal, Vol. 33, No. 10, pp 807—811 (2001).

9. Shentu, Baoqing; Nishide, Hiroyuki. Facilitated Oxygen Transport Membranes of Picket-fence Cobaltporphyrin Complexed with Various Polymer Matrixes Ind. Eng. Chem. Res. 2003, 42, 5954-5958

10. H. Shinohara, T. Arai, H. Nishide ,Oxygen binding to simplke cobalt porphyrins combined with polyvinylimidazole Macromol Symp., 186, 135-139 ( 2002)

11. Mariýa P. Bernal, Manuel Bardaj, Joaquiýn Coronas, Jesús Santamariý Facilitated transport of $O_2$ through alumina–zeolite composite membranes containing a solution with a reducible metal complex , Journal of Membrane Science 203 (2002) 209–213

12. A. Figoli_, W.F.C. Sager, M.H.V. Mulder Facilitated oxygen transport in liquid membranes:review and new concepts Journal of Membrane Science 181 (2001) 97–110

13. Wessling, R. A.; Gibbs, D. S.; DeLassus, P. T.; Howell, B. A. In Encyclopedia of Polymer Science and Technology; Kroschwitz, J. I., Ed.; John Wiley: New York, 1987; Vol. 17,

14. James Larminie and Andrew Dicks. Fuel Cell Systems Explained. Second Edition WILEY

15. D. E. Fain. Development of Inorganic Membranes for Gas Separation. Becthel Jacobs Company LLC report. (865) 574 9932.

16. H.Yasuda, Kj. Rosengren, Journal of. Applied. Polymer. Science, 14, 1970, 2839

17. W.J. Chen, C.R.Martin, Journal of. Membrane Science, 95, 1994, 51

18. J. Brandrup , E.H.Immergut, E.A. Grulke Polymer handbook 4th edition Wiley Interscience Pages – VI – 543

19. James P. Collman, John I. Brauman, Kenneth M. Doxsee, Thomas R. Halbert,Susan E. Hayes, and Kenneth S. Suslick Journal of the American Chemical Society 1 100:9 / April 26, 1978.

# Chapter 3: $CO_2$ Poisoning of Alkaline Fuel Cells

Chapter 3 contains three main parts:

(1) The methodology and laboratory set up used to collect data needed for the quantification of the phenomenon of carbon dioxide poisoning of the alkaline fuel cells. We used a methanol alkaline fuel cell, which was run in a $CO_2$ enriched atmosphere.

(2) Development of a phenomenological model to explain the results obtained in Part (1). The model is capable of predicting the performance of a methanol alkaline fuel cell when operated in a carbon dioxide enriched atmosphere (same model can be used for $H_2$ alkaline fuel cell also). It also enables us to calculate the time at which the electrolyte is needed to be flushed out to maintain the cell performance at a desired level.

(3) Fabrication and testing of polymer membranes to mitigate the $CO_2$ poisoning in alkaline fuel cells when operated with $CO_2$ contaminated $H_2$ ("dirty" hydrogen).

# Part 3.1 Quantification of Carbon dioxide Poisoning in Air Breathing Alkaline Fuel Cells

## Introduction

Carbon dioxide intolerance has impeded the development of alkaline fuel cells as an alternate power source. The $CO_2$ "poisoning" of the fuel cell electrolyte could come from the anode side (if "dirty" $H_2$ is used as fuel), from the cathode side (if air instead of pure $O_2$ is used as an oxidant), or from inside the electrolyte (if methanol is used as a fuel). In this work, a novel analytical approach is employed to study and quantify the carbon dioxide poisoning problem. Accelerated tests were carried out in an alkaline fuel cell using methanol as a fuel with different electrical loads and varying the concentration of carbon dioxide in a mixture of $CO_2/O_2$ used as oxidant. Two characteristic quantities, $t_{max}$, $R_{max}$ and $\alpha$ (discussed in section 3.2), were specified, which were shown to comprehensively define the nature and extent of the carbon dioxide poisoning in alkaline fuel cells. Section 3.2 focuses on the quantification and model development of the carbon dioxide poisoning in alkaline fuel cells.

## Experimental

### Accelerated $CO_2$ poisoning studies

Small alkaline fuel cells operating on methanol as a fuel were purchased from the Department of Chemistry at The University of Hong Kong (Model # HKU-002C). The cathode was an AC-65 air cathode manufactured by Alupower, Inc. [1]. The anode was fabricated on a nickel foam mat with a mixture of Pt/Co/Ni catalyst for methanol oxidation. The area of the anode was 18 cm$^2$ as compared to 10 cm$^2$ of the cathode. The anode and cathode were separated by approximately 0.8 cm of an alkaline electrolyte. The cell body was made of plexiglass, which can be hermetically sealed during operation. The fuel cells were soaked with deionized water for 24 hours prior to each cell run in order to ensure that the air-cathode had been properly soaked and wetted so as to maximize the active area for electrochemical reactions. For each cell run, a mixture of 30 ml of 1 M potassium hydroxide (purchased from Aldrich) and 3 ml of 99.8% anhydrous methanol (purchased from Aldrich) was used as the electrolyte. Figure 3.1 shows the polarization curves for these fuels at three different electrolyte concentrations. We notice that as the $K_2CO_3$ concentration increases in the electrolyte, both activation polarization and ohmic polarization increase, resulting in lower current output at a given voltage.

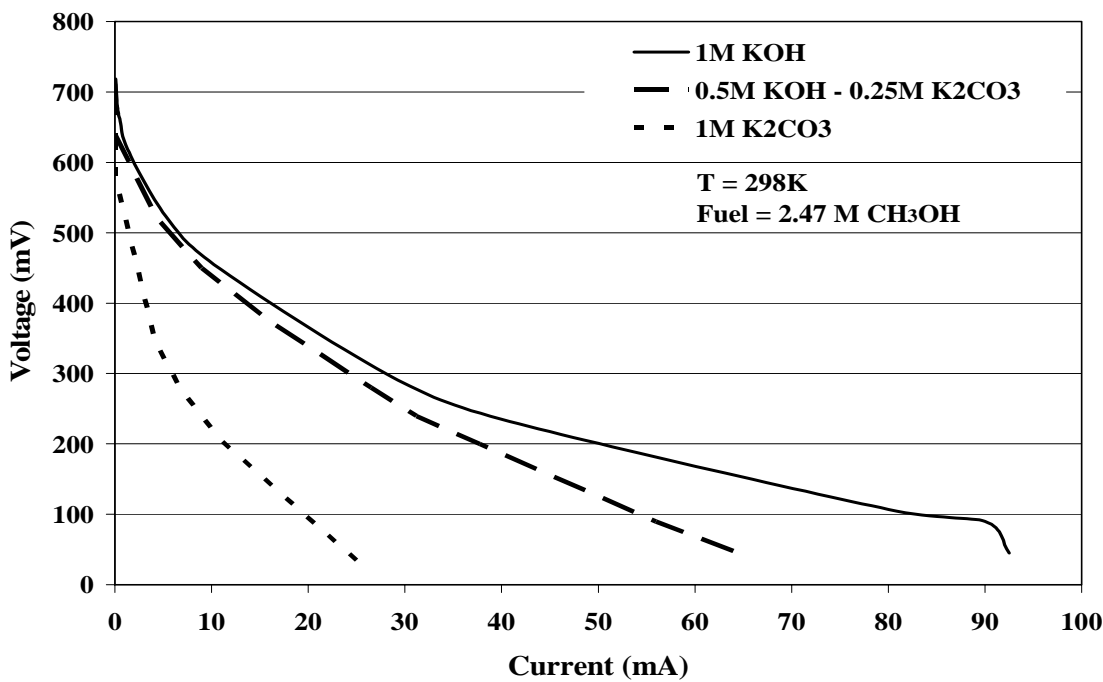

**Figure 3.1** Polarization curves of the fuel cells (HKU-002C) with different electrolytes at 298K.

For operating the fuel cell in atmospheres with different concentrations of oxygen and carbon dioxide, an air-tight plexiglass chamber was designed in which the cell could be placed and operated. A schematic of our experimental setup is shown in Figure 3.2. The chamber had an inlet for the gases mixed in different ratios and was maintained at a slightly positive pressure by dipping the outlet in 2 cm of water. This was done to prevent any gases in the atmosphere from leaking inside. Appropriate metal contacts were provided so the cell could be connected to an external load and an electrochemical data recorder which was placed outside. Oxygen and carbon dioxide were mixed in different ratios using a gas proportioner, model no. 7951-4-4, obtained from Specialty Gas Equipment (Ohio). The gas proportioner was capable of mixing gases in any proportion with an accuracy of 1 - 2%. Different mixtures of oxygen and carbon dioxide (99.99% purity purchased from MG Industries, PA) were introduced into the test chamber with the fuel cell operating inside. A load of known resistance was applied across the fuel cell and an electrochemical data recorder, Auto AC DSP purchased from ACM Instruments (UK), was used to measure and record the cell current for approximately 5 hours at an interval of 4 seconds each.

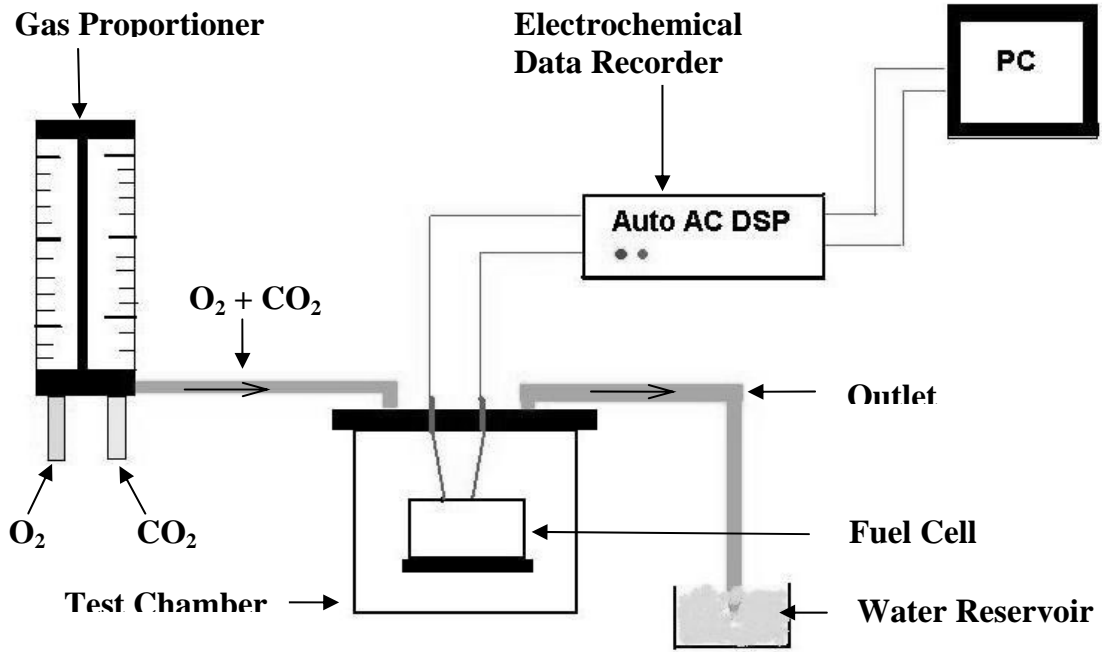

**Figure 3.2** Schematic of the experimental setup used for accelerated $CO_2$ poisoning experiments.

Cell runs with carbon dioxide levels in the atmosphere at 5%, 15%, 25%, 30%, 50% and 80% were carried out as described above. For every atmospheric composition, external loads of 1.3 Ω, 2.8 Ω, 5.2 Ω and 10.2 Ω were applied and the cell current was recorded. In this way, we generated a library of the fuel cell performances in different atmospheres and with different applied loads. The data gathered from the experiments, explained above, were used in quantification and model development of $CO_2$ observed in our fuel cells.

## Part 3.2. Phenomenological Model of Carbon dioxide Poisoning in Alkaline Fuel Cells

### Introduction

We developed a model which predicts $CO_2$ poisoning effect in AFC, with non-circulating electrolyte (KOH aq.) and methanol as the fuel, under conditions of accelerated poisoning. These accelerated conditions were created by operating the fuel cell in atmosphere enriched with $CO_2$ as discussed in the previous section. Based on the experimental data generated as described in section 3.1, we have defined three variable parameters, $t_{max}$, $R_{max}$ and $\alpha$ which accurately describe and predict the onset and magnitude of poisoning in AFCs under different operating conditions. The first two quantities are shown in Figure 3.3, which plots the current vs. time and current decay rate vs. time for a fuel cell operating in $CO_2$ enriched atmosphere (85%$O_2$-15%$CO_2$). **$t_{max}$** is the time at which the current decay rate is maximum while **$R_{max}$** is the value of current decay rate at this time. Although the current decay rate value ($R_{max}$) is negative, its magnitude (absolute value) is plotted in the figure below.

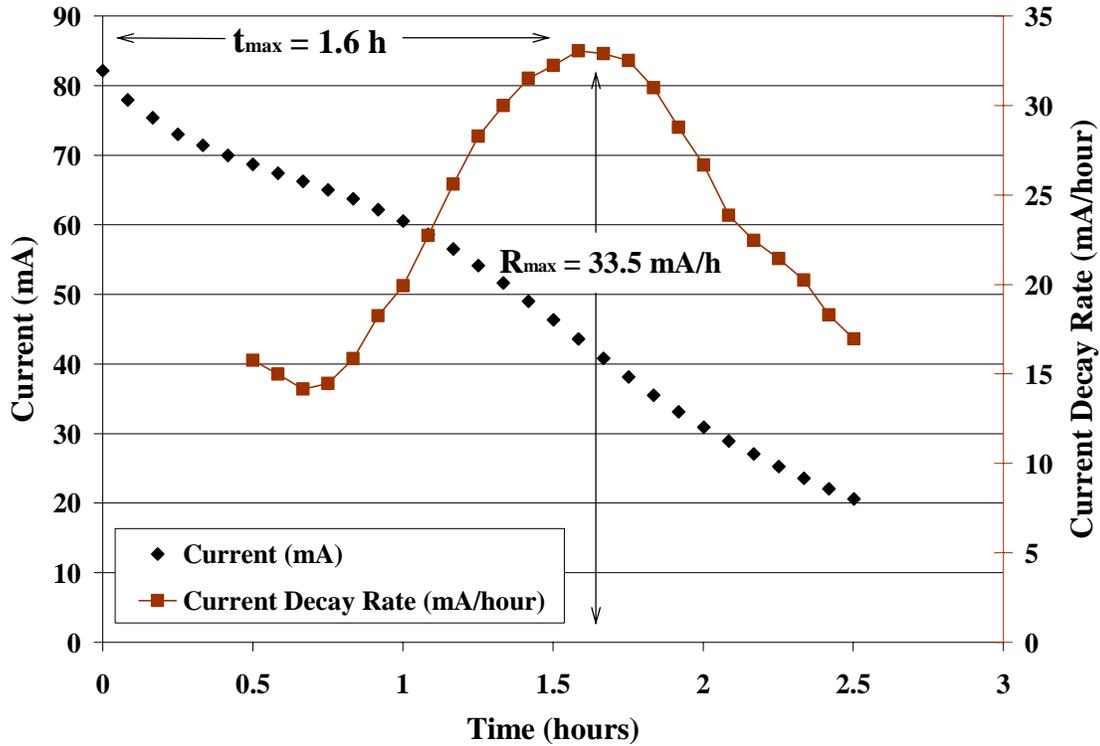

**Figure 3.3** Current vs. time and current decay rate vs. time plot of alkaline fuel cells operated in carbon dioxide enriched atmosphere.

The last quantity '$\alpha$' is the ratio of cell current at t = $t_{max}$ to the initial steady state current observed. The initial steady state current is generally taken as the current value observed after 600 seconds of the start of the experiment.

## How Does the Model Work?

As mentioned earlier, we defined three quantities $t_{max}$, $R_{max}$ and $\alpha$ which are characteristics of a fuel cell's performance in a carbon dioxide enriched atmosphere (for definitions refer section 1). For a given set of operating conditions, values of these characteristic quantities have a unique value. The idea is to find out the dependency of these quantities on the chosen operating parameters ($CO_2$ concentration and applied load). We carried out accelerated carbon dioxide poisoning experiments for different sets of atmospheric carbon dioxide concentration and applied load using the methodology and experimental setup discussed in section 3.1. Values of $t_{max}$, $R_{max}$ and $\alpha$ were calculated graphically and then their dependence on the operating parameters was determined. We found out certain unique trends in which the characteristic quantities vary with the operating parameters. For example, Variation of $t_{max}$ with the applied load was linear for all the atmospheric carbon dioxide concentrations (refer to Figure 3.4). Similarly, the variation of $R_{max}$ with the applied load was hyperbolic (refer to Figure 3.5).

The value of $\alpha$ lies between 0.5 and 0.8, depending on the operating conditions. Higher carbon dioxide concentration and lower applied load increases the value of $\alpha$. Using the plots in Figures 3.4 and 3.5, we could predict the value of $t_{max}$ and $R_{max}$ for any carbon dioxide concentration in the range of 10% to 50% by volume and for any load between 1Ω and 10Ω. Once the values of these characteristic quantities are known, the methodology to theoretically predict the current vs. time plot of a fuel cell is described in the next section.

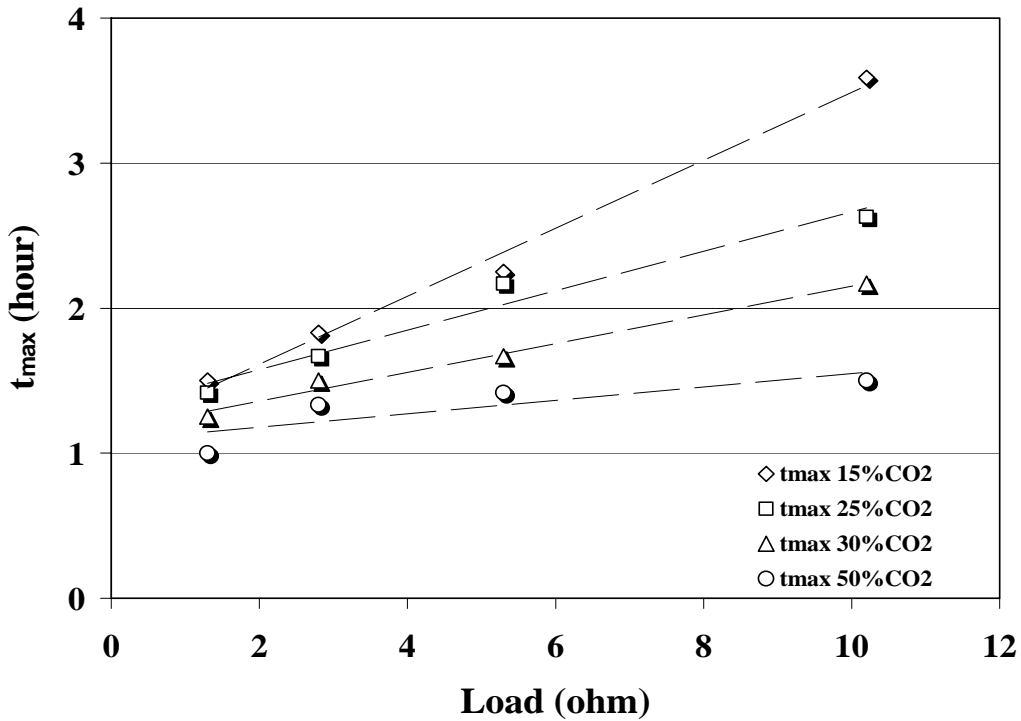

**Figure 3.4** Dependence of $t_{max}$ on applied Load at different $CO_2$ concentrations.

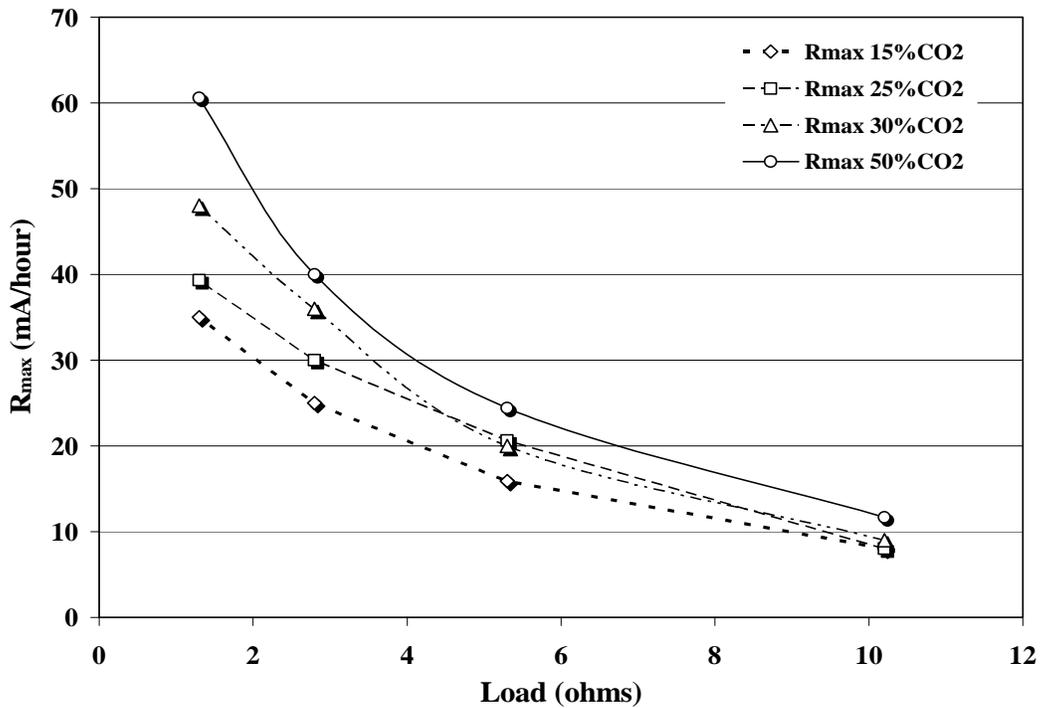

**Figure 3.5** Dependence of $R_{max}$ on applied Load at different $CO_2$ concentrations.

## Methodology:

The next question which arises is "how do we model the time dependency of current with the prior knowledge of $t_{max}$, $R_{max}$ and $\alpha$?" Suppose that operating conditions (*$CO_2$ concentration* and *applied load*) are specified. For these values of operating parameters we can accurately determine the values of our characteristics quantities ($t_{max}$, $R_{max}$ and $\alpha$) as described earlier in this section. Once these values are known, the following steps are carried out in the given order to determine the current vs. time curve for a fuel cell undergoing $CO_2$ poisoning:

On a two dimensional plot with X axis representing time 't' (in hours) and Y axis representing cell current 'i' (in mA)

1) Two straight lines are plotted with equations $i = i_{ss(KOH)}$ and $i = i_{ss(K2CO3)}$ (refer Figure 3.6). The first straight line corresponds to steady state current given by the cell when it has 100% KOH as the electrolyte. While the other straight line corresponds to case of 100% $K_2CO_3$ electrolyte. The values of steady state current at the specified load can be determined from the polarization curves of our fuel cells with pure KOH and $K_2CO_3$ as the electrolyte, as shown in figure 3.1.
2) The point corresponding to the coordinate ($t_{max}$, $\alpha \times i_{ss(KOH)}$) is plotted on the same graph (Figure 3.6). This point shows the state of the fuel cell at $t = t_{max}$ as we know that cell current at this time is α times the steady state current when the electrolyte is 100% KOH (refer to the definition of α).

3) A straight line is plotted, passing through this point and with a slope equal to $R_{max}$. This was done because we know that at $t = t_{max}$ the rate of current decay is equal to $R_{max}$ (Figure 3.6).

4) The straight line plotted in step 3 intersects with the two straight lines plotted in step 1 resulting in a 'Z' shaped plot. We choose 5 points on this 'Z' shaped plot in such a way that we could plot a 4$^{th}$ degree Lagrange's polynomial through these points. 1$^{st}$ point is chosen as $(0, i_{ss(KOH)})$, which marks the beginning of the experiment at $t = 0$ and 100% KOH as electrolyte. The 3$^{rd}$ point is chosen as $(t_{max}, \alpha \times i_{ss(KOH)})$ for the reason described in step 2. The 5$^{th}$ point is chosen as $(3 \times t_{max}, i_{ss(K2CO3)})$ because it marks the end of the experiment. The rationale of choosing the 5$^{th}$ point is that at $t = 3 \times t_{max}$ we are certain that the entire electrolyte has been fully converted to $K_2CO_3$ irrespective of the operating conditions. The 2$^{nd}$ and 4$^{th}$ points are chosen such that they are the midpoints of the 1$^{st}$ point and the point of intersection of line 1 and line 3 and the 5$^{th}$ point and the point of intersection of line 2 and line 3, respectively. The coordinates of the 2$^{nd}$ point and the 4$^{th}$ point are given by $([(1-\alpha)i_{ss(KOH)}/R_{max} + t_{max}], i_{ss(KOH)})$ and $([(i_{ss(K2CO3)} - \alpha i_{ss(KOH)} + R_{max}t_{max})/R_{max}], i_{ss(K2CO3)})$, respectively. Since we know the values of $t_{max}$, $R_{max}$, $\alpha$, $i_{ss(KOH)}$ and $i_{ss(K2CO3)}$, we could find out the numerical values of these 5 points.

5) From step 4 we have 5 points as

$1^{st}$ point: $(0, i_{ss(KOH)})$

$2^{nd}$ point: $([(1-\alpha)i_{ss(KOH)}/R_{max} + t_{max}], i_{ss(KOH)})$

$3^{rd}$ point: $(t_{max}, \alpha\, i_{ss(KOH)})$

$4^{th}$ point: $([(i_{ss(K2CO3)} - \alpha i_{ss(KOH)} + R_{max}t_{max})/R_{max}], i_{ss(K2CO3)})$

$5^{th}$ point: $(3t_{max}, i_{ss(K2CO3)})$

Through these 5 points, we can fit a Lagrange's 4$^{th}$ degree polynomial which would be the *modeled current vs. time curve* of a fuel cell operating in specified accelerated poisoning conditions. We wrote a MATLAB program, which, when given an input of the characteristic quantities ($t_{max}$, $R_{max}$ and α) would generate the modeled i vs. t curve (Figure 3.6).

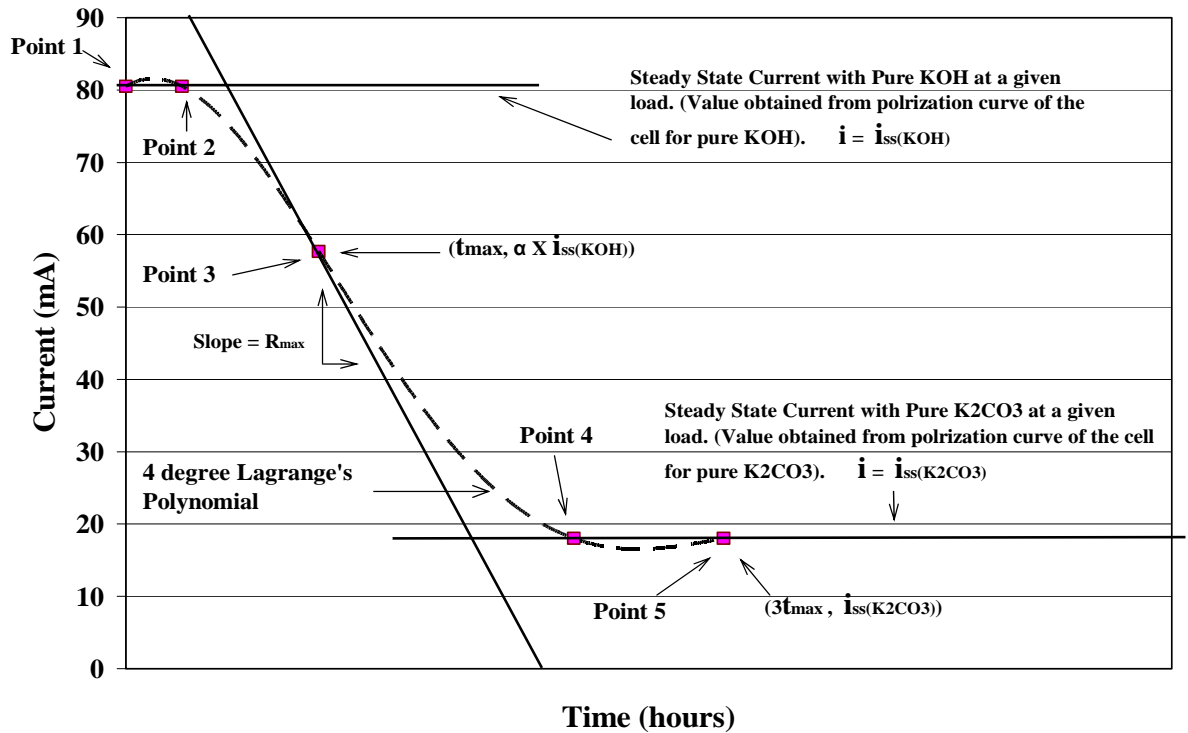

**Figure 3.6** Graphical representation of the methodology to obtain the Current vs. time curve of a fuel cell using the proposed model.

## Part 3.3: Polymer Membranes for Alkaline Fuel Cells Running on Dirty Hydrogen: Mitigation of the $CO_2$ Poisoning Effect.

### Introduction

One possible solution to the carbon dioxide poisoning in alkaline fuel cells is to filter out $CO_2$ from the reactants gas streams. The efficiency of filtration need not be 100% and would depend on the nature of the filter used to carry out the $CO_2$ separation. In this work we explored the possibility of using polymer membranes to mitigate the carbon dioxide poisoning in alkaline fuel cells. The experiments were carried out with the fuel cells operating in $CO_2$ contaminated hydrogen on the anode side or oxygen on the cathode side. In general, because the comparative size of the molecules of $H_2$, $CO_2$, and $O_2$, it is relatively easy to separate $CO_2$ from $H_2$ only by choosing a size selective polymer membrane with the correct pore size (Figure 3.9). However, it is extremely difficult to separate the $CO_2$ from the $O_2$ because of the similarity in their similar size. Accordingly, different types of membranes (Polymer-porphyrine membranes) are proposed to be used to separate $CO_2$ from $O_2$. We independently investigated the efficiency of our membranes on the anode side and cathode side. This part of the report presents the results of the polymer membrane study when the anode gas stream ($H_2$) was contaminated with 5 Vol% $CO_2$. Air (0.03 Vol% $CO_2$) was used at the cathode side.

Owing to the lesser $CO_2$ concentration, poisoning due to the cathode gas stream was assumed to be negligible.

It was found that polymer membranes of polystyrene were highly effective in alleviating the detrimental effect of carbon dioxide poisoning from the use of "dirty" hydrogen as a fuel. This reduction of $CO_2$ poisoning is translated in a very significant increment in the life of a fuel cell simply by placing a polystyrene membrane in the anode gas stream in such a way that the gases pass through it. To quantify the effect, we compared the current output of our cells at a given load when the anode gas stream was with and without the polymer membranes. The techniques used and the results achieved are discussed below.

## Experimental

### Materials

LabCell-50 alkaline fuel cells were purchased from Astris Energi, Inc. for the experiments. They operated with 2-5 M KOH (aq.) as the electrolyte, $H_2$ as the fuel on the anode side and $O_2$ as the oxidant on the cathode side. The active area of the electrodes was 50 $cm^2$ and the open circuit potential (OCP) was 0.9-1.0 V. All the experiments were conducted at room temperature.

Any polymeric material will separate gases to some extent. Proper selection of the polymetric material comprising the membrane is extremely important, as it determines the ultimate performance of the gas separation module. Module productivity for a given base material is determined by three factors: partial pressure difference across the membrane, membrane thickness and membrane surface area. The partial pressure difference is typically determined by the specifics of the application.

### Fuel Cell Testing

Alkaline fuel cells (LabCell 50 purchased from Astris Energi Inc) operating on air/hydrogen were used for the experimentation. The electrolyte was aqueous 2 molar KOH and was circulated through the cell using a peristaltic pump at a rate of 1 ml/s. Pressurized air and hydrogen at an absolute pressure of 24.6 PSI were fed to the fuel cell cathode and anode as the oxidant and fuel source, respectively.

In a typical experiment a load of 1 Ω was placed across the cell and the cell was operated on air/$H_2$ for 2 hours. The $H_2$ was contaminated by 5 Vol% $CO_2$ which was introduced in the anode gas stream by a gas proportioner (model no. 7951-4-4 purchased from Specialty Gas Equipment, OH). All gases were 99.99% pure and were purchased from MG Industries, PA. An electrochemical data recorder, Auto AC DSP purchased from ACM Instruments (UK), was used to measure and record the cell current output for 2 hours.

The current output for the fuel cells running on "dirty" hydrogen and air at a constant load of 1 Ω, were digitally recorded, both with and without the polystyrene membranes placed in the hydrogen gas stream. When a polystyrene membrane is placed in between the anode gas stream and the cell we expect some enrichment of $H_2$ in the permeated gas because of the size selective nature of polystyrene polymer [3]. We observed that our fuel cell's life increased appreciably when using the polystyrene membrane as compared to the case when there was no membrane in the anode gas stream. The results are discussed in detail in the next section.

## References

1. W. H. Hoge, "Air Cathodes and Materials Therefore", USP 4,885,217 (1989).
2. D. E. Fain. Development of Inorganic Membranes for Gas Separation. Becthel Jacobs Company LLC report. (865) 574 9932.

3. **J. Brandrup, E. H. Immergut, and E. A. Grulke**, Polymeric Handbook $4^{th}$ edition, John Wiley & Sons (2003)

# III. Findings, Conclusions and Recommendations

# Chapter 1. Findings, Discussion, and Recommendations

This second part of the report contains five main Parts:

**Part 1:** P*ermeability study* conducted on the polymer-porphyrine membranes. The membranes were designed to show oxygen selectivity by incorporating cobalt porphyrine in the polymer matrix.

**Part 2:** *Quantification* of $CO_2$ poisoning in alkaline fuel cells under conditions of accelerated poisoning. The quantification was based on only two operating parameters viz. %$CO_2$ in atmosphere and applied electric load.

**Part 3:** *Experimental validation* of the phenomenological model developed, which is capable of predicting a fuel cell's performance when operating in $CO_2$ enriched atmosphere. Data gathered in part 2 was used to validate the model, which proved to be significantly accurate.

**Part 4:** P*erformance study* of the fuel cells when polymer membranes (polystyrene, polyvinyl pyridine) were interfaced with the air cathode. The parameters defined in the *quantification study* of $CO_2$ poisoning (part 2) were used to compare the efficiencies of the polymer membranes in reducing the $CO_2$ poisoning in fuel cells.

**Part 5:** P*erformance study* of the fuel cells when polymer membranes (polystyrene) were interfaced with the anode stream. Unlike the results discussed in part 4, the hydrogen (anode) stream was contaminated with $CO_2$ ("dirty" hydrogen) and air was used on the cathode side.

These 5-parts are divided into **two chapters**. Chapter 1 presents the results and discusses the findings of the five parts, while Chapter 2 presents future recommendations for each part.

# Part 1. Results and Discussion of Membranes

## Membrane Permeation Studies:

The results of the permeation studies for the different designed membranes are shown below.

### (1) Polystyrene Membranes:

The permeability results for the different polystyrene membranes cast on microporous steel discs (pure polystyrene, 5% and 10% Co(acac) and 10% CoP) are given in Tables 1.1.1A to 1.1.1D.

**Table 1.1.1A:** Membrane permeability at different inlet pressures for pure polystyrene membrane

| Pressure(psi) | Flux Oxygen X $10^8$ moles/s | Pressure(psi) | Flux Nitrogen X $10^8$ moles/s | Pressure(psi) | Flux Carbondioxide X $10^8$ moles/s |
|---|---|---|---|---|---|
| 27.3 | 3.95 | 28.9 | 4.19 | 21.7 | 3.93 |
| 50.7 | 3.95 | 75.9 | 4.01 | 33.5 | 3.93 |
| 125.5 | 3.92 | 136.3 | 3.97 | 72.2 | 3.87 |

**Table 1.1.1B:** Membrane permeability at different inlet pressures for polystyrene membrane blended with 5wt% Co (acac)

| Pressure(psi) | Flux Oxygen X $10^8$ moles/s | Pressure(psi) | Flux Nitrogen X $10^8$ moles/s | Pressure(psi) | Flux Carbondioxide X $10^8$ moles/s |
|---|---|---|---|---|---|
| 21.2 | 6.51 | 20.5 | 6.44 | 27.2 | 6.57 |
| 30.5 | 6.49 | 30.5 | 6.43 | 35.9 | 6.52 |
| 81.1 | 6.22 | 41 | 6.45 | 72.9 | 6.33 |
| 113.5 | 6.06 | | | | |

**Table 1.1.1C:** Membrane permeability at different inlet pressures for polystyrene membrane blended with 10wt% Co (acac)

| Pressure(psi) | Flux Oxygen X $10^8$ moles/s | Pressure(psi) | Flux Nitrogen X $10^8$ moles/s | Pressure(psi) | Flux Carbondioxide X $10^8$ moles/s |
|---|---|---|---|---|---|
| 22.1 | 6.64 | 21.3 | 6.57 | 29.1 | 6.69 |
| 29.7 | 6.52 | 32.4 | 6.51 | 36.4 | 6.58 |
| 80.1 | 6.29 | 42.1 | 6.47 | 69.9 | 6.46 |

**Table 1.1.1D:** Membrane permeability at different inlet pressures for for polystyrene membrane blended with 10wt% CoP

| Pressure(psi) | Flux Oxygen X $10^8$ moles/s | Pressure(psi) | Flux Nitrogen X $10^8$ moles/s | Pressure(psi) | Flux Carbondioxide X $10^8$ moles/s |
|---|---|---|---|---|---|
| 21.4 | 9.94 | 20.1 | 7.67 | 31.1 | 10.94 |
| 30.7 | 9.85 | 30.8 | 7.58 | 33.3 | 10.77 |
| 82.2 | 9.64 | 41.4 | 7.27 | 71.5 | 10.45 |

The selectivity of the membrane for the two different gases was calculated by taking the ratio of molar fluxes for the two gases. The selectivity of various polystyrene membranes towards oxygen/nitrogen and oxygen/carbon dioxide at an upstream side pressure of 25 psi is summarized in Figure 1.1.1.

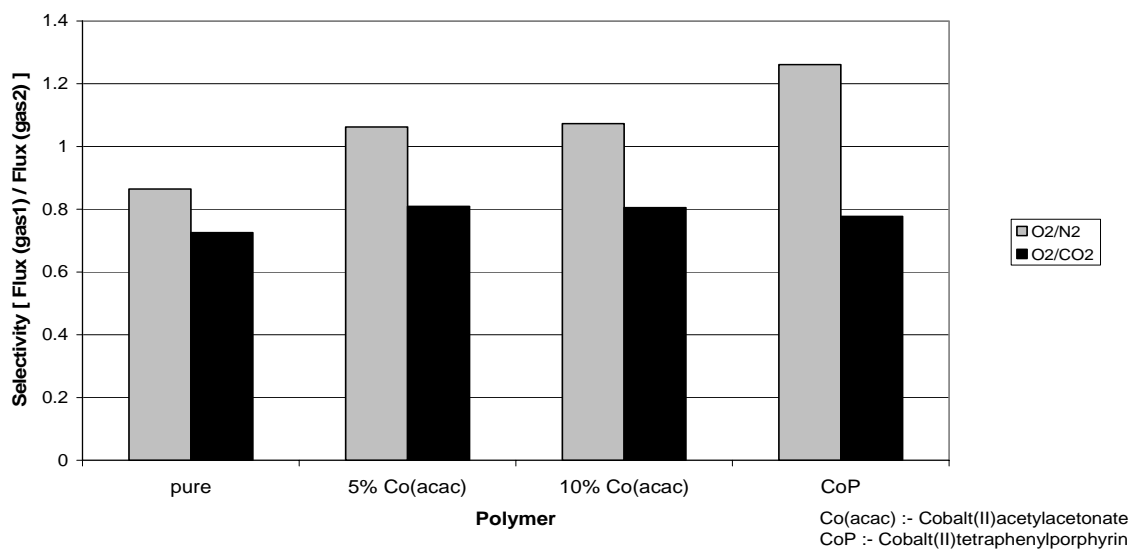

**Figure 1.1.1:** Gas selectivities for Polystyrene membranes blended with different Cobalt complexes (measured at an upstream side pressure of 25 psi)

As can be seen from fig 1.1.1, a pure polystyrene membrane has an $O_2/N_2$ and $O_2/CO_2$ selectivity of less than 1. This indicates that a pure polystyrene membrane has a higher permeability for $N_2$ and $CO_2$ than for oxygen. Introduction of the oxygen binding cobalt complexes slightly increased the $O_2/N_2$ selectivity to above 1, indicating that the membrane is becoming more permeable to oxygen. This indicates facilitated transport of oxygen across the membrane due to the cobalt chelate. However, very little change/

increase was observed in $O_2/CO_2$ selectivity. Moreover, all gas selectivities were close to 1, thereby, clearly indicating that this type of membrane was not very effective in separating gases under the test conditions.

### (2) PMMA Membranes:

The permeability results for different poly(methyl methacrylate) membranes casted on microporous steel discs (pure poly(methyl methacrylate), 5% and 10% Co(acac) and 10% CoP) are given in Tables 1.1.2A to 1.1.2D below.

**Table 1.1.2A:** Membrane permeability at different inlet pressures for for pure polymethylmethacrylate membrane

| Pressure(psi) | Flux Oxygen X $10^8$mols/s | Pressure(psi) | Flux Nitrogen X $10^8$mols/s | Pressure(psi) | Flux Carbondioxide X $10^8$mols/s |
|---|---|---|---|---|---|
| 20.4 | 1.53 | 20.4 | 1.77 | 20.6 | 2.11 |
| 30.5 | 1.47 | 30.5 | 1.73 | 30.7 | 2.07 |
| 80.4 | 1.3 | 40.5 | 1.66 | 70.7 | 1.97 |

**Table 1.1.2B:** Membrane permeability at different inlet pressures for polymethylmethacrylate membrane blended with 5wt% Co (acac)

| Pressure(psi) | Flux Oxygen X $10^8$moles/s | Pressure(psi) | Flux Nitrogen X $10^8$moles/s | Pressure(psi) | Flux Carbondioxide X $10^8$moles/s |
|---|---|---|---|---|---|
| 20.4 | 5.13 | 20.5 | 4.83 | 20.6 | 6.34 |
| 30.4 | 5.03 | 30.4 | 4.74 | 30.7 | 6.21 |
| 80.4 | 4.85 | 40.5 | 4.45 | 70.7 | 5.92 |

**Table 1.1.2C:** Membrane permeability at different inlet pressures for polymethylmethacrylate membrane blended with 10wt% Co (acac)

| Pressure(psi) | Flux Oxygen X $10^8$moles/s | Pressure(psi) | Flux Nitrogen X $10^8$moles/s | Pressure(psi) | Flux Carbondioxide X $10^8$moles/s |
|---|---|---|---|---|---|
| 19.9 | 5.29 | 21.1 | 4.93 | 20.1 | 6.57 |
| 30.6 | 5.11 | 30.4 | 4.83 | 30.3 | 6.42 |
| 81.6 | 4.93 | 41.6 | 4.63 | 71.7 | 6.07 |

**Table 1.1.2D:** Membrane permeability at different inlet pressures for polymethylmethacrylate membrane blended with 10wt% CoP

| Pressure(psi) | Flux Oxygen X $10^8$ moles/s | Pressure(psi) | Flux Nitrogen X $10^8$ moles/s | Pressure(psi) | Flux Carbondioxide X $10^8$ moles/s |
|---|---|---|---|---|---|
| 20.5 | 7.73 | 20.4 | 6.13 | 20.1 | 9.95 |
| 30.4 | 7.45 | 31.1 | 5.92 | 30.6 | 9.75 |
| 78.9 | 6.98 | 40.3 | 5.59 | 70.2 | 9.12 |

The selectivity of the membrane for the two different gases was calculated by taking the ratio of molar fluxes for the two gases. The selectivity of various polymethylmethacrylate membranes towards oxygen/nitrogen and oxygen/carbon dioxide at an upstream side pressure of 21 psi is given in Figure 1.1.2.

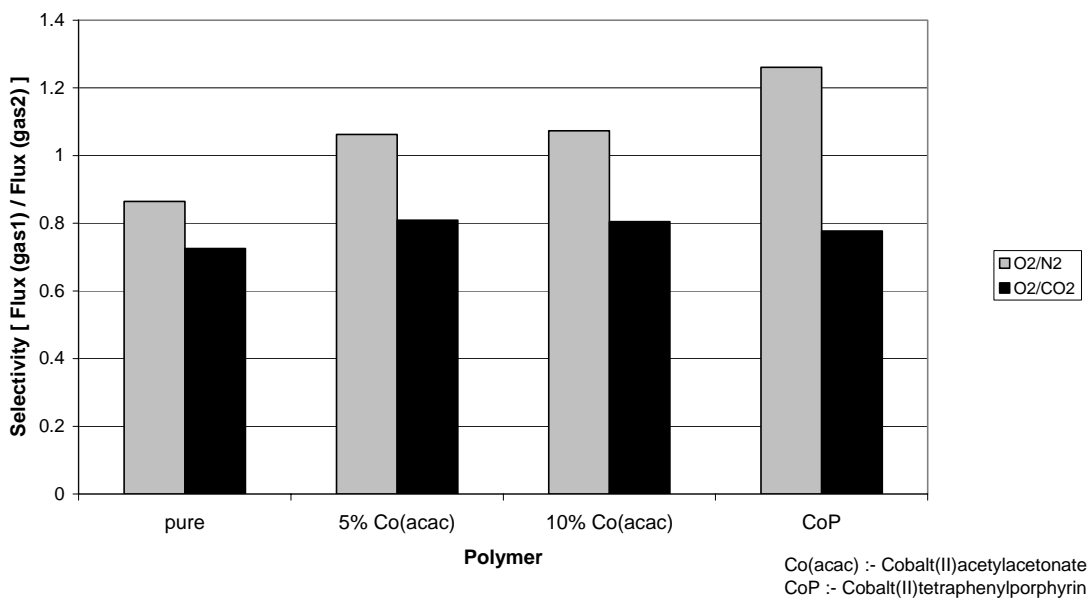

**Figure 1.1.2:** Gas Selectivities for Poly(methyl methacrylate) membranes blended with different Cobalt complexes (measured at an upstream side pressure of 21 psi)

The results were similar to those seen for polystyrene based membranes described earlier. Slight increases in $O_2/N_2$ selectivities were observed for membranes having the cobalt chelates. However this increase was not substantial and all measured gas selectivities were close to 1, thereby, indicating that the membrane was not effective in separating gases under the test conditions.

### (3) PDMS

Sylgard based poly(dimethylsiloxane)membranes failed to solidify with the cobalt complexes. We decided to go ahead and test these membranes as liquid membranes held in the pores of a solid support. However, the CVVP apparatus could not be used to study the selectivity of the sylgard based PDMS liquid membrane systems where the oxygen binding carrier complex is dissolved in a liquid/liquid-gel and held in a micro porous solid support matrix. Due to the large pressure difference between the upstream and the downstream chambers (~14 psi) the liquid/liquid-gel was forced out of the pores of the solid support and no data on selectivity was obtained.

### (4) Polyvinylpyridine Membranes

The permeability results for the different polyvinylpyridine membranes casted on microporous steel discs (pure and 10% CoP) are given in Tables 1.1.3A and 1.1.3B.

**Table 1.1.3A:** Membrane permeability at different inlet pressures for pure polyvinylpyridine Membranes

| Pressure(psi) | Flux | | |
|---|---|---|---|
| | Oxygen $X10^8$ moles/s | Nitrogen $X10^8$ moles/s | Carbon dioxide $X10^8$ moles/s |
| 16.4 | 33.8 | 34.8 | 35.3 |
| 20.4 | 48.9 | 52.4 | 51.6 |
| 24.4 | 65.5 | 71.8 | 69.2 |
| 28.4 | 83.3 | 92.1 | 88.7 |
| 32.4 | 102.5 | 116.5 | 108.9 |
| 36.4 | 120.4 | 137.6 | 129.2 |

**Table 1.1.3B:** Membrane permeability at different inlet pressures for polyvinylpyridine blended with 10 wt% CoP membranes

| Pressure(psi) | Flux | | |
|---|---|---|---|
| | Oxygen $X10^8$ moles/s | Nitrogen $X10^8$ moles/s | Carbon dioxide $X10^8$ moles/s |
| 16.4 | 76.9 | 77.9 | 73.7 |
| 20.4 | 106.2 | 106.5 | 101.9 |
| 24.4 | 137.9 | 137 | 132 |
| 28.4 | 173.8 | 172.5 | 163 |
| 32.4 | 208.4 | 208 | 195.4 |

The selectivity of the membrane for the two different gases was calculated by taking the ratio of molar fluxes for the two gases. The selectivity of pure polyvinylpyridine and 10% polyvinylpyridine membranes towards oxygen/nitrogen and oxygen/carbon dioxide at an upstream side pressure of 16 psi is given in Figure 1.1.3. Once again, all the gas selectivities measured are close to 1, thereby, indicating that this type of membrane is not very effective in separating gases.

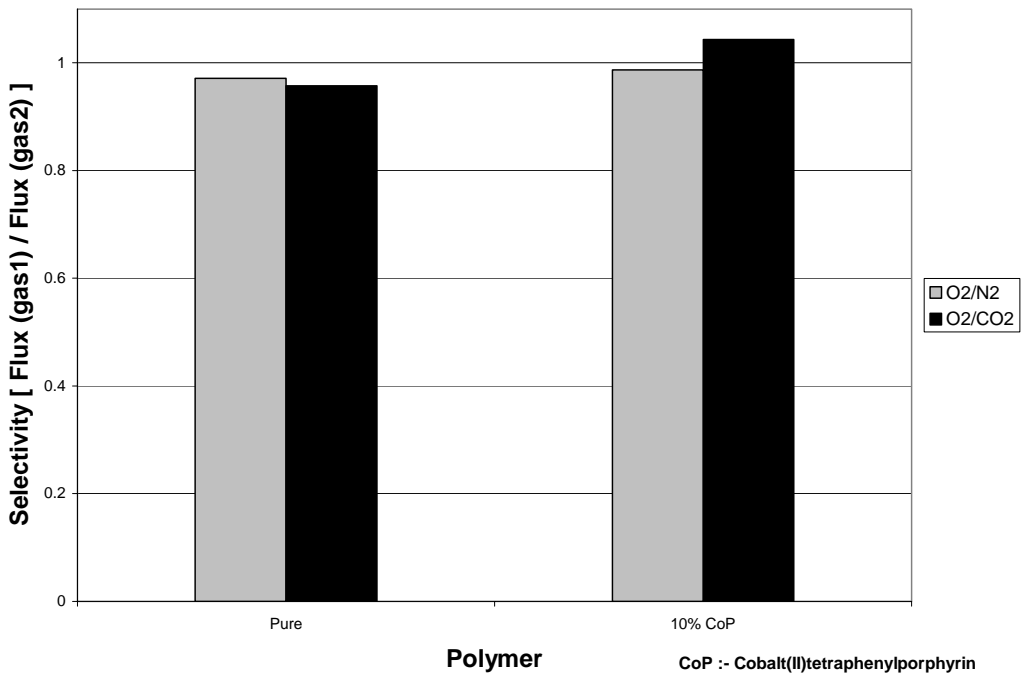

**Figure1.1.3:** Gas selectivity for Polyvinylpyridine membrane blended with Cobalt porphyrin (measured at an upstream side pressure of 16 psi).

The permeability tests discussed above show the results for a stand alone membrane not interfaced with the fuel cell. A large pressure difference between the two sides of the membrane resulted in a larger diffusion of the gases across the membrane, which swamped the effect of the selectivity, i.e. "facilitated transport" of gases. Hence, selectivity for most of the membranes studied yielded selectivity around 1, indicating that the membranes were not very effective in keeping out carbon dioxide under the test conditions. However, one trend seen in the CVVP testing of all membranes was that blending the cobalt complexes with the polymer increased the $O_2/N_2$ selectivity, albeit slightly. We think that this indicates that a positive contribution to selective oxygen permeability due to the cobalt complex is in fact occurring. However, this contribution could not be studied and quantified accurately by the CVVP technique. Therefore, a recommendation to overcome this problem is discussed in the last section of this report.

# Part 2. Results and Discussion on the Quantification of $CO_2$ Poisoning in Air Breathing, Methanol Fueled Alkaline Fuel Cell.

## Accelerated Testing: $t_{max}$ and $R_{max}$

As explained in Chapter 3.2 of the Activities section of this report, we defined two variables, $t_{max}$ (unit of time) and $R_{max}$ (units of current/time), to quantify the $CO_2$ "poisoning" in alkaline fuel cells. These two variables were used to compare and contrast the cell performances when operated in a $CO_2$ enriched atmosphere. To elucidate the physical meaning of $t_{max}$ and $R_{max}$, current vs. time plots of a fuel cell running (1) in air and (2) in a gas mixture of 85% $O_2$-15% $CO_2$ are shown in Figure 1.2.1. The plot of the fuel cell running in a carbon dioxide enriched atmosphere has a sigmoid shape, which is characteristic of the fuel cells undergoing carbon dioxide poisoning with a non-circulating electrolyte. At the beginning of the experiment when most of the electrolyte was a mixture of KOH and methanol, the current output was recorded as 83 mA (for HKU-002C fuel cells obtained from Hong Kong University). As the $CO_2$ poisoning began to affect the cell's electrolyte, current output decreased gradually for the first hour but then started falling rapidly at around 1.5 hours. At the end of the run, the current stabilized at 15 mA and remained stable no matter how long carbon dioxide enriched oxygen was fed into the fuel cell. At the time of 1.5 hours, the KOH had been completely converted to $K_2CO_3$ by the $CO_2$ reacting in the electrolyte  The time at which current decay rate is the highest is referred to as **$t_{max}$** and the value of current decay rate (dI/dt) corresponding to this time is **$R_{max}$**.

These characteristic quantities are graphically represented in Figure 1.2.2, in which rate of current decay is plotted with time. The value of the current decay rate was highest at 1.5 hours ($t_{max}$) and had a value of -33.5 mA/hr ($R_{max}$). Since the $CO_2$ "poisoning" would be negligible in a time span of 3-5 hours when the cell is running in air (0.03% $CO_2$), $t_{max}$ and $R_{max}$ were essentially absent as shown in Figure 1.2.2. Thus, the inverted bell shaped curve in Figure 1.2.2 is characteristic of a fuel cell being poisoned in an atmosphere enriched with $CO_2$ and was observed in all accelerated poisoning cell runs.

What is the rationale of quantifying $CO_2$ poisoning in terms of $t_{max}$ and $R_{max}$? If the values of $t_{max}$ and $R_{max}$ are known for an alkaline fuel cell undergoing poisoning, one can precisely predict the course of the cell current with the passage of time using the phenomenological model developed in this work (refer to section 3.2 in the Activities section).

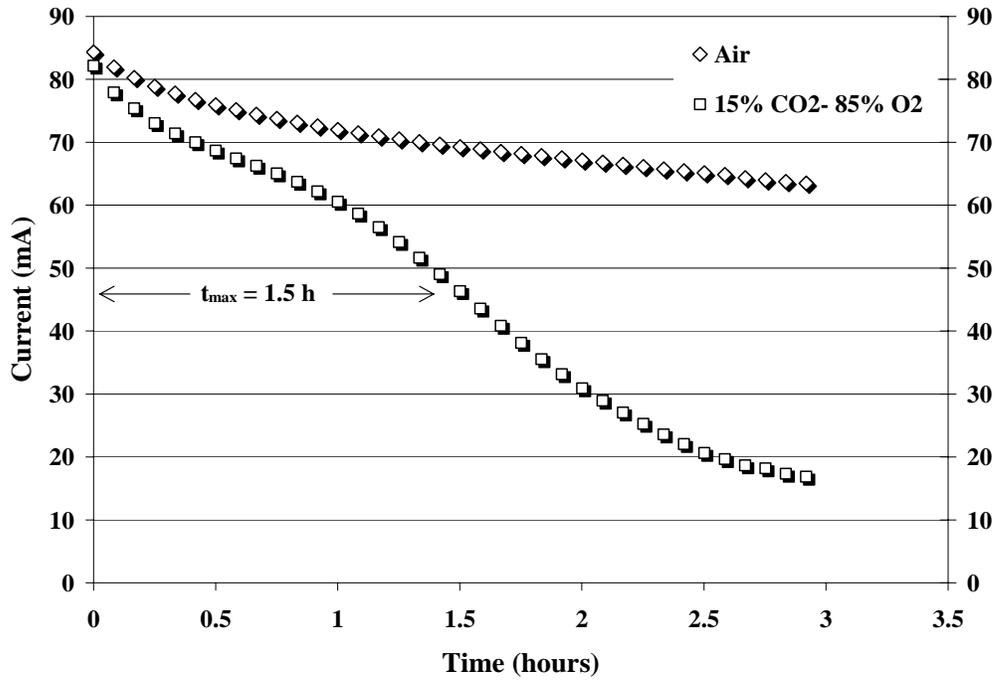

**Figure 1.2.1:** Current vs. Time plots for fuel cells running in air and 85% $O_2$–15% $CO_2$ at 298 K with an applied load of 1.3 Ω. (Composition of Air: 21Vol% $O_2$-0.03 Vol%$CO_2$).

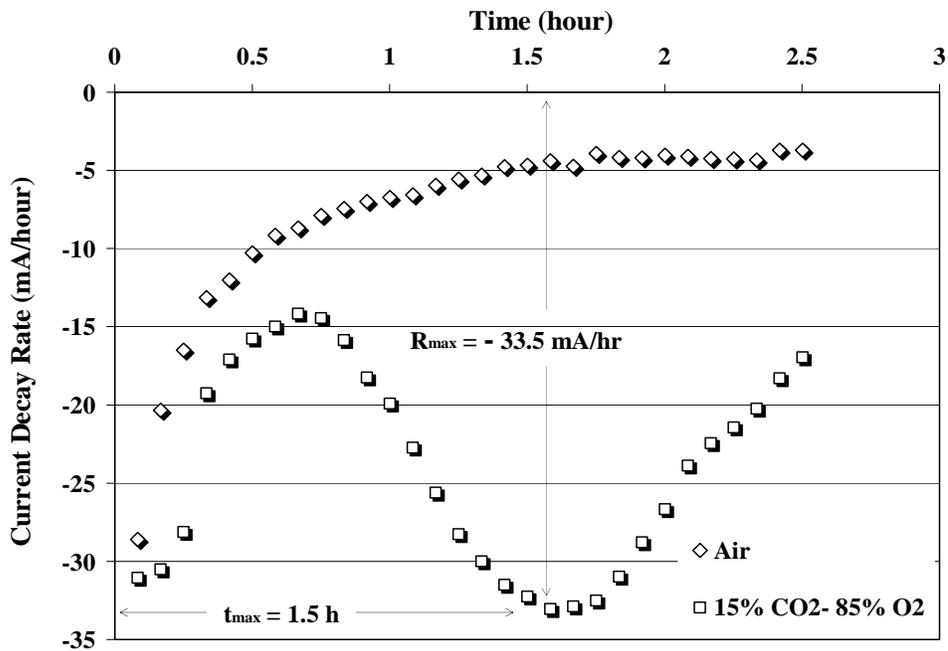

**Figure 1.2.2:** Current Decay Rate vs. Time plot for the fuel cells running on 100% $O_2$ and 85% $O_2$–15% $CO_2$ at 298 K with an applied load of 1.3 Ω

Thus, we found the dependence of $t_{max}$ and $R_{max}$ as a function of operating parameters, i.e. **atmospheric $CO_2$ concentration** and **cell applied load,** so that by knowing these quantities, one can predict the cell performance over a period of time in a $CO_2$ enriched atmosphere using the phenomenological model. In addition, the quantities $t_{max}$ and $R_{max}$ would also serve as benchmarks to compare the performance of a cell at different $CO_2$ "poisoning" or cell life scenarios. Any improvement in the performance at a given atmospheric composition and load is reflected on the values of these quantities.

For a given set of operating conditions, $t_{max}$ and $R_{max}$ have specific values but if the process of poisoning is delayed or in other words, the cell life is prolonged, the value of $t_{max}$ will go up and $R_{max}$ will go down.

## Dependence of $t_{max}$

### a) $t_{max}$ dependence on carbon dioxide concentration:

$t_{max}$ varies inversely with the rate of carbon dioxide poisoning. The value of $t_{max}$ would be smaller for a faster rate of poisoning. Accordingly, as the percentage of carbon dioxide in the atmosphere increases, the value of $t_{max}$ decreases. Figure 1.2.3 illustrates this concept by showing the dependence of $t_{max}$ on the atmospheric concentration of carbon dioxide at $298^oK$. As shown in Figure 1.2.3, $t_{max}$ varied quasi-linearly with the carbon dioxide concentration between 10% and 30%. As the concentration went higher than 30%, the relation lost linearity and eventually stabilized at 0.8 hours. The deviation from the linearity above 30% can be explained by the fixed permeability of the carbon dioxide through the air cathode. Even though the amount of carbon dioxide in the atmosphere increased, its entry into the cell (electrolyte) was limited by the permeability of the air cathode. On the other hand, at carbon dioxide concentrations below 10%, the plot again deviated from linearity and touched the Y-axis asymptotically. The reason for the sharp increase in $t_{max}$ values below 10% is that as the carbon dioxide concentration approaches zero, a fuel cell would take an infinitely long time to degrade (considering only $CO_2$ poisoning as the degradating factor). Consequently, $t_{max}$ would approach infinity.

### b) $t_{max}$ dependence on applied load:

Interestingly, $t_{max}$ was also found to be a strong function of the load applied on the fuel cell. Figure 1.2.4 shows the dependence of $t_{max}$ on the applied load at different concentrations of carbon dioxide in the atmosphere. As before, all experimental parameters, other than the applied load and the carbon dioxide concentration in the chamber, were kept constant. For every $CO_2$ concentration, as the applied load on the cell was decreased, the value of $t_{max}$ went down. At 15% concentration of carbon dioxide in the atmosphere, the value of $t_{max}$ was 3.6 hours with a load of 10.2 Ω as compared to 1.5 hours with a load of 1.3 Ω. Small loads seemed to expedite the carbon dioxide poisoning in the fuel cells. A smaller load would withdraw a higher current from the fuel cell. It would mean that the rate at which oxygen is being consumed at the cathode is also high. A higher consumption rate would lead to higher intake of oxygen from the atmosphere. Since, carbon dioxide is also present in the atmosphere, the rate of carbon dioxide intake would go up simultaneously, leading to faster cell degradation.

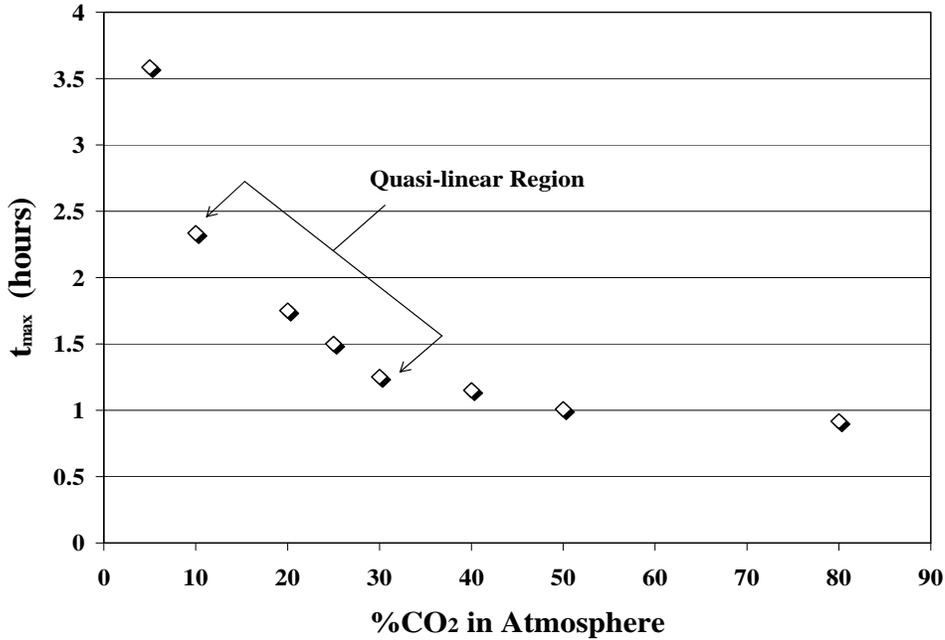

**Figure 1.2.3:** Dependence of $t_{max}$ on the carbon dioxide concentration in the atmosphere at 298 K with an applied load of 1.3 $\Omega$.

## Dependence of $R_{max}$

### a) Rmax dependence on CO2 concentration:

The dependence of $R_{max}$ on the concentration of carbon dioxide in the atmosphere is straight forward.  Higher concentrations of carbon dioxide would lead to higher poisoning rates and, hence, larger $R_{max}$ values. Figure 1.2.5 shows the variation in $R_{max}$ with atmospheric carbon dioxide concentration when the applied load was 1.3 $\Omega$. It is observed that $R_{max}$ varied quasi-linearly with carbon dioxide concentration between 5% and 30% in the atmosphere.  As the concentration of carbon dioxide was increased further, the linearity was lost and $R_{max}$ stabilized at approximately -65 mA/hour.  This observation indicates that at very high concentrations of carbon dioxide (> 60%), $R_{max}$ essentially becomes constant at a specified load.  The reason for this departure from linearity at high $CO_2$ concentrations is similar to that observed with $t_{max}$ in the preceding section, i.e. limited permeability of the air cathode.  On the other hand, when the concentration of carbon dioxide approached zero, the $R_{max}$ value also tended to zero. It can be confirmed by extrapolating the curve in Figure 1.2.5 towards 0% concentration values. This suggests that at infinitesimally small carbon dioxide concentrations, $R_{max}$ would be infinitesimally small and, hence, carbon dioxide poisoning would be too negligible to measure.

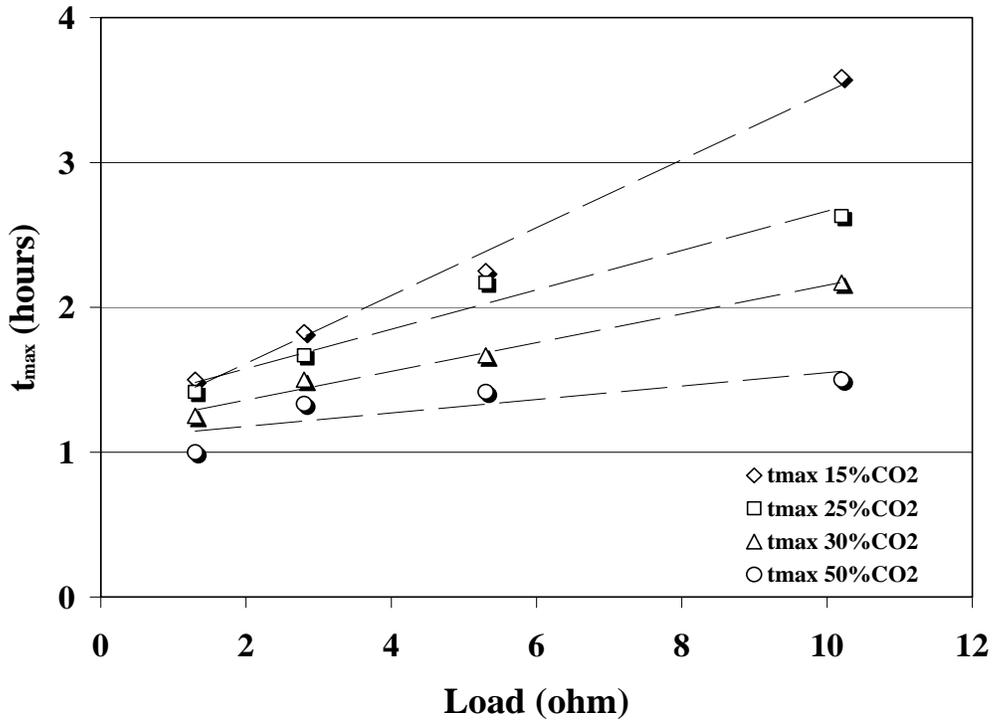

**Figure 1.2.4:** Dependence of $t_{max}$ on the applied load at different atmospheric compositions and 298 K

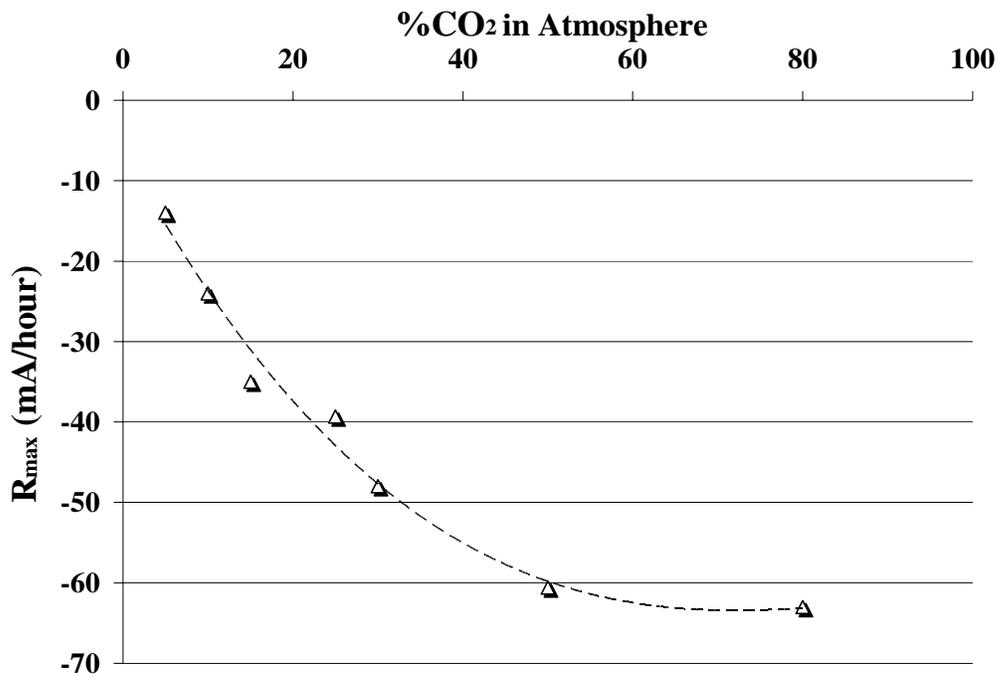

**Figure 1.2.5:** Dependence of $R_{max}$ on the atmospheric carbon dioxide concentration at 298 K with an applied load of 1.3 Ω.

### b) $R_{max}$ dependence on applied load:

$R_{max}$ was also found to be strongly influenced by the applied load, as in the case with $t_{max}$. Figure 1.2.6 shows that $R_{max}$ varied with the applied load in a hyperbolic fashion. It should be noted that in figure 1.2.6 the absolute value of $R_{max}$ is plotted. In reality, the $R_{max}$ value is negative because of the decay of the current. Reasons for higher absolute values of $R_{max}$ at smaller loads are the same as for lower $t_{max}$ values at smaller loads. But unlike $t_{max}$, $R_{max}$ became an even stronger function of the applied load at very high concentrations of carbon dioxide (>50%). Figure 1.2.6 also indicates that as the carbon dioxide concentration was reduced, the dependence of $R_{max}$ on the applied load decreased and, at extremely small concentrations, this dependence would cease to exist. This is opposite of the behavior of $t_{max}$, which became sensitive to the applied load at lower carbon dioxide concentrations.

Therefore, we suggest that at low carbon dioxide concentrations (< 2%), the poisoning is defined by $t_{max}$ as $R_{max}$ becomes largely invariable. In the same way, at high carbon dioxide concentrations (> 60%), the poisoning is defined by $R_{max}$ as $t_{max}$ becomes constant. And at intermediate concentrations of carbon dioxide, we need to specify both quantities in order to completely define the carbon dioxide poisoning in the AFC. Thus, the above studies lead us to conclude that for any future testing of AFCs using the accelerated poisoning methodology, the best region to study the cell is between 10%–30% of carbon dioxide in the atmosphere.

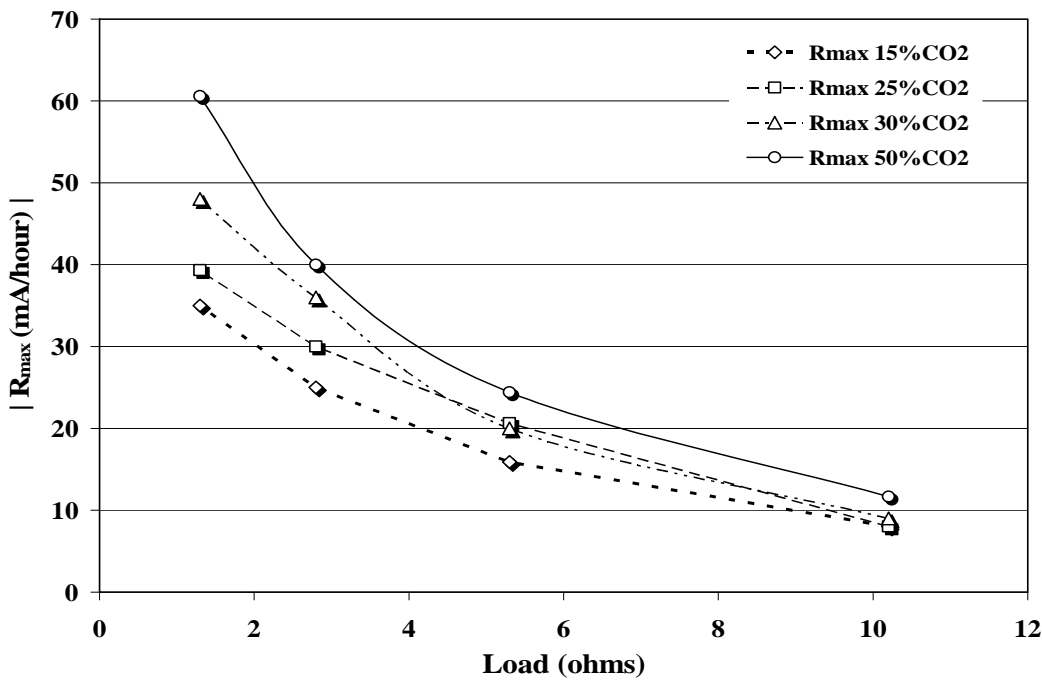

**Figure 1.2.6** Dependence of $R_{max}$ (absolute value) on the applied load at different atmospheric compositions and 298 K

## $CO_2$ Production by Internal Oxidation of Methanol

One possible complication in interpreting the results could arise because of the in situ carbon dioxide production by the oxidation of the methanol. Since methanol was used as the fuel, carbon dioxide was produced as a by-product of the anodic oxidation of methanol according to the following reaction.

$$CH_3OH + 6OH^- \rightarrow CO_2 + 5H_2O + 6e^-$$

The above reaction shows that 1 mole of methanol produces 1 mole of carbon dioxide during complete oxidation. Thus, a passage of current would cause production of carbon dioxide at the anode and, hence, contribute to the carbon dioxide poisoning in the fuel cells. In addition, another source of carbon dioxide inside the fuel cells is the parasitic oxidation of methanol at the cathode due to methanol reactions. Therefore, it is necessary to quantify the cumulative production of carbon dioxide by the oxidation of methanol and delineate its contribution to the total carbon dioxide poisoning measured; i.e. we need to separate the contribution of the $CO_2$ poisoning when the $CO_2$ contamination is coming from the cathode and when the $CO_2$ is coming from the methanol reactions.

By decreasing the applied load (allowing higher current flows), we would increase the contribution of internally generated carbon dioxide (methanol oxidation reactions) in the total carbon dioxide poisoning effect. In order to relate the amount of $K_2CO_3$ formed by $CO_2$ contamination of the alkaline electrolyte (see Chapter 1 of "Project Activities" section) solely due to the internal methanol oxidation to the total amount of $K_2CO_3$ formed when contaminated with $CO_2$ oxidant is used at the air cathode, we ran our fuel cells in a 100% $O_2$ environment for 3.33 hours. Consequently, the conversion of KOH to $K_2CO_3$ will be solely due to the $CO_2$ generated internally by methanol oxidation. We started each run with 1M of KOH fresh solution as the electrolyte. At the end of the experiment, the electrolyte was sampled out and its composition was determined using *Winkler's* method of volumetric hydroxide/carbonate estimation using *bromocresol green* and *phenolphthalein* as the indicators [1]. $N_2$ was bubbled for 2-3 minutes through the sampled electrolyte before each titration was carried out. This was done in order to get rid of any dissolved carbon dioxide in the electrolyte. Dissolved carbon dioxide might affect the titration results by forming traces of carbonic acid ($H_2CO_3$). Results of this composition analysis are presented in Figure 1.2.7, which plots $K_2CO_3$ formed or KOH remaining as a function of the applied load after a run of 3.33 hours in a 100% $O_2$ environment.

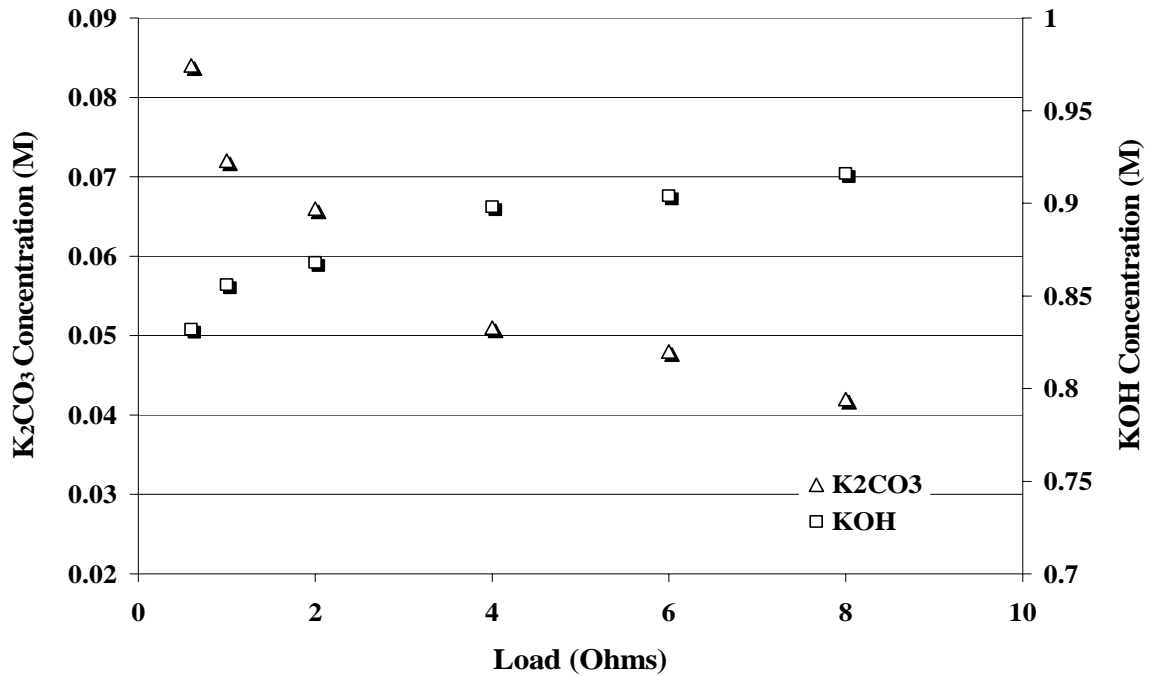

**Figure 1.2.7** Dependence of $K_2CO_3$ formed/KOH remaining with the applied load after a fuel cell is run for 12,000 seconds (3.33 hours) in 100% $O_2$ atmosphere

The time dependence of $K_2CO_3$ formed due to methanol oxidation at a specific applied load was determined (Figure 1.2.8). Figure 1.2.8 shows a linear dependence of $K_2CO_3$ formed with time at an applied load of 1Ω and an atmosphere of 100% $O_2$. The straight line describing this time dependence passes through the origin as shown in Figure 1.2.8. Similar linearity in the time dependence of $K_2CO_3$ concentration was assumed for all the other applied loads for further calculations. In order to determine the contribution of internally generated carbon dioxide in the net carbon dioxide poisoning, it was necessary to know the time at which all of the KOH converted to $K_2CO_3$ for any operating condition. Using the Winkler's method of volumetric titration for the electrolyte composition analysis, it was found that at the time equal to twice of $t_{max}$ (t = 2 X $t_{max}$) all of the KOH was converted to $K_2CO_3$ irrespective of the operating conditions. Thus, the amount of $K_2CO_3$ formed by methanol oxidation only up to t = 2 X $t_{max}$ was needed to be determined. Since we knew the value of $t_{max}$ for different $CO_2$ concentrations and applied loads, we calculated the amount of $K_2CO_3$ produced by internally generated carbon dioxide using the data given in Figures 1.2.7 and 1.2.8. From Figure 1.2.7, we determined the amount of $K_2CO_3$ formed at any load after 3.33 hours. This value of $K_2CO_3$ concentration was then scaled down using the fact that the variation of $K_2CO_3$ concentration with time is linear and the resulting straight line passes through the origin as shown in Figure 1.2.8.

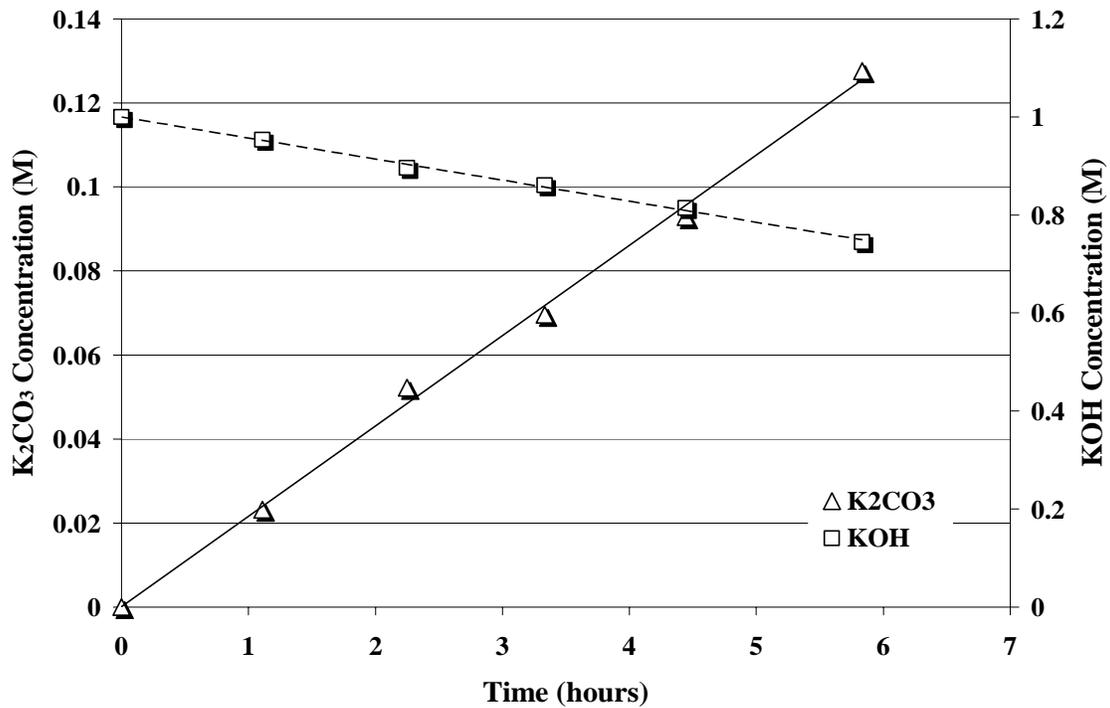

**Figure 1.2.8** Time dependence of $K_2CO_3$ concentration resulting from methanol oxidation at an applied load of 1Ω and 100% $O_2$ atmosphere.

This final value represented the fraction of $K_2CO_3$ formed by internally generated carbon dioxide in the total $K_2CO_3$ formed due to overall carbon dioxide poisoning (internally and from the air cathode contamination when pure $O_2$ was not used as an oxidant). The results of the quantification of the contribution of methanol oxidation in the net (total) $CO_2$ poisoning in our study are presented in Figure 1.2.9.

We observe in Figure 1.2.9 that the contribution of methanol oxidation in the total $K_2CO_3$ production decreases with the increase in carbon dioxide concentration in the atmosphere and the applied load. For the range of operating conditions in our experiments (Load = 1 to 10Ω and %$CO_2$ in Atmosphere = 10% to 50%), the contribution of the internally generated carbon dioxide to the total carbon dioxide poisoning was determined to be between 6% and 14% of the total $CO_2$ poisoning depending if a 90% or 10% $O_2/CO_2$ gas mixture (at the air cathode) was used (Figure 1.2.9).

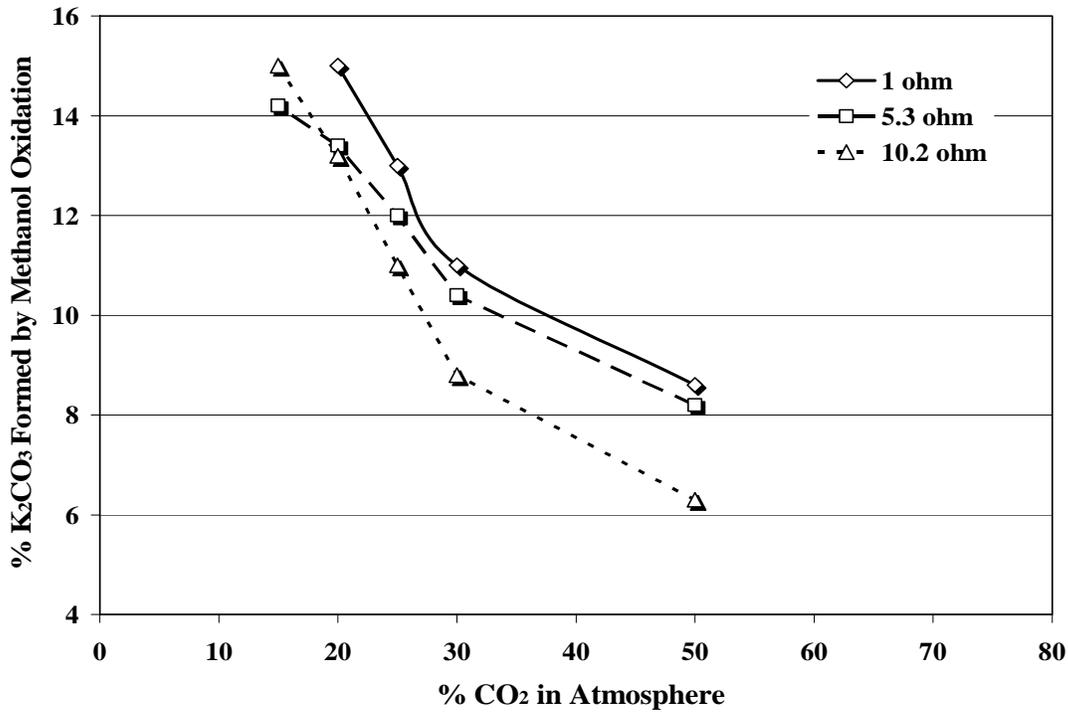

**Figure 1.2.9** Percentage of $K_2CO_3$ formed by internally generated $CO_2$ in the net $K_2CO_3$ produced due to carbon dioxide poisoning.

### Electrolyte Compositional Analysis

In the previous sections, we related parameters defining carbon dioxide poisoning ($t_{max}$ and $R_{max}$) with the operating parameters ($CO_2$ concentration in the air cathode oxidant stream and applied load). The variation in a cell performance is primarily attributed to the change in the composition of its electrolyte with the progress in poisoning. In this section, the dependence of the current output of a cell on the compositional changes of its electrolyte due to carbon dioxide poisoning is elucidated. To understand the relationship between the cell current and electrolyte composition, we carried out dynamic and static analyses of the electrolyte.

## Steady State Electrolyte Testing

We made mixtures of KOH (aq) and $K_2CO_3$ (aq), with concentrations ranging from 0.0 M to 1.0 M. Using each mixture as an electrolyte, we measured the cell steady state current produced by the fuel cell under a load of 1.3 Ω and operated in a 100% $O_2$ oxidant in the air cathode. The concentration of methanol in the electrolyte was maintained at 2.5 M. The mixture combinations used to define the alkaline electrolyte, along with the cell current when a load of 1.3 Ω was applied, are listed in Table 1.2.1.

**Table 1.2.1:** Composition of different mixtures in terms of KOH and $K_2CO_3$ concentrations and the steady state currents obtained using the mixtures as electrolytes.

| Serial number | KOH molarity | $K_2CO_3$ molarity | Steady State Current (mA) |
|---|---|---|---|
| Mix. 1 | 1.0 | 0.00 | 60.0 |
| Mix. 2 | 0.87 | 0.065 | 59.1 |
| Mix. 3 | 0.8 | 0.10 | 58.2 |
| Mix. 4 | 0.6 | 0.20 | 53.9 |
| Mix. 5 | 0.5 | 0.25 | 47.6 |
| Mix. 6 | 0.4 | 0.30 | 41.3 |
| Mix. 7 | 0.3 | 0.35 | 35.0 |
| Mix. 8 | 0.2 | 0.40 | 25.2 |
| Mix. 9 | 0.1 | 0.45 | 18.1 |
| Mix. 10 | 0.0 | 0.50 | 15.2 |

Figure 1.2.10 shows the plot of cell steady state current as a function of the different KOH concentrations in the electrolyte mixture. It can be seen that decreasing the KOH concentration decreases the cell current output, as expected. But the cell current did not drop suddenly until the KOH concentration was reduced from 1 M to about 0.6 M. The cell undergoing $CO_2$ poisoning continues to give the desired current output until the KOH concentration in the electrolyte reached a value (≈0.6 M for our fuel cell system). The dependence of the cell current on the KOH concentration explains the sigmoidal current vs. time plots obtained for fuel cells being $CO_2$-poisoned (Figure 1.2.1).

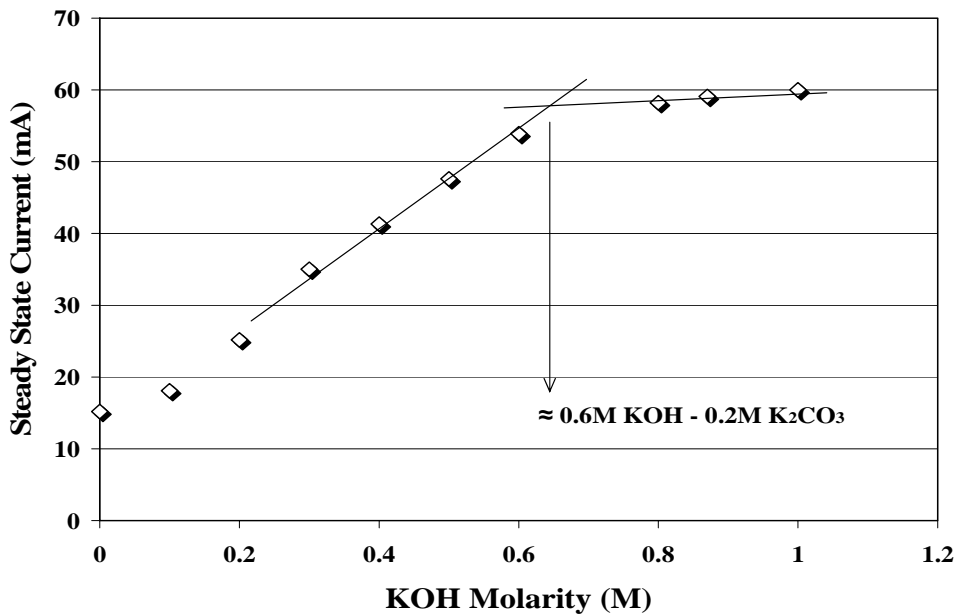

**Figure 1.2.10:** Variation of steady state current output of a fuel cell, at a load of 1.3 Ω, with the KOH concentration in its electrolyte. Operating atmosphere is maintained at 100% $O_2$

We called this analysis steady state because in each case, while measuring the steady state current, the electrolyte composition across the cell was uniform and the electrolyte in the vicinity of the anode and cathode was the same as in the bulk of the cell. This will not be true in the case of a fuel cell undergoing "dynamic" carbon dioxide poisoning due to $CO_2$ poisoning by the cathode side, because the majority of $K_2CO_3$ would be formed at the cathode. Therefore, at any time, the concentration of $K_2CO_3$ around the cathode would be higher than the anode. Thus, we also carried out "dynamic" or non steady state electrolyte analysis to better understand the current dependence on the compositional changes of the electrolyte.

## Dynamic Testing

In dynamic electrolyte analyses, we started with a fuel cell having 1 M KOH, electrolyte and operating in a 70%$O_2$-30%$CO_2$ atmosphere fed by the air cathode to the cell and with an applied load of 1.3 Ω. The fuel cell was stopped after different time intervals of operation and the composition of the electrolyte was sampled and determined using *Wrinkler's method of titration* [1]. Table 1.2.2 summarizes the results of the "dynamic" electrolyte analysis.

**Table 1.2.2:** Composition of the electrolyte in a fuel cell, undergoing poisoning, at different time intervals and the current output at that time.

| Time (seconds) | Composition | KOH Conversion Rate (M/hour) | Current Output (mA) |
|---|---|---|---|
| 250 | 0.87 M KOH-0.65 M $K_2CO_3$ | 1.87 | 59.0 |
| 500 | 0.78 M KOH-0.11 M $K_2CO_3$ | 1.29 | 54.9 |
| 1000 | 0.70 M KOH-0.15 M $K_2CO_3$ | 0.58 | 54.0 |
| 2000 | 0.56 M KOH-0.22 M $K_2CO_3$ | 0.50 | 52.5 |
| 3000 | 0.44 M KOH-0.28 M $K_2CO_3$ | 0.43 | 50.1 |
| 4500 | 0.28 M KOH-0.36 M $K_2CO_3$ | 0.38 | 46.8 |

Table 1.2.2 shows the conversion rate of KOH to $K_2CO_3$ is higher in the beginning, at a value of 1.87 M/hour, as compared to 0.38 M/hour towards the end. The decrease in the conversion rate of KOH can be attributed to the limited mass transfer of the reaction product ($K_2CO_3$) from the reaction site (cathode) to the bulk. Current output of the cell also changed, accordingly, with the change in KOH concentration. Figure 1.2.11 shows the plot of the current output with KOH concentration observed in dynamic electrolyte analysis. It is compared with a similar plot in Figure 1.2.10 obtained from steady state analysis.

A difference in the results obtained from steady state (static) and dynamic analysis can be observed in this plot at lower concentrations of KOH in the electrolyte. At the same composition of electrolyte, the current output given by a cell is higher in dynamic analysis than in static. This can be explained by the non-uniform concentration distribution of $K_2CO_3$ between the electrodes in a dynamically degrading fuel cell, as explained earlier. Although the overall concentration of $K_2CO_3$ was the same in both steady state and "dynamic" analysis, $K_2CO_3$ concentration in the electrolyte was not uniform in the "dynamic" case. The concentration of KOH around the anode is higher in the "dynamic" case. Since it is the kinetics of methanol oxidation at the anode which is lower at higher $K_2CO_3$ concentrations [2, 3], the cell current output on the dynamic test is higher than in the "static" case. The difference between results in the "static" and dynamic" cases becomes zero if the electrolyte between the electrodes is stirred or if the distance between the cathode and anode is reduced.

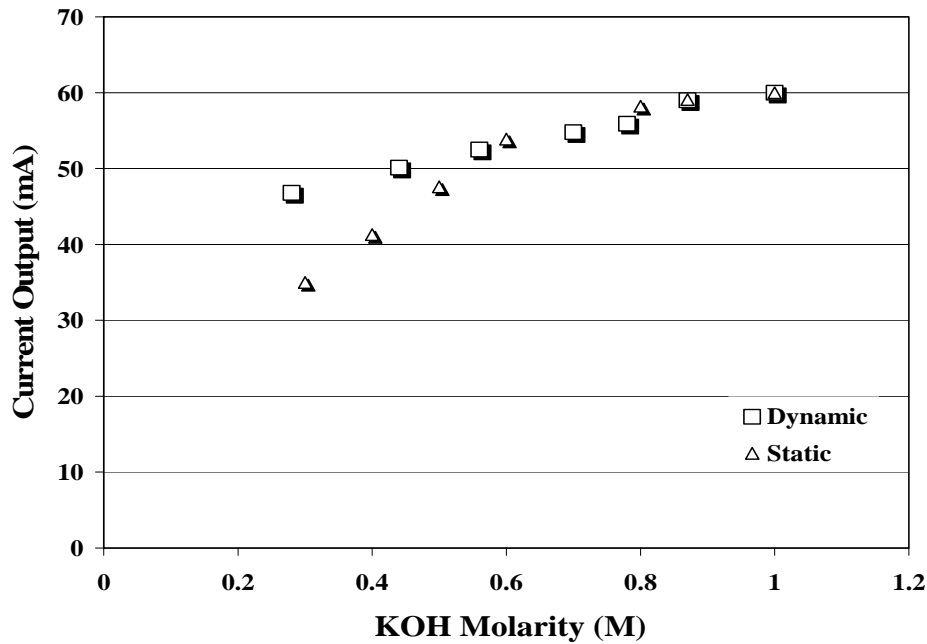

**Figure 1.2.11:** Plot of Current output vs. KOH concentration for static and dynamic electrolyte analysis at a load of 1.3 $\Omega$ and in 100% $O_2$ atmosphere.

## Electrode Polarization and Electrolyte Conductivity Studies

The static and dynamic analysis of the electrolyte showed a strong dependence of current output of a fuel cell on its electrolyte composition. The conversion of KOH to $K_2CO_3$ by carbon dioxide poisoning makes the oxidation of fuel at the anode difficult. This conclusion is supported in the potentiostatic polarization curves of the anode and cathode for a fuel cell running on 1M KOH and 0.5M $K_2CO_3$ separately, as shown in Figure 1.2.12. Methanol concentration in both cases was 2.47M. Figure 1.2.12 suggests that the reason for low current output given by a cell in the presence of $K_2CO_3$ is the sluggish kinetics of methanol fuel oxidation at the anode. The kinetics of oxygen reduction at the cathode remains unchanged for both KOH and $K_2CO_3$ electrolytes.

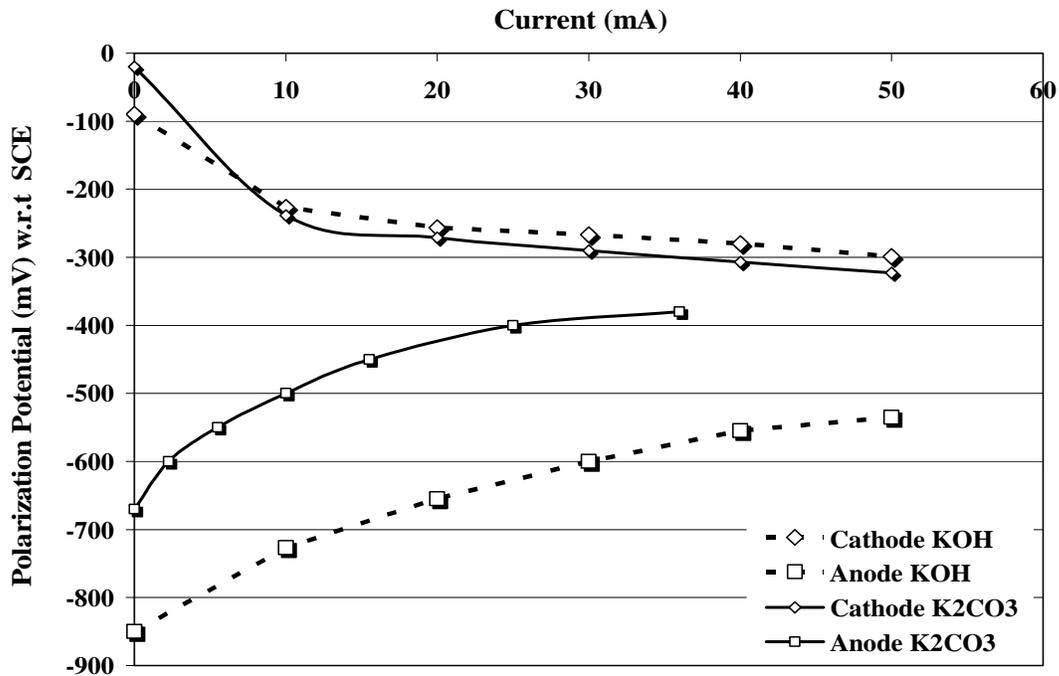

**Figure 1.2.12** Anodic and Cathodic potentiostatic polarization curves of a fuel cell with 1M KOH and 0.5M $K_2CO_3$ as electrolytes.

Another factor which may decrease the current output of a fuel cell with $K_2CO_3$ as the electrolyte is the increased ohmic loss (refer polarization curves for HKU-002C Fuel Cells) because of the lower conductivity of $CO_3^{(2-)}$ ions as compared to $OH^{(-)}$ ions [4]. Figure 1.2.13 plots the conductivities of mixtures, tabulated in Table 1.2.1, with the KOH concentration in the mixtures. The dependence of the conductivity of the electrolyte on KOH concentration was found to be linear. Therefore, lesser conductivity at lower mole fractions of KOH as compared to $K_2CO_3$ in the electrolyte also lowers the overall performance of a fuel cell as the carbon dioxide poisoning proliferates in it.

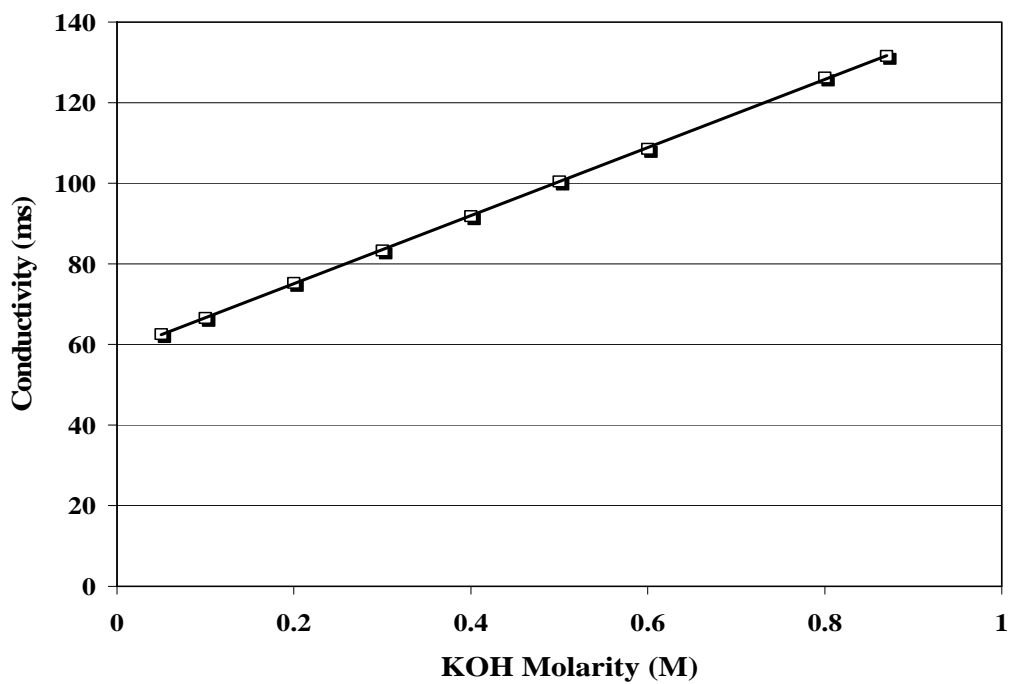

**Figure 1.2.13:** Variation of electrolyte conductivity with KOH concentration.

## References

1. D.A. Skoog, D.M. West, F.J. Holler, Fundamentals of Analytical Chemistry, seventh ed., Saunders College Publishing, pp. 257-262.

2. E.H. Yu, K. Scott, R.W. Reeve, Journal of Electroanalytical Chemistry, **547** (2003) 17-24.

3. E.J. Cairns, Handbook of Fuel Cells- Fundamentals Technology and Applications, first ed., 2003 pp 301-309.

4. D.R. Lide, CRC Handbook of Chemistry and Physics, eighty third ed., CRC press publication.

# Part 3. Result and discussion of $CO_2$ Poisoning Model

We developed a model (as described in section 3.2 of "Project Activities") to accurately determine a fuel cell's performance in a carbon dioxide enriched atmosphere. The model is capable of determining the time when the electrolyte should be changed in the cell such that the cell's life and efficiency are optimized.

For this model to be of any importance, it was necessary to validate the model results with the experimental results. For this purpose, we plotted a current vs. time curve of a fuel cell for specified operating conditions using our model. Then under the same conditions, we ran our fuel cells to get the experimental data. We observed a close matching of the values calculated with the model when compared to the data obtained experimentally. Figure 1.3.1 and Figure 1.3.2 compare the results of our model with the experimental data at operating conditions specified as (50%$O_2$-50%$CO_2$ and 2.7Ω) and (85%$O_2$-15%$CO_2$ and 5.3Ω), respectively.

The model created was validated and the results were satisfactory.

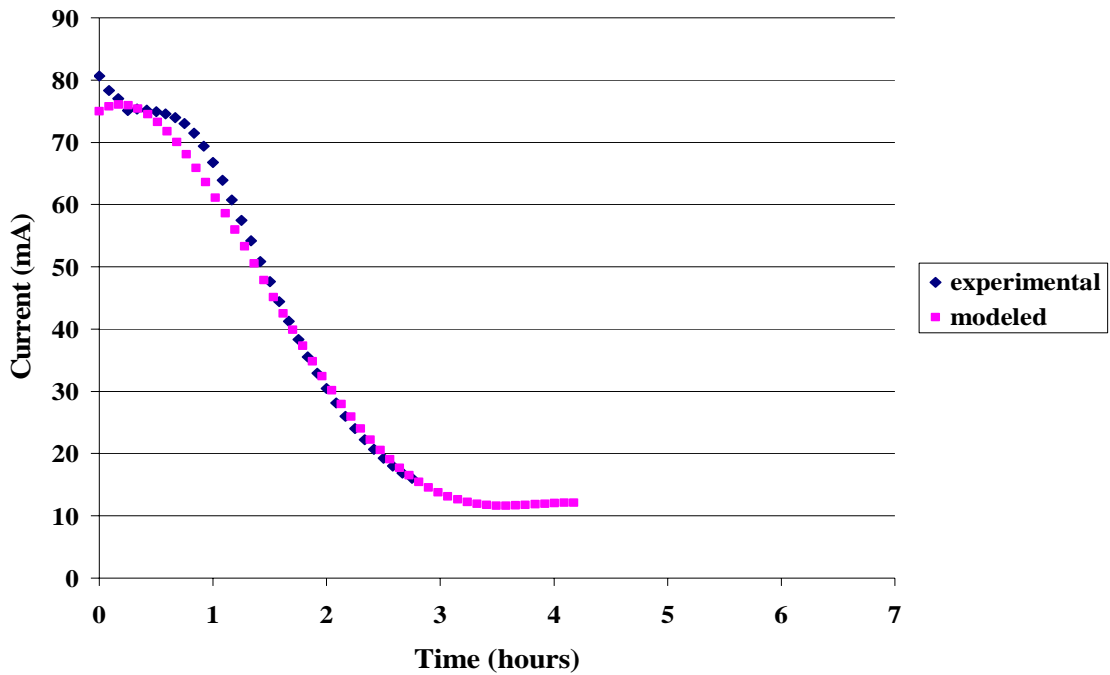

**Figure 1.3.1:** Comparison of model results with the experimental data at 50%$O_2$-50%$CO_2$ atmosphere and 2.7Ω load.

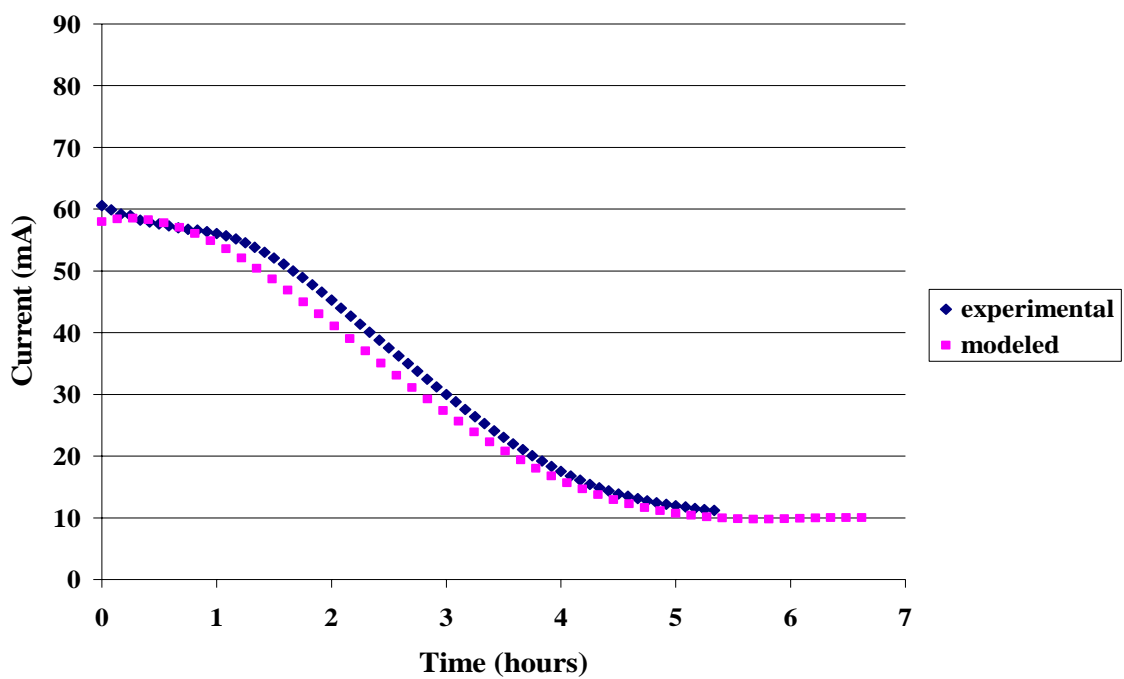

**Figure 1.3.2:** Comparison of model results with the experimental data at 85%$O_2$-15%$CO_2$ atmosphere and 5.3Ω load.

## Part 4. Evaluation of Polymer Membranes Interfaced with the Air Cathode of an AFC

After understanding the reasons of the $CO_2$ poisoning and developing a methodology and a model to evaluate the subsequent degradation, we went ahead and evaluated the cell performances when a polymeric membrane was placed upstream of the air cathode and the alkaline methanol fuel cell was fed with a $CO_2/O_2$ mixture as an oxidant. The gas mixture was feed to the cell by the air cathode. We show, herewith, the results obtained when Polystyrene and Polyvinyl Pyridine (PVP) membranes were interfaced with the air cathode.

This procedure was followed for fuel cells with and without the polymer membranes. Membranes of polyvinyl pyridine **(4)** and polystyrene **(1)** were fabricated on a filter paper so as to give the membranes required mechanical stability.

The polymers were dissolved in toluene at room temperature and then poured over 1.97''x 2.76'' filter paper for casting. The amount of solution poured was directly proportional to the required thickness of the membranes. The solvent was then evaporated in nitrogen until an absolutely dry membrane was obtained, 250 μm thick for a polystyrene membrane and 250 thick for a polyvinyl pyridine membrane.

These membranes were then mounted in the air-cathodes for the electrochemical testing, as described earlier. The air-cathode of one fuel cell was unmasked in the carbon dioxide rich atmosphere while the other two were coupled with polyvinyl pyridine and polystyrene membranes.

The membranes were intended to be tested for their efficiency in dealing with the carbon dioxide poisoning problem in alkaline fuel cells. To test these polymer membranes, a similar experimental setup was used as shown in Figure 3.2 (Refer Project Activities). Results of the membrane evaluation are summarized in Figure 1.4.1. Figure 1.4.1 shows that a pure polystyrene membrane gives the fuel cell better protection from $CO_2$ poisoning than pure PVP membrane. The decay in the current is less severe in the case of the polystyrene membrane, as shown in Figure 1.4.1, than in the case of the PVP membrane. The same observation is corroborated by the plot of the current decay rate vs. time in Figure 1.4.2. The characteristic quantity $t_{max}$ has gone up from 3.42 hours, in the case of a pure PVP membrane, to 6.0 hours, in the case of a polystyrene membrane. Both these quantities are way more than in the case of fuel cell with no membrane where the $t_{max}$ value is just 1.42 hours (Figure 1.4.2). Therefore, placing a pure polystyrene membrane in the fuel cell increases the life of the fuel cell by 4.3 times as compared to the fuel cell left uncovered in the high $CO_2$ atmosphere.

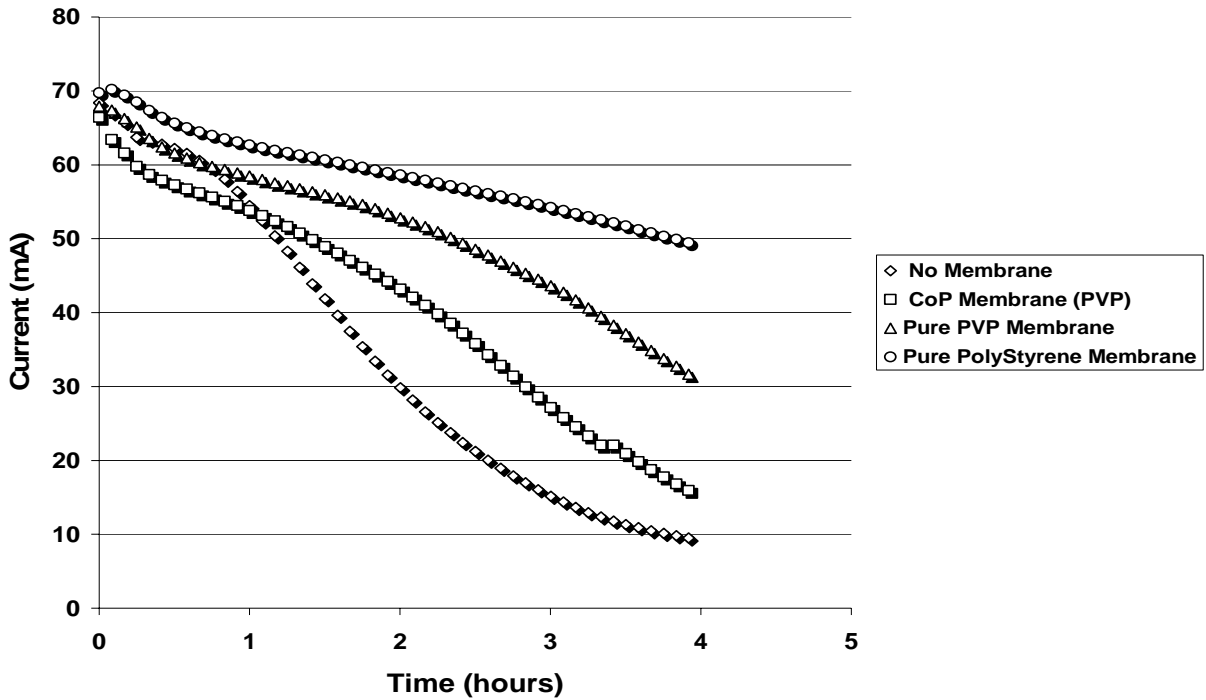

**Figure 1.4.1:** Current vs. Time plots for the fuel cells with no membrane, CoP membrane, Pure PVP membrane and Pure Polystyrene membrane. The atmosphere was maintained at 70%$O_2$-30%$CO_2$ and the load applied was 1.3 Ω.

An important thing to note here is that a polystyrene membrane is not enriching the air as compared to a PVP membrane. It is just that the polystyrene membrane we used in our experiments was more efficient in cutting down the diffusion of $CO_2$ in the fuel cell as compared to the PVP membrane. It could either be because of the thickness of the polystyrene membrane or because of the denser (high crosslink) structure of it.

Therefore, it is concluded after doing a series of experiments, that both pure Polystyrene and PVP membranes reduce the carbon dioxide poisoning problem by lowering the diffusion of the gases (both $O_2$ and $CO_2$) to the air cathode. The diffusion of any gas through a solid polymer membrane would be less due to physical barriers for gas transport. The greater the thickness of this physical barrier, the less the diffusion of gases across it.

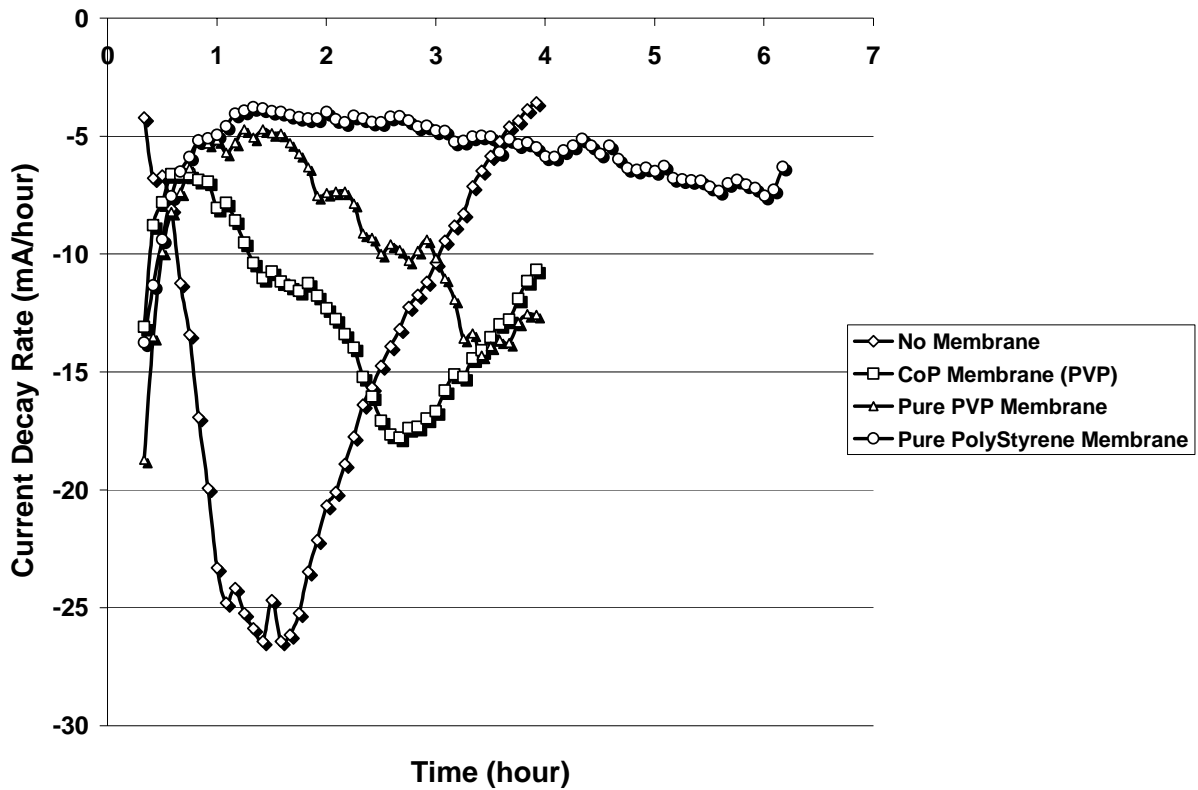

**Figure 1.4.2:** Current Decay Rate vs. Time plots for the fuel cells with no membrane, CoP membrane, Pure PVP Membrane and Pure Polystyrene membrane. The atmosphere was maintained at 70%$O_2$-30%$CO_2$ and the load applied was 1.3 Ω.

# Part 5. Results and Discussion of $CO_2$ Poisoning When an AFC is fueled with "Dirty" Hydrogen.

The alkaline fuel cells, as described in Chapter 3.3 of the "Project Activities" section of the report, operating on air/hydrogen uses an aqueous 2 M KOH electrolyte feed to the cell by using a peristaltic pump at a rate of 1 ml/s. In a typical experiment, a load of 1 ohm was placed across the cell and the cell was operated on air/pure-$H_2$ for 2000 sec. to allow the cell current to stabilize. We show the variation of the cell current with time due to the electrolyte $CO_2$ poisoning due to the fact that "dirty" hydrogen instead of pure $H_2$ was fed to the cell at the anode side as a fuel. "Dirty" hydrogen was defined as a gas mixture of $O_2$ and $CO_2$ at different weight percentiles.

## Variation of fuel cell current output with time

The current vs. time plots for a fuel cell with and without the polystyrene membrane placed upstream of the cell gas fuel running on 5% $CO_2$/95% $H_2$ contaminated "dirty" hydrogen for a load of 1 Ω is shown in Figure 1.5.1.

When $CO_2$ is presented in the $H_2$ stream, we noticed a sharp current decrease at the time that the $CO_2$ was injected in the gas stream (0.5 hrs). The sharp change in cell current is more noticeable when a membrane is not present and less noticeable when a double membrane is placed upstream of the cell anode. These results could be explained as the immediate blockage of the anodic membrane due to the formation of potassium carbonate at the membrane pores and due to the fact that a 2 M KOH concentration in the electrolyte was used [1, 2]. A different answer to the sharp cell current drop when the cell becomes in contact with $CO_2$ is non-specific adsorption of $CO_2$ at the catalytic sites and is discussed later [3]. We elucidate in this research program the reasons for this sharp change in cell current, the results are shown in the next sections of this Chapter.

After this initial sharp drop, the plots exhibit a sigmoid shape which is characteristic of the poisoning effect on the cell electrolyte [4]. After that immediate cell current drop due to the partial blockage of the anode membrane pores, equilibrium is reached in the pores between the potassium carbonate precipitated at the pores and the electrolyte composition.

The fuel cell starts with an initial current output of 640 - 670 mA which drops rapidly to 550 – 600 mA as soon as $CO_2$ is introduced into the $H_2$ stream. As the $CO_2$ poisoning proceeds, the alkaline KOH electrolyte is slowly converted to $K_2CO_3$ resulting in a gradual loss in cell output from about 550 mA to 430 mA. When all the KOH has been converted to $K_2CO_3$, the current levels out to a constant value of about 430 mA. For a fuel cell operating on dirty $H_2$, complete conversion of the electrolyte from KOH to $K_2CO_3$ occurs within an hour for a cell without the polystyrene membrane. For a fuel cell having the protective polystyrene membrane to minimize the $CO_2$ contaminant, the $CO_2$ poisoning process is delayed to 1.5 hours when 1 coat membrane is used. The process is delayed for 2.5 hours when a thicker 2 coat membrane is used.

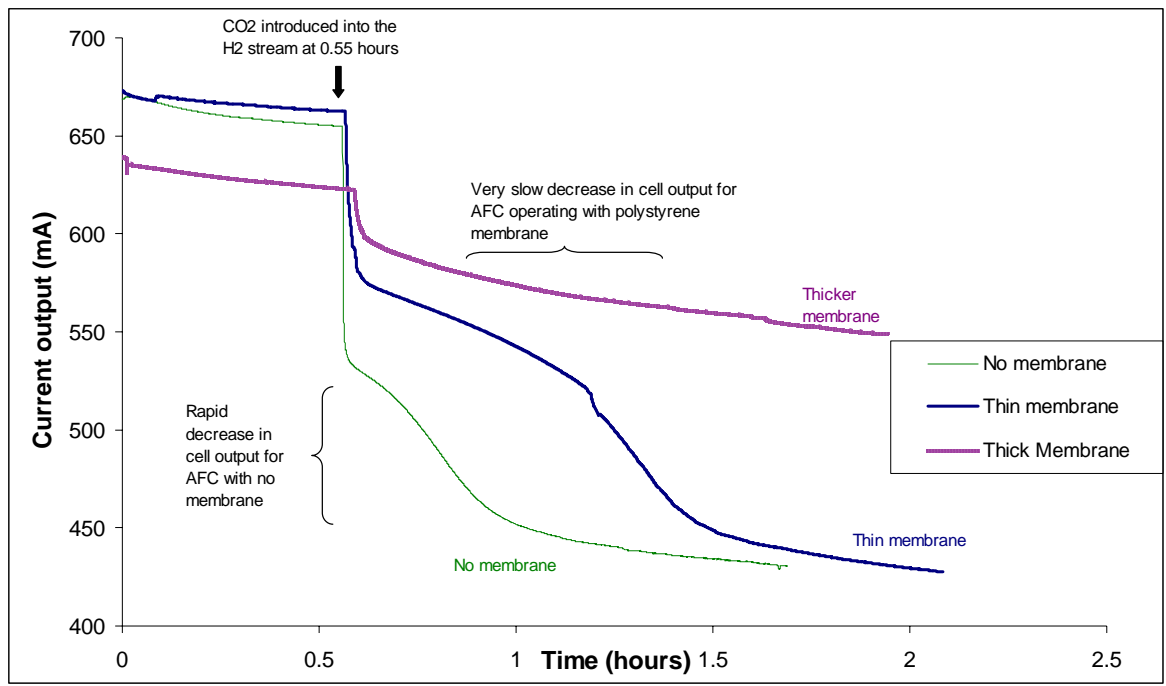

**Figure 1.5.1.** Comparison of the current output vs. time for an alkaline fuel cell with and without membranes operating on 5 % $CO_2$ contaminated "dirty" hydrogen as fuel

Figure 1.5.1 shows that the fuel cell operating with the polystyrene membrane on the anode "dirty" fuel side performed better than the cell without the membrane. After 2 hours of operation, a cell without a membrane has a current output of 440 mA across a 1 Ω load while a current of 560 mA is measured when a thick membrane is used to separate the $H_2$ from the "dirty" fuel contaminated with $CO_2$. The current is obtained in a cell with an active electrode area of 50 $cm^2$. In both cases, the initial current decrease when we introduce $CO_2$ into the $H_2$ stream and the gradual current decrease in cell output due to the poisoning of the electrolyte were significantly decreased by introducing the polystyrene membrane in the dirty $H_2$ fuel stream.

In our work on carbon dioxide poisoning studies on air breathing methanol fuel cells, we defined two variable parameters, $t_{max}$ and $R_{max}$, which accurately describe the onset and magnitude of poisoning in AFCs under different operating conditions [4]. The time at which the current decay rate was highest is referred to as $t_{max}$ and the value of the current decay rate (dI/dt) at this time is $R_{max}$. A shorter $t_{max}$ and a larger magnitude of $R_{max}$ correspond to a faster poisoning effect. These characteristic quantities for the AFC undergoing poisoning in dirty hydrogen are graphically represented in Figure 1.5.2.

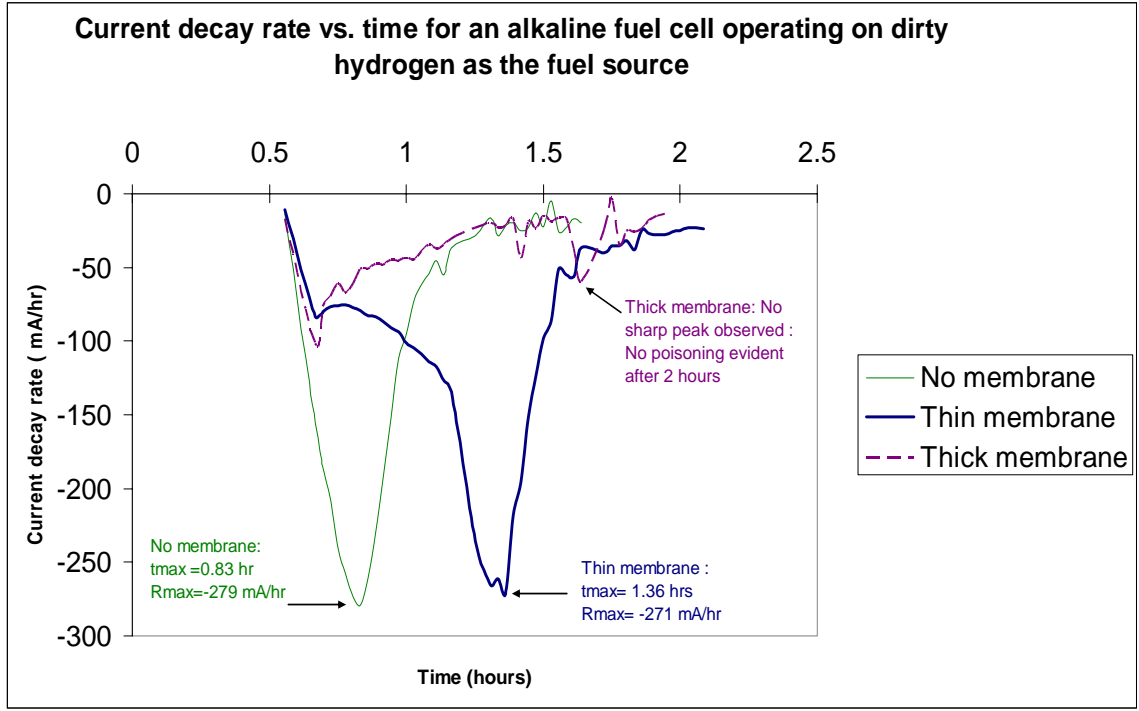

**Figure 1.5.2:** Current decay rate of an alkaline fuel cell with and without membranes operating on 5 % $CO_2$ contaminated "dirty" hydrogen as fuel. Only data points after 0.6 hours (100 seconds after injection of the $CO_2$ contaminant) are included in the plot above.

As can be seen in Figure 1.5.2, $R_{max}$ and $t_{max}$ values for the cell with a membrane indicate slower poisoning as compared to a cell with no membrane. The time of maximum decay, i.e. $t_{max}$, is shifted from 0.83 hours for an alkaline fuel cell without a membrane to 1.36 hours for a cell with the 1 coat polystyrene membrane. The current decay plot for the fuel cell with the 2 thicker coats of polystyrene essentially shows no sharp peaks corresponding to $t_{max}$ and $R_{max}$, thereby indicating that the poisoning process is very slow.

## Effect of Inert gas on Fuel Cell output

While operating the AFC, we observed a sharp instantaneous drop in cell current output not related to the alkaline electrolyte poisoning occurring as soon as $CO_2$ was introduced into the $H_2$ stream. This drop is reversible and the current returns to its initial value as soon as the $CO_2$ is switched off from the $H_2$ stream (Figure 1.5.3). This effect is also observed when nitrogen is introduced into the $H_2$ stream in place of $CO_2$ (Figure 1.5.4).

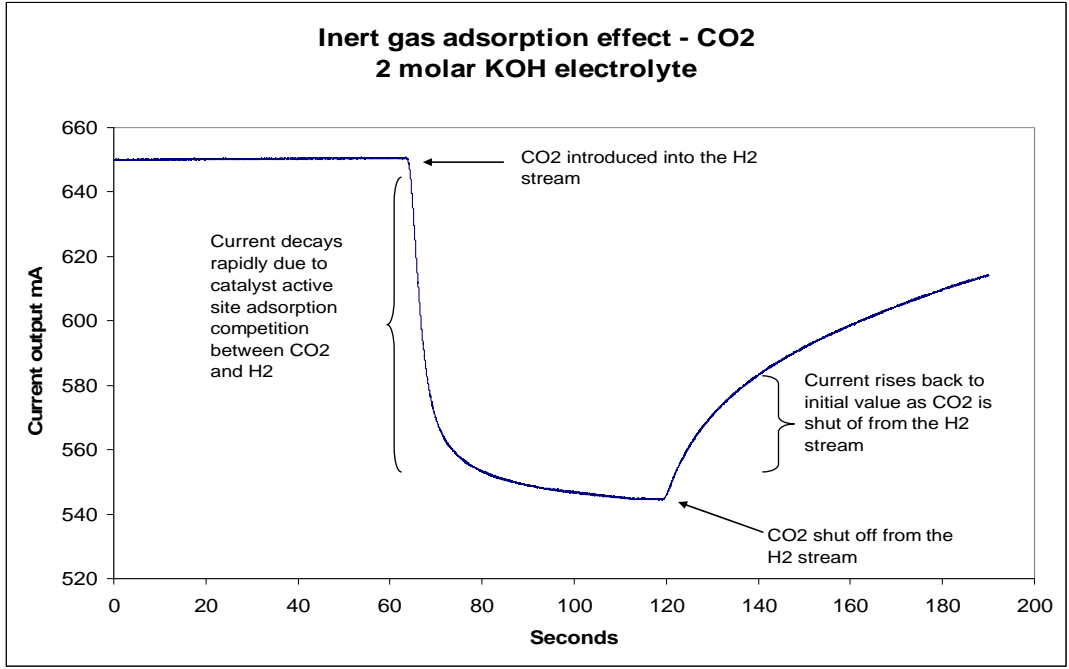

**Figure 1.5.3:** Reversible instantaneous drop in cell current on the introduction of $CO_2$ into the $H_2$ fuel stream for an alkaline fuel cell with 2 M KOH electrolyte.

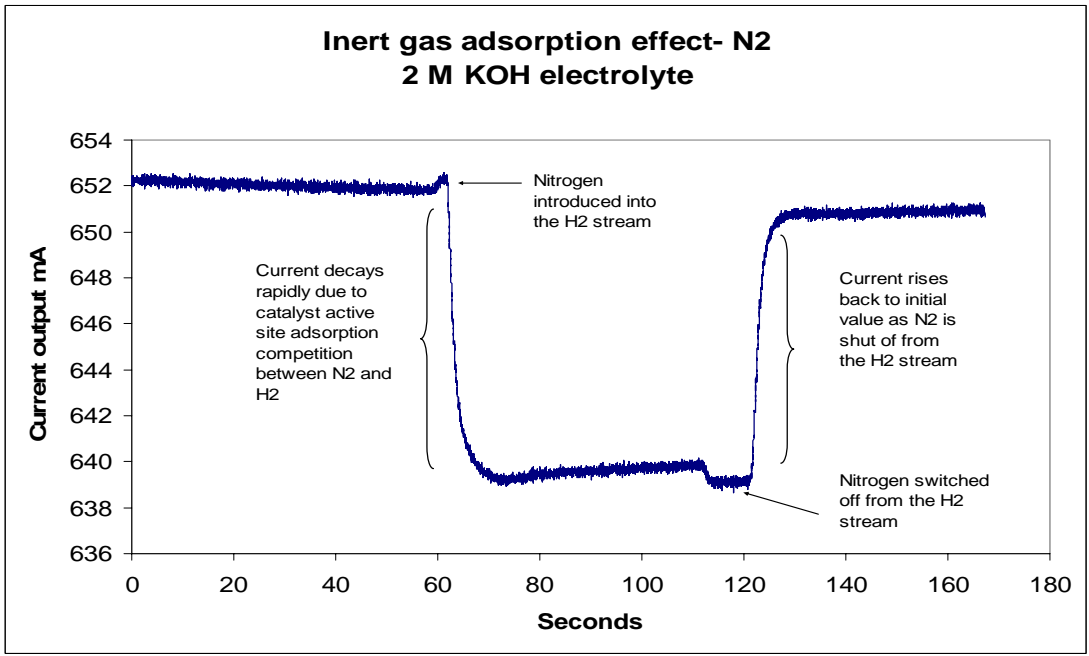

**Figure 1.5.4:** Reversible instantaneous drop in cell current on the introduction of $N_2$ into the $H_2$ fuel stream for an alkaline fuel cell with 2 M KOH electrolyte.

This sharp drop is unlikely to be due to electrolyte poisoning, i.e. electrolyte conversion from KOH to $K_2CO_3$, since it is almost instantaneous (within ~ 15 seconds of injecting $CO_2$) and the electrolyte has had no time to react with the $CO_2$. Hence, this effect should be present irrespective of whether the electrolyte is pure KOH (zero poisoning) or pure $K_2CO_3$ (100% poisoning). To prove this point, fuel cells were run with pure 1 M $K_2CO_3$ as the electrolyte with $CO_2$ (Figure 1.5.5) and $N_2$ (Figure 1.5.6) as the contaminants in the $H_2$ fuel stream. Sharp instantaneous drops in cell current outputs were observed in both cases as soon as the contaminant gas was introduced into the $H_2$ fuel stream. The cell output decrease in both cases was reversible and the current returned to its initial value as soon as the contaminant source was shut off.

This current decrease can be attributed to competition for the catalyst's active site between the fuel, i.e. $H_2$ and the inert gas such as $CO_2$ or $N_2$. In the case of $CO_2$, two additional factors result in this initial decrease to be larger in magnitude than that of nitrogen. First, $CO_2$ is more compressible and has a higher boiling point than nitrogen. This would result in a stronger adsorption of $CO_2$ at the catalytic active sites than $N_2$. Secondly, a reverse water gas shift reaction occurs at the catalyst active sites upon $CO_2$ adsorption which results in transient formation of CO. This CO formation poisons the catalytic active sites and interferes with the normal electrochemical reduction of hydrogen.

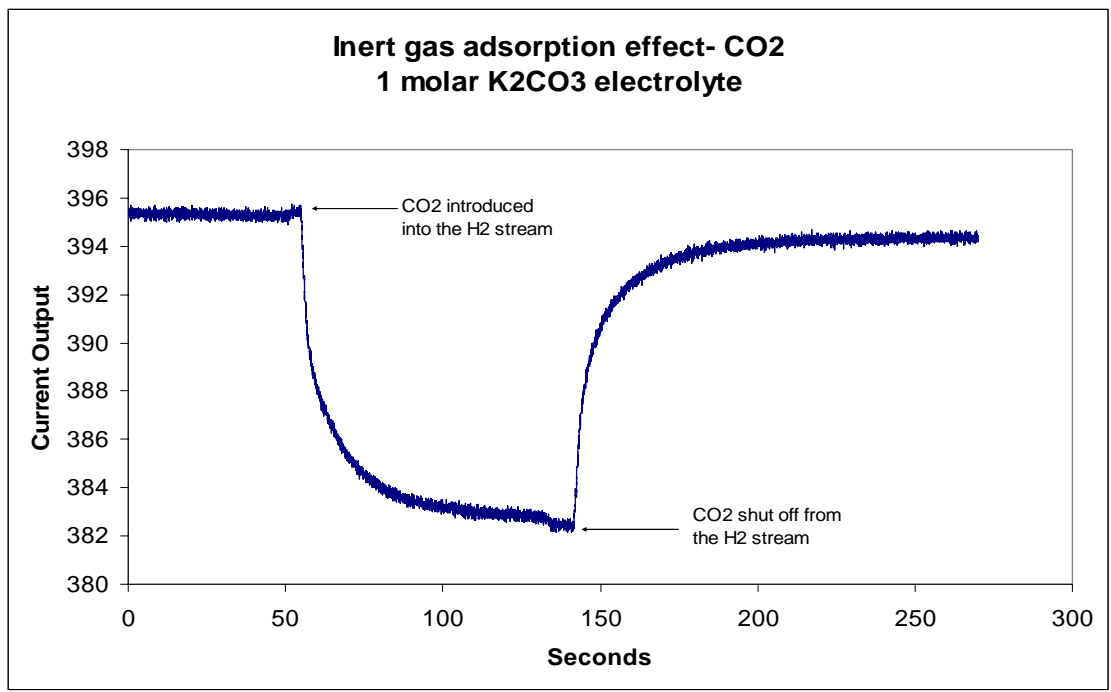

**Figure 1.5.5:** Reversible instantaneous drop in cell current on the introduction of $CO_2$ into the $H_2$ fuel stream for an alkaline fuel cell with 1 M $K_2CO_3$ electrolyte.

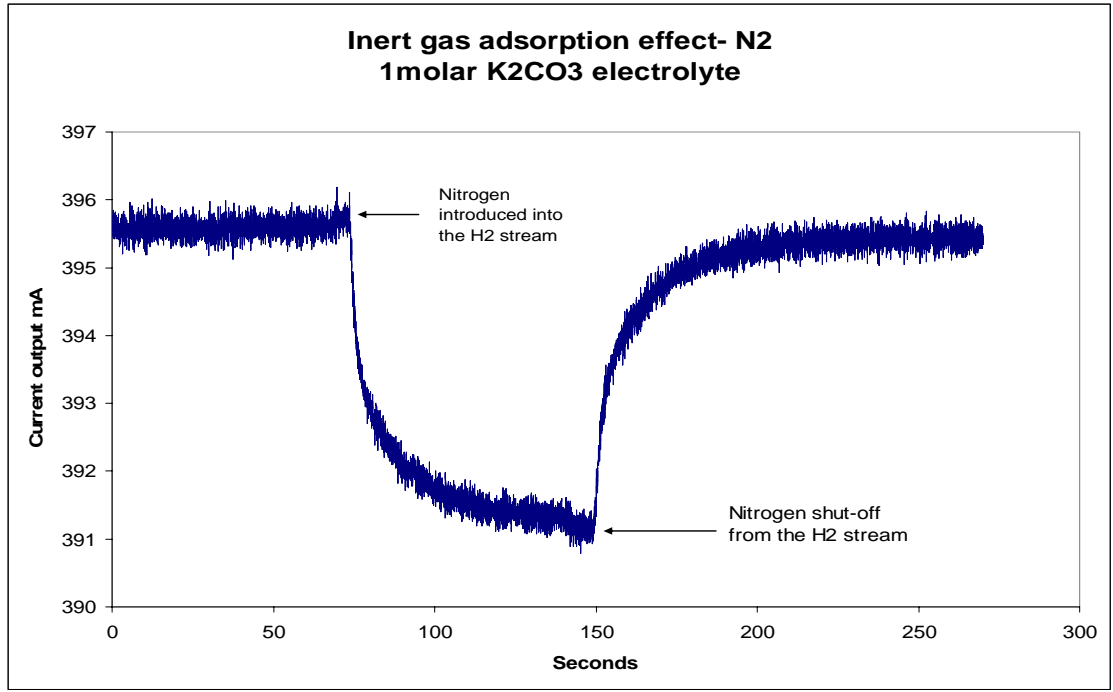

**Figure 1.5.6**: Reversible instantaneous drop in cell current on the introduction of $N_2$ into the $H_2$ fuel stream for an alkaline fuel cell with 1 M $K_2CO_3$ electrolyte.

## References

1. Kordesch K., Gunter S., *Fuel cells and their applications*, Wiley-VCH, Berlin, Germany (1996).

2. K.V., Kordesch, Outlook for Alkaline Fuel Cell Batteries, From Electrocatalysis to Fuel Cells, Seattle, WA, 1972, pp.157.

3. "The influence of carbon dioxide on PEM fuel cell anodes", F.A. de Bruijn, D.C. Papageorgopoulos, E.F. Sitters, G.J.M. Janssen, Journal of Power Sources, (2002), **110**, 117-124

4. "Quantification of carbon dioxide poisoning in air breathing alkaline fuel cells", A. Tewari, V. Sambhy, A. Sen, M.U. Macdonald, Journal of Power Sources, (2005), accepted for publication

# Chapter 2. Conclusions and Recommendation

## Part 1. Membrane Permeability Studies and CVVP Drawbacks

After testing a number of polymer membranes blended with different oxygen-binding complexes, we came to the conclusion that the CVVP Gas Permeation Apparatus is not an ideal instrument for studying facilitated transport across membranes due to the large pressure difference between the two sides of the membrane( >15 psi). The total permeability of a gas through a polymer membrane is the sum of its physical diffusion (1$^{st}$ term of the equation 1), and its chemical diffusion (2nd term of the equation 1).

$$P_{O2} = K_D D_D + D_C C_C K / (1 + K\ pO_2) \qquad (1)$$

Where, $P_{O2}$ is the total oxygen permeability, $K_D$ is the solubility of oxygen (ln $K_D$ = M + 0.016 $T_c$, where M is the parameter depending on the polymer (free volume, interactions) and $T_c$ is the critical temperature of the gas), $D_D$ equals the physical diffusion coefficient ($D_D$ = A exp(-$E_a$/RT), where A is a constant, $E_a$ is the activation energy, R is the gas constant, and T equals the temperature), $D_C$ is the chemical diffusion coefficient, $C_C$ is the saturated amount of specifically absorbed oxygen, K is the oxygen binding equilibrium constant, and $pO_2$ is the partial pressure of oxygen in the upstream side.

A large pressure difference between the two sides of the membrane results in a large physical diffusion of the gases across the membrane (1$^{st}$ term of the equation), which might swamp the effect of the selective facilitated transport of gases (2nd term of the equation). Hence, selectivity for most of the membranes studied came out to be around 1, indicating that the membranes were not as effective in keeping out carbon dioxide under the test conditions. To study the permeability of the liquid membrane systems we developed, we fabricated a gas permeation apparatus coupled to a Gas Chromatograph or an electrochemical oxygen sensor which will not have such a large pressure difference between the inlet and the permeate sides of the liquid membrane, hence, allowing us to study selectivity in liquid membrane systems. A schematic diagram of such a system is given in Figure2.1.1.

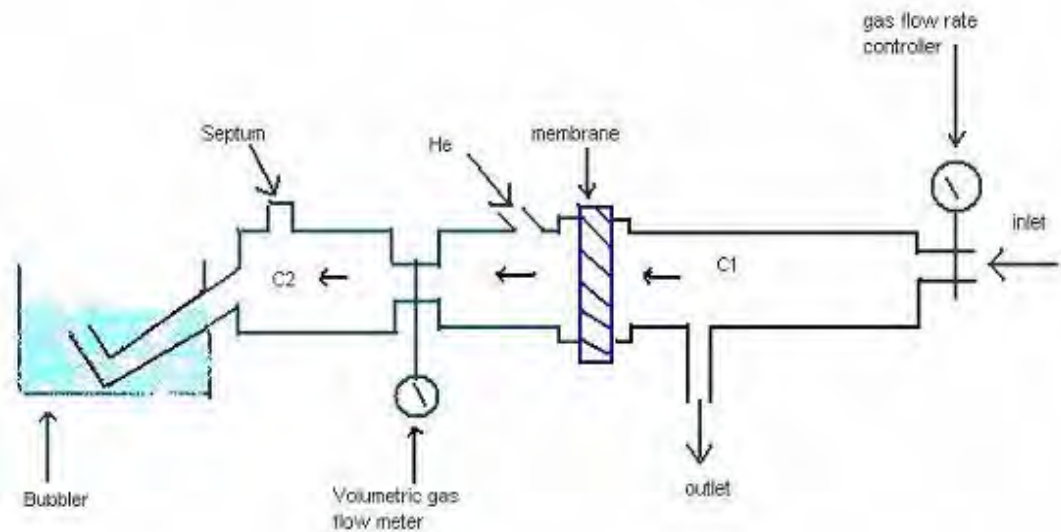

**Figure 2.1.1:** Schematic representation of the apparatus designed to measure membrane permeability.

The electrochemical testing of the fuel cell demonstrated the advantages of using membranes on the upstream flux of air to the air cathode, and the possibility of using air instead of pure oxygen as an oxidant in the alkaline fuel cell in order to extend their operational life and lower their maintenance.

## Part 2. Conclusions on the quantification study of the $CO_2$ poisoning in alkaline fuel cells (fueled by methanol) under accelerated poisoning conditions.

Quantification of carbon dioxide poisoning in Alkaline Fuel Cells is of great importance because it enables us to predict the current output of a fuel cell running in a carbon dioxide enriched atmosphere, when the operating conditions are specified. In this study, a novel method to study and quantify the carbon dioxide poisoning in AFCs, with non-circulating electrolyte, under accelerated poisoning conditions was developed.

Two important parameters, i.e. $t_{max}$ and $R_{max}$, describing the poisoning affect under accelerated testing conditions, were defined. A lower value of $t_{max}$ or a higher value of $R_{max}$ would signify a higher rate of carbon dioxide poisoning. With the knowledge of these quantities, carbon dioxide poisoning can be defined, provided the other operating parameters, viz. temperature, electrolyte concentrations etc., remain the same. Therefore, the poisoning was studied and quantified on the basis of $t_{max}$ and $R_{max}$. $t_{max}$ (time at which rate of current decay is maximum) was found to decrease with the increase in concentration of carbon dioxide in the atmosphere and increase with the applied load. On the contrary, $R_{max}$ (maximum current decay rate) was found to decrease with the applied load and increase with the concentration of carbon dioxide. At very low concentrations (< 2%) $t_{max}$ alone defined the carbon dioxide poisoning as $R_{max}$ became a constant. For the same reason, $R_{max}$ represented the poisoning completely at high carbon dioxide concentrations (> 60%). Lower applied loads were found to expedite the poisoning in AFCs. The reasons attributed to this are higher carbon dioxide uptake by the fuel cell and larger production of carbon dioxide by the anodic oxidation of methanol. A detailed study of the contribution of methanol oxidation to the total carbon dioxide poisoning for the loads and gas mixture percentile used in the experiment showed that the amount of internally generated $CO_2$ was from 6% to 14% to the total $CO_2$ (coming from the air cathode (reach mixture of $O_2/CO_2$) and coming from the methanol reactions internally).

The effect of electrolyte composition on the current output of the AFC was also studied. It was found that a significant decrease in the cell current starts to occur only after about 60% of KOH has been converted to $K_2CO_3$. This yields important practical information about the time to replenish the electrolyte in the AFC such that no change in power output takes place.

Also the potentiostatic polarization study of the cathode and anode in the presence of KOH (aq) and $K_2CO_3$ (aq) electrolytes was carried out. It indicated that the primary reason for decreases in the cell current during poisoning was sluggish methanol oxidation kinetics at the anode in the presence of carbonate. Better catalyst systems which are tolerant to carbonate as the electrolyte may, in future, allow the use of AFCs in air by reducing the carbon dioxide poisoning problem.

Of great importance will be the evaluation of methanol oxidation on different catalyst surfaces in order to increase and speed up the rate of the electrochemical oxidation of methanol and to decrease and slow down the chemical oxidation of methanol that occurs as a parasitic reaction at the cathode (cross over).

## Part 3. Conclusion on Predictive Capability of the Model

Our model predicts the variation of the cell current with time when the operating conditions are specified. Current output of a cell under such accelerated poisoning condition depends on various operating parameters. Following is the list of such operating parameters that affect a fuel cell's performance in our tests:

1) Atmospheric carbon dioxide concentration.
2) Load applied across the fuel cells electrodes.
3) Operating temperature.
4) Electrolyte (KOH) concentration.
5) Methanol concentration in the electrolyte.

For the sake of simplicity, we modeled the time dependence of cell current on the basis of only the first two operating parameters. We showed that this simple model gives accurate results for the alkaline cells fueled with methanol. It is also belived that a similar phenomenological model can be developed for a fuel cell operating with $H_2$ fuel. It is because $H_2$ fuel cells show a similar "sigmoid" current decay curve as observed in methanol fueled alkaline cells.

Therefore, we recommend extending the model to account for the cell performances when the alkaline fuel cell is being fueled with $H_2$ and the $CO_2$ contamination occurring from the anode side. More operating parameters (as listed above) can be taken into account in the modified model. Thus, to deal with the possible complexity of the model (more input variables), we intend to develop a neural network based model.

## Part 4. Air Cathode Membranes Electrochemical Testing

We recommend in a future work to first evaluate the membranes independently of the cell. After selecting the most promising one among the newly designed membranes, we would test them in conjugation with the fuel cells. Contaminations from both cathode and anode stream would be taken in account.

We recommend to extend our membrane work to a more exhaustive and systematic work. There are several parameters and variables in the designing and making of the membranes that were not investigated in this research program due to the time and budget limitations.

## Part 5. $CO_2$ Poisoning AFC when Fuelled with "Dirty" Hydrogen.

To separate $H_2$ from a mixture of $CO_2$ and $H_2$ is less challenging than to separate $O_2$ and $CO_2$. The reason being, there is a large size difference between the $H_2$ and the $CO_2$ molecule, which is marginal between $O_2$ and $CO_2$ molecules.

It is evident that interfacing "off the shelf" polystyrene membranes which have intrinsic gas selectivity towards $H_2$ significantly improved the performance of the cell running in "dirty" hydrogen. It should be emphasized that the results presented here are preliminary and significant further improvement is possible. Additional membrane optimizations like thickness, geometry and use of more selective polymers are being done and should yield significantly better results. In addition, ceramic based nanoporous membranes can also be fabricated and tested for their efficiencies in separating $H_2$ from $CO_2$.